\documentclass[10pt]{article}
\usepackage{amsfonts}
\usepackage{amsmath}
\DeclareMathSymbol{\mlq}{\mathord}{operators}{``}
\DeclareMathSymbol{\mrq}{\mathord}{operators}{`'}
\usepackage{amssymb}
\usepackage[T1]{fontenc}
\usepackage{cancel}
\usepackage[all]{xy}
\usepackage{mathalfa}
\usepackage[mathscr]{euscript}
\usepackage{xcolor}
\usepackage{setspace}
\numberwithin{equation}{section}
\newtheorem{Thm}{Theorem}[section]
\newtheorem{Cor}{Corollary}[section]
\newtheorem{Def}{Definition}[section]
\newtheorem{Lemma}{Lemma}[section]
\newtheorem{Prop}{Proposition}[section]
\newtheorem{Remark}{Remark}[section]
\newtheorem{Example}{Example}[section]
\def\N{\mathbb{N}}
\def\Z{\mathbb{Z}}

\def\C{\mathbb{C}}
\def\K{\mathscr{K}}

\def\P{\mathbb{P}}

\def\z{\mathbf{z}}

\def\fd{\hfill$\square$}
\newcommand{\prs}[2]{\mathop{\langle#1,#2\rangle}}
\date{}
\begin{document}
\title
{\emph{WDVV solutions associated with  the genus one holomorphic differential}}
\author
{{\small\bf Chaabane REJEB}\footnote{Universit\'{e} de Sherbrooke, CANADA.
Email: chaabane.rejeb@usherbrooke.ca, chaabane.rejeb@gmail.com}}
\maketitle

\begin{abstract}
Consider the  Hurwitz space $\mathcal{H}_1(n_0,\dots,n_m)$  of genus one ramified coverings  of fixed degree with $m+1$ prescribed poles of  order $n_0+1,\dots,n_m+1$, respectively.  Based on a recent  formula proved in \cite{Rejeb23}, we derive  an explicit solution to the WDVV equations associated with the Dubrovin-Frobenius manifold structure on $\mathcal{H}_1(n_0,\dots,n_m)$ induced by the normalized holomorphic differential $\phi$. The resulting  solution is a function of  $2+2m+\sum_{j=0}^mn_j$ variables and  is written in terms of Bell polynomials, Eisenstein series  as well as  Weierstrass functions. In addition, by means of flat coordinates related to $\phi$, we construct Landau–Ginzburg superpotentials for these Frobenius manifold structures.

\bigskip\noindent MSC (2020) primary: 53D45, 35C05; secondary: 35G20, 58J60.

\end{abstract}

\tableofcontents

\section{Introduction}
In this paper, our focus lies on providing  explicit examples of genus one   solutions to the  Witten-Dijkgraaf-E.Verlinde-H.Verlinde (WDVV) associativity  equations \cite{DVV, Witten}:
\begin{equation}\label{WDVV}
F_{\alpha}F_1^{-1}F_{\beta}=F_{\beta}F_1^{-1}F_{\alpha}, \quad\quad \alpha,\beta=1,\dots,N,
\end{equation}
where $F=F(t^1,\dots,t^N)$ and $F_{\alpha}$ denotes the Hessian  matrix of $\partial_{t^{\alpha}}F$ for each $\alpha$ with the requirement  that the matrix  $F_1$ is constant and nondegenerate.  The Geometric framework behind the  WDVV equations has been  developed  by Dubrovin, culminating in the remarkable theory of Frobenius manifolds \cite{Dubrovin2D}.\\
A  Hurwitz space  $\mathcal{M}=\mathcal{H}_g(n_0,\dots,n_m)$ of dimension $N$ in genus $g$ is the moduli space of branched coverings of genus $g$ of the Riemann sphere $\P^1$ with  $N$ finite simple branch points and  a prescribed ramification profile over the point at infinity of fixed branching degrees $n_0+1,\dots,n_m+1$. The $N$ simple branch points can be used as natural local parameters on the complex manifold $\mathcal{M}$. In \cite{Dubrovin2D}, Dubrovin constructed a family of $N$ semi-simple Frobenius manifold  structures on $\mathcal{M}$, each of them is induced by a specific choice of particular differential $\omega_0$ (called primary) defined on the underlying Riemann surfaces. The prepotential $\mathbf{F}_{\omega_0}$ of the obtained  Hurwitz-Frobenius manifold determined by $\omega_0$ arises as an example of quasi-homogeneous solutions to the WDVV equations (\ref{WDVV}). Nevertheless, calculating  the explicit form of WDVV solutions $\big\{\mathbf{F}_{\omega_0}\big\}$  presents a notably intricate task.\\
This work aims to provide an answer to this question in the special case of the Hurwitz space  $\mathcal{H}_1(n_0,\dots,n_m)$  in genus one and where the primary differential $\omega_0=\phi$ is the normalized holomorphic differential, but   without any restriction on the combinatorial parameters $(n_0,\dots,n_m)$. The expression for the obtained WDVV prepotential $\mathbf{F}_{\phi}$ associated with the holomorphic primary differential $\phi$ is stated in Theorem \ref{Main result} below.\\
Before moving on to  the  ideas of our approach, let us review the known results in this direction. The first explicit example of a genus one prepotential $\mathbf{F}_{\phi}$  was obtained by Dubrovin \cite{Dubrovin2D} in the particular case where the Hurwitz space is  that of the Weierstrass elliptic curves. Moreover,  Bertola realized  the $\phi$-Frobenius manifold structure on the Hurwitz space $\mathcal{H}_1(n)$ as that on the orbit space of Jacobi group of type $A_n$ and computed for $n=2,3,4$ the corresponding WDVV solution \cite{Bertola1, Bertola2}. Other explicit examples are found in \cite{Miguel} and arose from the four dimensional  Hurwitz-Frobenius manifold structure and its deformation  on the space $\mathcal{H}_1(0,0)$. The method employed to calculate the mentioned examples uses a Dubrovin bilinear pairing formula written in terms of objects defined on the underlying Riemann surfaces. We refer to Lecture 5 in \cite{Dubrovin2D} (and \cite{Vasilisa}) for details. \\
Recently, the author  established in \cite{Rejeb23} a new alternative formula to calculate the WDVV prepotential  of any Hurwitz-Frobenius manifold of dimension $N$ in genus $g$. The structure of this formula is based on a precise duality relation between flat coordinates and  on the degrees of the Dubrovin primary differentials. The concept of the degree of a primary differential $\omega_0$ arises in a natural way  as the power of $\lambda$ (where $\lambda$ defines a  branched covering of the Riemann sphere) in the integral representation formula of $\omega_0$ with respect to the canonical symmetric bidifferential $W(P,Q)$. Here we would like to point out that, as it was observed in \cite{Vasilisa} (see also \cite{Rejeb23}), the symmetric bidifferential $W(P,Q)$ offers a new viewpoint to redescribe the ingredients of the Frobenius manifold structure on the Hurwitz space $\mathcal{M}$ in genus $g$.
The  Hurwitz-Frobenius manifolds  in genus one and their ingredients, including  the genus one bidifferential $W(P,Q)$, primary differentials and  the corresponding  prepotentials $\big\{\mathbf{F}_{\omega_0}\big\}$ are presented in the next section. The new  method  for  calculating  the WDVV solutions $\mathbf{F}_{\omega_0}$ is given by formula (\ref{Prepotential}) below.\\
The first key element  of our approach  to obtaining  the explicit expression for the prepotential $\mathbf{F}_{\phi}$ is to rewrite the new formula (\ref{Prepotential})  in the special case of the holomorphic differential which is of degree $0$. This leads to formula (\ref{Prep-phi1}) below which shows that only  few terms are unknown and (most of them) occur as the first coefficients of the Laurent expansions of some meromorphic differentials near the poles of a  genus one branched covering.\\
The calculation of these terms using  flat coordinates associated with the holomorphic differential $\phi$ requires several technical steps and  their expressions involve Weierstrass functions, Bell polynomials and Eisenstein series. In particular, we construct   Landau-Ginzburg (LG) superpotentials of the Hurwitz space $\mathcal{H}_1(n_0,\dots,n_m)$.

\medskip

This paper is organized as follows. In the next section we collect the properties of the main ingredients used in this work. This includes  reviews on Bell polynomials, Weierstrass functions and   the Dubrovin-Frobenius manifold structures on  the considered Hurwitz space $\mathcal{H}_1(n_0,\dots,n_m)$. Section 3 is technical, it is devoted to calculating the explicit form of the coefficients of the Laurent expansions of Abelian differentials of the second and third kind whose poles are among those of a fixed covering belonging to the Hurwitz space $\mathcal{H}_1(n_0,\dots,n_m)$.
Section 4 focuses on the main result of this work and some applications. The last section deals with the WDVV solutions associated with the deformed Frobenius-Hurwitz  manifold determined by the so-called holomorphic $q$-differential, with $q$ being a fixed complex parameter.

\section{Preliminaries}
\subsection{Bell polynomials}
This subsection deals with some properties of partial ordinary Bell polynomials that will be repeatedly used in this paper. The reader is referred to \cite{Bell,Chara, Comtet} for details. \\
Let $n\geq k\geq0$ be two  integers. The partial ordinary  Bell polynomial $\mathcal{B}_{n,k}$ is  defined by \cite{Bell,Chara, Comtet}
\begin{equation}\label{Bell-partial}
\mathcal{B}_{n,k}:=\mathcal{B}_{n,k}(x_1,\dots,x_{n-k+1})
:=k!\sum_{\bigtriangleup(n,k)}\bigg(\prod_{\mu=1}^{n-k+1}\frac{x_{\mu}^{j_{\mu}}}{j_{\mu}!}\bigg),
\end{equation}
where the summation is over the set $\bigtriangleup(n,k)$ of all partitions of $n$ into $k$ parts, that is, over  all nonnegative integers $j_1,\dots,j_{n-k+1}$ satisfying
\begin{equation*}
\begin{split}
&j_1+\dots+j_{n-k+1}=k;\\
&j_1+2j_2+\dots+(n-k+1)j_{n-k+1}=n.
\end{split}
\end{equation*}
The partial ordinary  Bell polynomials have the generating function:
\begin{equation}\label{Bell2}
\bigg(\sum_{j=1}^{\infty}x_jt^j\bigg)^k=\sum_{n=k}^{\infty}\mathcal{B}_{n,k}(x_1,\dots,x_{n-k+1})t^n.
\end{equation}
In particular, we have the following special cases of partial ordinary Bell polynomials:
\begin{equation}\label{Bell-Examples}
 \forall\ n\geq 1,\quad \mathcal{B}_{0,0}=1;\quad \quad \mathcal{B}_{n,0}=0;\quad \quad \mathcal{B}_{n,1}=x_n;\quad \quad \mathcal{B}_{n,n}=x_1^n;\quad \quad \mathcal{B}_{n,2}=\sum_{j=1}^{n-1}x_jx_{n-j}.
\end{equation}
Moreover, (\ref{Bell2}) implies that partial ordinary Bell polynomials enjoy the following recursion relation:
\begin{equation}\label{Bell-rec0}
\mathcal{B}_{n+k,k}(x_1,\dots,x_{n-k+1})=\sum_{j=0}^{\min(n,k)}\binom{k}{j}x_1^{k-j}\mathcal{B}_{n,j}(x_2,\dots,x_{n-j+2}).
\end{equation}
The partial ordinary Bell polynomials are closely related to the (better-known) partial exponential Bell polynomials $\mathbf{B}_{n,k}$:
\begin{equation}\label{Bell-OE}
\mathbf{B}_{n,k}(x_1,\dots,x_{n-k+1})=\frac{n!}{k!}\mathcal{B}_{n,k}\Big(\frac{x_1}{1!},\dots,\frac{x_{n-k+1}}{(n-k+1)!}\Big).
\end{equation}
The nth (exponential) complete Bell polynomial is defined by  $\mathbf{B}_0=1$ and
\begin{equation}\label{Bell-complete}
\mathbf{B}_n(x_1,\dots,x_n)=\sum_{k=1}^n\mathbf{B}_{n,k}(x_1,\dots,x_{n-k+1})=\sum_{k=1}^n\frac{n!}{k!}\mathcal{B}_{n,k}\Big(\frac{x_1}{1!},\dots,\frac{x_{n-k+1}}{(n-k+1)!}\Big).
\end{equation}
The partial ordinary (and exponential) Bell polynomials $\mathcal{B}_{n,k}$ ($\mathbf{B}_{n,k}$) are homogeneous of degree $k$:
\begin{equation}\label{Bell-homog}
\mathcal{B}_{n,k}(cx_1,\dots,cx_{n-k+1})=c^k\mathcal{B}_{n,k}(x_1,\dots,x_{n-k+1}), \quad \quad \forall\ c.
\end{equation}
On the other hand, according to formulas (11.11) and (11.12) in \cite{Chara} (see also \cite{Comtet} (Chapter 3)), it is known that the partial exponential  Bell  polynomials satisfy the recurrence relations:
\begin{align*}
&\mathbf{B}_{n+1,k+1}(x_1,\dots,x_{n-k+1})=\sum_{j=k}^{n}\binom{n}{j}x_{n+1-j}\mathbf{B}_{j,k}(x_1,\dots,x_{j-k+1});\\
&\mathbf{B}_{n+1,k+1}(x_1,\dots,x_{n-k+1})=\frac{1}{k+1}\sum_{j=k}^{n}\binom{n+1}{j}x_{n+1-j}\mathbf{B}_{j,k}(x_1,\dots,x_{j-k+1}).
\end{align*}
By (\ref{Bell-OE}), we easily see that the analogues of these formulas for the partial ordinary Bell polynomials take the following form:
\begin{align}
&\label{Bell-rec1}\mathcal{B}_{n+1,k+1}(x_1,\dots,x_{n-k+1})=\frac{k+1}{n+1}\sum_{j=k}^n(n+1-j)x_{n+1-j}\mathcal{B}_{j,k}(x_1,\dots,x_{j-k+1});\\
&\label{Bell-rec2}\mathcal{B}_{n+1,k+1}(x_1,\dots,x_{n-k+1})=\sum_{j=k}^nx_{n+1-j}\mathcal{B}_{j,k}(x_1,\dots,x_{j-k+1}).
\end{align}
In particular,
\begin{equation}\label{Bell-rec3}
\mathcal{B}_{n+1,k+1}(x_1,\dots,x_{n-k+1})=\frac{k+1}{n+2+k}\sum_{j=k}^n(n+2-j)x_{n+1-j}\mathcal{B}_{j,k}(x_1,\dots,x_{j-k+1}).
\end{equation}

\medskip

In order to make formulas simpler  during the upcoming  sections, we introduce the following rational function involving partial ordinary Bell polynomials.
\begin{Def}
Let $\mu,k$  be two positive integers satisfying $\mu-k+1\neq 0$. Let  $\mathcal{R}_{\mu,k}:=\mathcal{R}_{\mu,k}(x_1,\dots,x_{\mu+1})$ be the  rational function  defined by
\begin{equation}\label{R-function}
\mathcal{R}_{\mu,k}:=\frac{1}{(\mu-k+1)}\sum_{n=0}^{\mu}\left\{\Big((\mu+1-n)x_{\mu+1-n}\Big)\sum_{\ell=0}^n\binom{-k}{\ell}x_1^{-k-\ell}
\mathcal{B}_{n,\ell}(x_2,\dots,x_{n-\ell+2})\right\},
\end{equation}
where $\displaystyle\binom{-k}{\ell}:=(-1)^{\ell}\binom{k+\ell-1}{\ell}$ is the generalized binomial coefficient.
\end{Def}
We have the following remarks:
\begin{enumerate}
\item The function $\mathcal{R}_{\mu,k}$ is polynomial in the variables $x_2,\dots,x_{\mu+1}$.
\item Because of the homogeneity  property (\ref{Bell-homog}) of partial Bell polynomials, we observe that the function $\mathcal{R}_{\mu,k}$ satisfies:
\begin{equation}\label{R-homog}
\forall\ c\neq 0,\quad \quad \mathcal{R}_{\mu,k}(cx_1,\dots,cx_{\mu+1})=c^{1-k}\mathcal{R}_{\mu,k}(x_1,\dots,x_{\mu+1}).
\end{equation}
\end{enumerate}

\begin{Lemma} Let $\mu,k$  be two positive integers satisfying  $\mu-k+1\neq 0$. Then the multivariate rational function (\ref{R-function}) is also given by
\begin{equation}\label{Bell-rec4}
\begin{split}
\mathcal{R}_{\mu,k}&=\sum_{\ell=0}^{\mu-1}\frac{1}{\ell+1}\binom{-k}{\ell}x_1^{-k-\ell}\mathcal{B}_{\mu,\ell+1}(x_2,\dots,x_{\mu-\ell+1}).
\end{split}
\end{equation}
\end{Lemma}
\emph{Proof:} When $\mu=1$, then we easily check that (\ref{Bell-rec4}) is valid. Assume that $\mu\geq 2$. To simplify, let us introduce the following notation
$$
\mathcal{B}_{n,\ell}:=\mathcal{B}_{n,\ell}(y_1,\dots,y_{n-\ell+1}),\quad \quad \text{with} \quad y_j=x_{j+1}.
$$
Then from  expression (\ref{R-function}) for $\mathcal{R}_{\mu,k}$, we can write
\begin{align*}
&(\mu-k+1)x_1^k\mathcal{R}_{\mu,k}\\
&=(\mu+1)y_{\mu}+\sum_{n=1}^{\mu}\sum_{\ell=1}^n\binom{-k}{\ell}(\mu+1-n)y_{\mu-n}y_0^{-\ell}\mathcal{B}_{n,\ell}(y_1,\dots,y_{n-\ell+1})\\
&=(\mu+1)y_{\mu}+y_0\sum_{\ell=1}^{\mu}\binom{-k}{\ell}y_0^{-\ell}\mathcal{B}_{\mu,\ell}
+\sum_{n=1}^{\mu-1}\sum_{\ell=1}^n\binom{-k}{\ell}(\mu+1-n)y_{\mu-n}y_0^{-\ell}\mathcal{B}_{n,\ell}\\
&=(\mu+1)\mathcal{B}_{\mu,1}+\sum_{\ell=0}^{\mu-1}\binom{-k}{\ell+1}y_0^{-\ell}\mathcal{B}_{\mu,\ell+1}
+\sum_{\ell=1}^{\mu-1}\binom{-k}{\ell}y_0^{-\ell}\bigg(\sum_{n=\ell}^{\mu-1}(\mu+1-n)y_{\mu-n}\mathcal{B}_{n,\ell}\bigg)\\
&=(\mu+1-k)\mathcal{B}_{\mu,1}+\sum_{\ell=1}^{\mu-1}\binom{-k}{\ell+1}y_0^{-\ell}\mathcal{B}_{\mu,\ell+1}
+\sum_{\ell=1}^{\mu-1}\binom{-k}{\ell}\frac{\mu+\ell+1}{\ell+1}y_0^{-\ell}\mathcal{B}_{\mu,\ell+1}\\
&=(\mu+1-k)\mathcal{B}_{\mu,1}-\sum_{\ell=1}^{\mu-1}\binom{-k}{\ell}\frac{k+\ell}{\ell+1}y_0^{-\ell}\mathcal{B}_{\mu,\ell+1}
+\sum_{\ell=1}^{\mu-1}\binom{-k}{\ell}\frac{\mu+\ell+1}{\ell+1}y_0^{-\ell}\mathcal{B}_{\mu,\ell+1}\\
&=(\mu+1-k)\sum_{\ell=0}^{\mu-1}\binom{-k}{\ell}\frac{\mathcal{B}_{\mu,\ell+1}}{(\ell+1)y_0^{\ell}},
\end{align*}
where we used the fact that  $\mathcal{B}_{\mu,1}(y_1,\dots,y_{\mu})=y_{\mu}$ in the third equality and applied recurrence relation (\ref{Bell-rec3}) in the forth equality.

\fd

\subsection{Weierstrass functions and Eisenstein series}
In this subsection,  basic properties  of the Weierstrass functions are presented. There exist several texts on the subject of elliptic functions and related topics. For instance, we refer to the monographs \cite{Akhiezer, Apostol, Chandra, Erdelyi, Du Val, Forsyth, Watson}.\\
Let $\tau$ be a complex number with strictly positive imaginary part and $\mathbb{L}$ be the lattice $\mathbb{L}:=\Z+\tau\Z$ generated by $1$ and $\tau$. The non-normalized Eisenstein series  are defined on the upper half-plane by
\begin{equation}\label{Eisenstein G}
G_{2k}=G_{2k}(\tau)=\sum_{w\in \mathbb{L},w\neq0}w^{-2k},\quad \quad \quad k\geq 2,
\end{equation}
and $G_{2k+1}\equiv0$. It is known that $G_{2k}$ is a modular form of weight $2k$, which means that
\begin{equation}\label{G-modular}
G_{2k}\Big(\frac{a\tau+b}{c\tau+d}\Big)=(c\tau+d)^{2k}G_{2k}(\tau),\quad \quad \quad \forall\
\begin{pmatrix}
  a & b \\
  c & d
\end{pmatrix}
\in SL(2,\Z).
\end{equation}
Expression (\ref{Eisenstein G}) can be extended to $k=1$ in the following way:
\begin{equation}\label{G2-def}
G_2=G_2(\tau)=2\sum_{n=1}^{\infty}\frac{1}{n^2}+\sum_{n\in \Z, n\neq 0}\sum_{m\in \Z}\frac{1}{(m+n\tau)^2},
\end{equation}
but then the series is not absolutely convergent. The Eisenstein series $G_2$ is a quasi-modular form of weight 2 \cite{Zagier}, since it satisfies
\begin{equation}\label{G2-modular}
G_{2}\Big(\frac{a\tau+b}{c\tau+d}\Big)=(c\tau+d)^{2}G_{2}(\tau)-{2{\rm i}\pi}c(c\tau+d).
\end{equation}
It is known  that for each integer $k\geq 1$, the Eisenstein series $G_{2k}$ is a  periodic function of period $1$ and  has the following  Fourier expansion (see for instance \cite{Apostol} (Section 3.10) and \cite{Chandra} (Chapter VI)):
\begin{equation}\label{Eisentein-Fourier}
G_{2k}(\tau)=2\sum_{n=1}^{\infty}\frac{1}{n^{2k}}+2\frac{(2{\rm i}\pi)^{2k}}{(2k-1)!}\sum_{n=1}^{\infty}\sigma_{2k-1}(n)e^{2{\rm}in\pi\tau},
\end{equation}
where $\sigma_k(n)$ is the sum of the k-th powers of all the positive divisors of $n$.\\
The $\mathbb{L}$-Weierstrass  functions $\wp$, $\zeta$ and $\sigma$ are respectively defined  by:
\begin{align}
\begin{split}\label{p-function}
&\wp(v)=\wp(v|\tau):=\frac{1}{v^2}+\sum_{w\in \mathbb{L},w\neq0}\Big(\frac{1}{(v-w)^2}-\frac{1}{w^2}\Big);
\end{split}\\
\begin{split}\label{zeta-function}
&\zeta(v)=\zeta(v|\tau):=\frac{1}{v}+\sum_{w\in \mathbb{L},w\neq0}\Big(\frac{1}{v-w}+\frac{1}{w}+\frac{v}{w^2}\Big);\\
\end{split}\\
\begin{split}\label{sigma-function}
&\sigma(v)=\sigma(v|\tau):=v\prod_{w\in \mathbb{L},w\neq0}\big(1-\frac{v}{w}\big)\exp\Big(\frac{v}{w}+\frac{v^2}{2w^2}\Big).
\end{split}
\end{align}
The function $\wp$ is elliptic (doubly periodic) with respect to the lattice $\mathbb{L}$ and the functions $\zeta$ and $\sigma$ are  quasi-elliptic since they satisfy (\ref{zeta-sigma periods}) below. \\
Here is the list of properties of the three functions that will be important in what follows.
\begin{itemize}
\item[1.] Differential equations for the $\wp$-function:
\begin{align}
\begin{split}\label{p-diff equation}
&\wp'^2(v)=4\wp^3(v)-g_2\wp(v)-g_3;
\end{split}\\
\begin{split}\label{pi-function equation}
&\wp''(v)=6\wp^2(v)-\frac{g_2}{2};\\
&\wp'''(v)=12\wp'(v)\wp(v);\\
&\wp^{(4)}(v)=120\wp^3(v)-18g_2\wp(v)-12g_3,
\end{split}
\end{align}
where  $g_2$ and $g_3$ are the Weierstrass invariants given by the  Eisenstein series $G_4$ and $G_6$:
\begin{equation}\label{g2-g3}
g_2=g(\tau):=60G_4,\quad \quad\quad
g_3=g_3(\tau)=140G_6.
\end{equation}
\item[2.] The Weierstrass function $\zeta$ satisfies:
\begin{equation}\label{zeta-derivative}
\zeta'(v):=-\wp(v).
\end{equation}
\item[3.] The function $\zeta$  is the logarithmic derivative of $\sigma$:
\begin{equation}\label{zeta-sigma}
\sigma'(v)/\sigma(v)=\zeta(v).
\end{equation}
\item[4.] We have the following modular properties:
\begin{equation}\label{psz-modular}
\begin{split}
&\sigma\Big(\frac{v}{c\tau+d}\Big|\frac{a\tau+b}{c\tau+d}\Big)=(c\tau+d)^{-1}\sigma(v|\tau);\\
&\zeta\Big(\frac{v}{c\tau+d}\Big|\frac{a\tau+b}{c\tau+d}\Big)=(c\tau+d)\zeta(v|\tau);\\
&\wp\Big(\frac{v}{c\tau+d}\Big|\frac{a\tau+b}{c\tau+d}\Big)=(c\tau+d)^{2}\wp(v|\tau),
\end{split}
\quad \quad\quad  \forall\ \begin{pmatrix}
  a & b \\
  c & d
\end{pmatrix}
\in SL(2,\Z).
\end{equation}
\item[5.] The functions $\zeta$ and $\sigma$ satisfy the following quasi-periodicity properties:
\begin{equation}\label{zeta-sigma periods}
\begin{split}
&\zeta(v+2\omega_j)=\zeta(v)+2\zeta(\omega_j);\\
&\sigma(v+2\omega_j)=-\sigma(v)e^{2\zeta(\omega_j)v+2\omega_j\zeta(\omega_j)},
\end{split}
\quad \quad \quad j=1,2,3,
\end{equation}
where $\omega_1=1/2$, $\omega_2=\tau/2$ and $\omega_3=\omega_1+\omega_2$ are the half periods.
\item[6.] The periods $\omega_1=1/2$, $\omega_2=\tau/2$ and the quasi-periods of $\zeta$ are related by the Legendre relation:
\begin{equation}\label{Legendre}
2\omega_2\zeta(\omega_1)-2\omega_1\zeta(\omega_2)=\tau\zeta(1/2)-\zeta(\tau/2)={\rm{i}}\pi.
\end{equation}
\item[7.] Addition theorems for the functions $\wp$ and $\zeta$:
\begin{align}
&\label{addition-pi}\wp(u+v)=-\wp(u)-\wp(v)+\frac{1}{4}\bigg(\frac{\wp'(u)-\wp'(v)}{\wp(u)-\wp(v)}\bigg)^2;\\
&\label{addition-zeta}\zeta(u+v)=\zeta(u)+\zeta(v)+\frac{\wp'(u)-\wp'(v)}{2\big(\wp(u)-\wp(v)\big)}.
\end{align}
\item[8.]  The Laurent expansion of the Weierstrass functions $\wp$ and $\zeta$ near the origin are given respectively by (see \cite{Apostol} (Theorem 1.11) and \cite{Chandra}):
\begin{align}
&\label{Laurend-p}\wp(v)=\frac{1}{v^2}+\sum_{\ell=1}^{\infty}(2\ell+1)G_{2\ell+2}v^{2\ell}=\frac{1}{v^2}+\sum_{\ell=2}^{\infty}(\ell+1)G_{\ell+2}v^{\ell};\\
&\label{Laurent zeta}\zeta(v)=\frac{1}{v}-\sum_{\ell=1}^{\infty}G_{2\ell+2}v^{2\ell+1}=\frac{1}{v}-\sum_{\ell=3}^{\infty}G_{\ell+1}v^{\ell},
\end{align}
where $G_{2\ell+1}\equiv0$ and $G_{2\ell}$, $\ell\geq2$, are the non-normalized Eisenstein  series (\ref{Eisenstein G}).
Note that equalities (\ref{Laurend-p}) and (\ref{Laurent zeta}) are valid whenever $v$ satisfies
$$
0<|v|<\rho:=\min\big\{|w|:\quad w\in \mathbb{L},w\neq0\big\}.
$$
\item[9.] The  Jacobi (odd) function $\theta_1$ is defined  by
\begin{equation}\label{Jacobi}
\theta_1(v)=\theta_1(v|\tau):=-i\sum_{n\in \Z}(-1)^ne^{{\rm{i}}\pi\tau(n+1/2)^2+2{\rm{i}}\pi(n+1/2)v}.
\end{equation}
Weierstrass functions are related to the Jacobi $\theta_1$-function by (see \cite{Erdelyi}, Section 13.20, and \cite{Akhiezer, Watson}):
\begin{align}
\begin{split}\label{sigma-theta1}
&\sigma(v|\tau)=e^{\zeta(1/2|\tau)v^2}\frac{\theta_1(v|\tau)}{\theta'_1(0|\tau)};
\end{split}\\
\begin{split}\label{zeta-theta1}
&\zeta(v|\tau)=\frac{\sigma'(v|\tau)}{\sigma(v|\tau)}=2\zeta(1/2|\tau)v+\frac{\theta'_1(v|\tau)}{\theta_1(v|\tau)};
\end{split}\\
\begin{split}\label{pi-theta1}
&\wp(v|\tau)=-\zeta'(v|\tau)=-2\zeta(1/2|\tau)-\frac{\theta''_1(v|\tau)}{\theta_1(v|\tau)}+\bigg(\frac{\theta'_1(v|\tau)}{\theta_1(v|\tau)}\bigg)^2.
\end{split}
\end{align}
\end{itemize}
\subsection{Frobenius manifold structures on  Hurwitz spaces in genus one}
\subsubsection{Frobenius manifolds}
\begin{Def}\label{F algebra} A $\C$-algebra $(\mathcal{A},\circ,e)$ supplied with a nondegenerate and symmetric inner product  $\prs{\cdot}{\cdot}:\mathcal{A}\times \mathcal{A}\longrightarrow\C$ is a Frobenius algebra if
\begin{description}
\item[i)] the algebra $(\mathcal{A},\circ,e)$  is associative, commutative with unity $e$;
\item[ii)] the multiplication  $``\circ"$ is compatible  with  $\prs{\cdot}{\cdot}$, in the sense that
$$
\prs{x\circ y}{z}=\prs{x}{y\circ z},\quad \quad \forall\ x,y,z\in \mathcal{A}.
$$
\end{description}
A Frobenius algebra $(\mathcal{A},\circ,e, \prs{\cdot}{\cdot})$ is called semi-simple if it contains no nilpotent element.
\end{Def}
\begin{Def}\label{Frob Man def} Let $\mathcal{M}$ be a complex  manifold of dimension N. A semi-simple Frobenius manifold structure on $\mathcal{M}$ is a data of $\big(\mathcal{M},\circ,e,\prs{\cdot}{\cdot},E\big)$ such that each tangent space $T_t\mathcal{M}$ is a semi-simple Frobenius algebra varying  analytically over $\mathcal{M}$ with the additional properties:
\begin{itemize}
\item[\emph{\textbf{F1)}}] the metric $\prs{\cdot}{\cdot}_t$ on $\mathcal{M}$ is flat  (but not necessarily real and  positive);
\item[\emph{\textbf{F2)}}] the unit vector field $e$ is flat with respect to the Levi-Civita connection $\nabla$ of the flat metric $\prs{\cdot}{\cdot}$, i.e. $\nabla e=0$;
\item[\emph{\textbf{F3)}}] the tensor $\big(\nabla_{v}\mathbf{c}\big)(x,y,z)$ is symmetric in four vector fields $v,x,y,z\in T_t\mathcal{M}$, where  $\mathbf{c}$ is the symmetric 3-tensor
$$
\mathbf{c}(x,y,z):=\prs{x\circ y}{z},\quad \forall\ x,y,z\in T_t\mathcal{M};
$$
\item[\emph{\textbf{F4)}}] the vector field $E$ is covariantly linear, i.e. $\nabla\nabla E=0$, and
\begin{align}
\begin{split}\label{Euler vector def}
&[E,e]=-e,\\
&[E,x\circ y]-[E,x]\circ y-x\circ[E,y]=x\circ y,\\
&\big(Lie_E\prs{\cdot}{\cdot}\big)(x,y):=E\prs{x}{y}-\prs{[E,x]}{y}-\prs{x}{[E,y]}=(2-D)\prs{x}{y},
\end{split}
\end{align}
for some constant $D$, called the charge of the Frobenius manifold. The vector field $E$ is called Euler vector field.
\end{itemize}
\end{Def}

\medskip

From \cite{Dubrovin2D}, we know that each Frobenius manifold $\big(\mathcal{M},\circ,e,\prs{\cdot}{\cdot},E\big)$ gives rise to a quasi-homogeneous solution $F$ to
the WDVV equations (1.1), called the prepotential of the Frobenius manifold. For completeness, let us briefly describe the construction of the  WDVV prepotential $F$.
First, since the metric $\prs{\cdot}{\cdot}$ is flat by \textbf{F1)}, it follows that  locally there exist flat coordinates $\{t^{\alpha}: \ 1\leq\alpha\leq N\}$ such  that the matrix  $(\eta_{\alpha\beta})=\big(\prs{\partial_{t^{\alpha}}}{\partial_{t^{\beta}}}\big)_{\alpha\beta}$ is constant. Second,  the symmetry property of the 4-tensor $\nabla\mathbf{c}$ in the condition \textbf{F3)} implies that  there is a function $F=F(t^1,\dots,t^{N})$ such that its third  derivatives (referred to as the 3-point correlation functions in topological field theory) give the 3-tensor $\mathbf{c}$:
\begin{equation}\label{Introd-prep def}
\partial_{t^{\alpha}}\partial_{t^{\beta}}\partial_{t^{\gamma}}F=\mathbf{c}\big(\partial_{t^{\alpha}},\partial_{t^{\beta}},\partial_{t^{\gamma}}\big).
\end{equation}
In particular, the prepotential  $F$ is unique up to addition of quadratic polynomial functions of $t^1,\dots,t^N$.
Third, due to  the flatness property  of the unit vector field $e$ (i.e. $\nabla e=0$),  by making a linear change of flat coordinates,  the unit field $e$ can be taken of the form $e=\partial_{t^1}$. As a consequence, the entries of the constant Gram matrix of the flat metric $\prs{\cdot}{\cdot}$ satisfy
$$
\eta_{\alpha\beta}:=\prs{\partial_{t^{\alpha}}}{\partial_{t^{\beta}}}=\partial_{t^1}\partial_{t^{\alpha}}\partial_{t^{\beta}}F.
$$
Now observing that the Frobenius algebra multiplication $``\circ" $ is given by
$$
\partial_{t^{\alpha}}\circ \partial_{t^{\beta}}=\eta^{\gamma\varepsilon}\prs{\partial_{t^{\alpha}}\circ\partial_{t^{\beta}}}{\partial_{t^{\varepsilon}}}\partial_{t^{\gamma}}
=\eta^{\gamma\varepsilon}\mathbf{c}\big(\partial_{t^{\alpha}},\partial_{t^{\beta}},\partial_{t^{\varepsilon}}\big)\partial_{t^{\gamma}},\quad \quad \quad \text{with}\quad
(\eta^{\alpha\beta})=\big(\eta_{\alpha\beta}\big)^{-1},
$$
(summation over the repeated indices is assumed) and using (\ref{Introd-prep def}), we conclude that the associativity property of the law $``\circ" $ is equivalent to the WDVV equations (\ref{WDVV}) for $F$.\\
Lastly, let us  deal with  the quasi-homogeneity of the function $F$ which occurs as a consequence of the requirements in (\ref{Euler vector def}).  Indeed, (\ref{Euler vector def}) yields that the Lie derivative, along the Euler vector field $E$, of the symmetric 3-tensor $\mathbf{c}$ satisfies:
\begin{equation}\label{Intro-Lie deriv-c}
\begin{split}
\big(Lie_E.\mathbf{c}\big)(x,y,z)&:=E.\mathbf{c}(x,y,z)-\mathbf{c}\big([E,x],y,z\big)-\mathbf{c}\big(x,[E,y],z\big)-\mathbf{c}\big(x,y,[E,z]\big)\\
&=(3-D)\mathbf{c}(x,y,z).
\end{split}
\end{equation}
Writing $E$ in the form  $E=\sum_{\epsilon}E^{\epsilon}(t)\partial_{t^{\epsilon}}$, with $t=(t^1,\dots,t^N)$ and making use of the requirement  $\nabla\nabla E=0$,  we obtain  $\partial_{t^{\alpha}}\partial_{t^{\beta}}E^{\epsilon}=0$, for all $\alpha,\beta,\epsilon$. From this fact and relations (\ref{Introd-prep def}) and (\ref{Intro-Lie deriv-c}), we arrive at
$$
\big(Lie_E.\mathbf{c}\big)\big(\partial_{t^{\alpha}},\partial_{t^{\beta}},\partial_{t^{\gamma}}\big)=\partial_{t^{\alpha}}\partial_{t^{\beta}}\partial_{t^{\gamma}}E.F=
(3-D)\partial_{t^{\alpha}}\partial_{t^{\beta}}\partial_{t^{\gamma}}F.
$$
Thus the prepotential $F$ is a quasi-homogeneous function of degree $3-D$, up to an addition of quadratic polynomials in $t^1,\dots,t^N$. Here we point out that according to Dubrovin \cite{Dubrovin2D} (Lecture 1),  a generalized quasi-homogeneity property is considered in the context of Frobenius manifolds. More precisely,  a function $f$ is called quasi-homogeneous  of degree $\nu_f$ with respect to the Euler vector field $E=\sum_{\alpha}\big(d_{\alpha}t^{\alpha}+r_{\alpha}\big)\partial_{t^{\alpha}}$ if
\begin{equation}\label{quasihomog-def}
\textstyle E.f=\nu_ff+\sum_{\alpha,\beta}A_{\alpha\beta}t^{\alpha}t^{\beta}+\sum_{\alpha}B_{\alpha}t^{\alpha}+C.
\end{equation}
\subsubsection{Hurwitz spaces in genus one}
Let $m$ and $n_0,\dots,n_m$ be nonnegative integers, with $\sum_{j=0}^m(n_j+1)\geq 2$. The  Hurwitz space $\mathcal{H}_1(n_0,\dots,n_m)$ in genus one  is the moduli space  of equivalence classes of branched coverings $(\C/{\mathbb{L}}, \lambda)$, where
\begin{itemize}
\item[1.] $\C/{\mathbb{L}}$ is the  complex torus corresponding to the lattice $\mathbb{L}=\Z+\tau\Z$, with $\tau$  being a complex number such that $\Im\tau>0$;
\item[2.] $\lambda: \C/{\mathbb{L}}\longrightarrow \P^1$ is a meromorphic function of degree $\sum_{j=0}^m(n_j+1)$ having $m+1$ prescribed poles $\infty^0,\dots,\infty^m \in \C/{\mathbb{L}}$ of order $n_0+1,\dots,n_m+1$, respectively. The meromorphic function $\lambda$ can be identified with an $\mathbb{L}$-elliptic (doubly periodic) function on the complex plane;
\item[3.] The $N$ zeros $P_1,\dots,P_N$ of the differential $d\lambda(P)$ are simple and their finite $\lambda$-images $\lambda_1:=\lambda(P_1),\dots,\lambda_N:=\lambda(P_N)\in \C$ are distinct. The points $P_1,\dots,P_N$ are known as the simple ramifications points of the covering $(\C/{\mathbb{L}}, \lambda)$ and their $\lambda$-projections $\lambda_1,\dots, \lambda_N \in \C$ are called simple  branch points.
\end{itemize}
Two pairs $(\C/{\mathbb{L}}, \lambda)$ and $(\C/{\widetilde{\mathbb{L}}}, \widetilde{\lambda})$ are called equivalent if there is a biholomorphic map
$h: \C/{\mathbb{L}}\longrightarrow \C/{\widetilde{\mathbb{L}}}$ such that $\lambda=\widetilde{\lambda}\circ{h}$.\\
According to the Riemann-Hurwitz formula the number $N$ of finite branch points is determined by means of the number $m+1$ of poles of $\lambda$ as well as their orders $n_0+1, \dots, n_m+1$. More precisely, we have
$$
N=2+2m+\sum_{i=0}^mn_i.
$$
Furthermore, from \cite{Clebsch, Hurwitz, Fulton}, it is known that the already described  Hurwitz space $\mathcal{M}=\mathcal{H}_1(n_0,\dots,n_m)$ is a connected complex manifold of dimension $N$ and the simple and finite branch points of a branched covering from $\mathcal{M}$ serve as a system of local coordinates on $\mathcal{M}$. They also used as a set of canonical coordinates in the construction of the semi-simple Frobenius manifold structures on $\mathcal{M}$ \cite{Dubrovin2D}.
\subsubsection{Genus one bidifferential}
 The bidifferential on the complex torus $\C/{\mathbb{L}}$ is  defined by
\begin{equation}\label{W-g1}
W(P,Q)=\Big(\wp(z_P-z_Q|\tau)+2\zeta(1/2|\tau)\Big)dz_Pdz_Q,
\end{equation}
where $P=z_P+\mathbb{L}, Q=z_Q+\mathbb{L}\in \C/{\mathbb{L}}$, $z_P,z_Q\in \C$ and $\wp, \zeta$ are the Weierstrass functions defined by (\ref{p-function})-(\ref{zeta-function}).
According to the already listed properties of the functions $\wp$ and $\zeta$, we  observe that the  bidifferential $W(P,Q)$ satisfies the following properties:
\begin{itemize}
\item[i)] It is a symmetric with respect to $P$ and $Q$: $W(P,Q)=W(Q,P)$.
\item[ii)] It has a pole of second order on the diagonal with biresidue 1:
\begin{equation}\label{W-asymptotic}
W(P,Q)\underset{P\sim Q}=\left(\big(z_P-z_Q\big)^{-2}+\mathcal{O}(1)\right)dz_Pdz_Q.
\end{equation}
This follows directly from the Laurent expansion (\ref{Laurend-p}) of the function $\wp$ near $0$.
\item[iii)] Using the quasi-periodicity properties (\ref{zeta-sigma periods}) of the  function $\zeta$ as well as the Legendre identity (\ref{Legendre}), we can see that periods of the differential $W(P,\cdot)$, with $P$ fixed,  are given by:
\begin{equation}\label{W-periods}
\oint_{Q\in a}W(P,Q)=0 \quad\quad \text{and}\quad \quad \oint_{Q\in b}W(P,Q)={2{\rm i}\pi}dz_P,
\end{equation}
where $\{a,b\}$ is a canonical homology  basis on $\C/{\mathbb{L}}$ such that the cycles $a$ and $b$ are identified to be the segments $a=[x,x+1]$ and $b=[x,x+\tau]$.
\end{itemize}
In the complex manifold $\mathcal{M}=\mathcal{H}_1(n_0,\dots,n_m)$, the dependence of the bidifferential $W(P,Q)$ on the branch points $\{\lambda_j\}_j$ of a covering $(\C/{\mathbb{L}}, \lambda)$ is specified  by Rauch's variational formulas \cite{Fay92, Rauch, Yamada}:
\begin{equation}\label{Rauch}
\partial_{\lambda_j}W(P,Q)=\frac{1}{2}W(P,P_j)W(P_j,Q).
\end{equation}
Here $W(P,P_j)$ denotes the evaluation of the bidifferential $W(P,Q)$  at $Q=P_j$ with respect to the standard local parameter $\mathbf{x}_j(Q)=\sqrt{\lambda(Q)-\lambda_j}$ near the ramification point $P_j$:
$$
W(P,P_j):=\frac{W(P,Q)}{d\mathbf{x}_j(Q)}\Big|_{Q=P_j}.
$$

\subsubsection{Dubrovin's primary differentials  and the corresponding Frobenius manifold structures}
Let $[(\C/{\mathbb{L}}, \lambda)]$ be a point  of the Hurwitz space $\mathcal{H}_1(n_0,\dots,n_m)$.  The genus one  primary  differentials on the torus $\C/{\mathbb{L}}$ are conveniently expressed  by means  of the  bidifferential $W(P,Q)$ (\ref{W-g1}) as follows \cite{Dubrovin2D, Vasilisa}:
\begin{equation}\label{Primary-e}
\begin{split}
&\textbf{1.}\quad \phi_{\mathbf{t}^{i,\alpha}}(P):=\frac{\sqrt{n_i+1}}{\alpha}\underset{\infty^i}{{\rm res}}\
\lambda(Q)^{\frac{\alpha}{n_i+1}}W(P,Q),\quad
\begin{array}{ll}
&i=0,\dots,m,\\
&\alpha=1,\dots,n_i;
\end{array}\\
&\textbf{2.}\quad \phi_{\mathbf{v}^i}(P):=\displaystyle \underset{\infty^i}{{\rm res}\ }\lambda(Q)W(P,Q), \quad \quad\quad  i=1,\dots,m;\\
&\textbf{3.}\quad \phi_{\mathbf{s}^i}(P):=\displaystyle \int_{\infty^0}^{\infty^i}W(P,Q), \quad \quad \quad \quad i=1,\dots,m;\\
&\textbf{4.}\quad \phi_{\mathbf{r}}(P):=\displaystyle\frac{1}{2{\rm{i}}\pi}\oint_{b}W(P,Q)=dz_P;\\
&\textbf{5.}\quad \phi_{\mathbf{u}}(P):=\displaystyle\oint_{a}\lambda(Q)W(P,Q).
\end{split}
\end{equation}
Here the points $\infty^0,\dots,\infty^m\in \C/{\mathbb{L}}$ are the assigned $m+1$ poles of the covering $\lambda$ and $n_0+1,\dots,n_m+1$ are their orders, respectively and $\{a,b\}$ is  a canonical homology  basis on $\C/{\mathbb{L}}$ with $a=[x,x+1]$ and $b=[x,x+\tau]$.\\
All the  differentials listed in (\ref{Primary-e}) share  the following particular integral representation formula with respect to the bidifferential $W(P,Q)$:
\begin{equation}\label{Int-rep-W}
\omega_0(P)=c_0\int_{\ell_0}\big(\lambda(Q)\big)^{d_0}W(P,Q),
\end{equation}
where $c_0$ is a nonzero  complex number,  $d_0\geq0$ and $\ell_0$ is a contour on the surface $\C/{\mathbb{L}}$ assumed to satisfy  the following two conditions:
\begin{description}
\item[i)] $\ell_0$ does not passe through any of the ramification points $P_j$ of the covering $(\C/{\mathbb{L}}, \lambda)$;
\item[ii)] the $\lambda$-projection $\lambda(\ell_0)\subset\P^1$ of $\ell_0$ is independent of the branch points $\{\lambda_j\}_j$, where by this we mean that the contour $\lambda(\ell_0)$ does not change under small variations of $\lambda_1,\dots,\lambda_N$.
\end{description}
These conditions imply that derivation, with respect to $\lambda_j$, under the integral sign in (\ref{Int-rep-W}) is permitted and due to the Rauch variational formulas (\ref{Rauch}), we have
$$
\partial_{\lambda_j}\omega_0(P)=\frac{1}{2}\omega_0(P_j)W(P,P_j),\quad \quad \text{with}\quad \quad \omega_0(P_j):=\frac{\omega_0(P)}{d\mathbf{x}_j(P)}\Big|_{P=P_j}.
$$
\begin{Def}\label{degree-diff} \cite{Rejeb23} A differential $\omega_0$ of the form (\ref{Int-rep-W}) is called quasi-homogeneous differential and
the corresponding nonnegative number  $d_0$  is called the quasi-homogeneous degree of $\omega_0$.
\end{Def}
The degrees of the primary differentials (\ref{Primary-e}) are given by
\begin{equation}\label{degrees}
\begin{split}
&deg\big(\phi_{\mathbf{t}^{i,\alpha}}\big)=\frac{\alpha}{n_i+1};\\
&deg\big(\phi_{\mathbf{v}^i}\big)=deg\big(\phi_{\mathbf{u}}\big)=1;\\
&deg\big(\phi_{\mathbf{s}^i}\big)=deg\big(\phi_{\mathbf{r}}\big)=0.
\end{split}
\end{equation}

\medskip

\noindent $\bullet$ \textbf{Darboux-Egoroff metrics and the corresponding flat coordinates:}  To each  differential $\omega_0$ among the list (\ref{Primary-e}), Dubrovin associated a Darboux-Egoroff metric (i.e. diagonal with diagonal terms generated by a potential)
\begin{equation}\label{eta-def}
\eta(\omega_0):=\frac{1}{2}\sum_j\big(\omega_0(P_j)\big)^2(d\lambda_j)^2=
\sum_j\bigg(\underset{P_j}{{\rm res}}\frac{\omega_0(P)^2}{d\lambda(P)}\bigg)(d\lambda_j)^2
\end{equation}
which is defined on the open subset $\mathcal{M}_{\omega_0}$ of the Hurwitz space $\mathcal{H}_1(n_0,\dots,n_m)$ determined by the conditions
\begin{equation}\label{open}
\omega_0(P_j)\neq 0, \quad\quad  \forall \ j.
\end{equation}
The rotation coefficients of the metric $\eta(\omega_0)$ do not depend on the chosen primary differential $\omega_0$ and are given by \cite{Vasilisa}
$$
\beta_{ij}=\frac{1}{2}W(P_i,P_j)=\frac{W(P,Q)}{d\mathbf{x}_i(P)d\mathbf{x}_j(P)}, \quad\quad i\neq j, \quad \mathbf{x}_j=\sqrt{\lambda(P)-\lambda_j}.
$$
Moreover, the following $N$ functions
\begin{equation}\label{Flat basis}
\begin{split}
&\mathbf{t}^{i,\alpha}(\omega_0):=\frac{\sqrt{n_i+1}}{\alpha} \underset{\infty^i}{{\rm res}}\  \lambda(P)^{\frac{\alpha}{n_i+1}}\omega_0(P),\quad
\begin{array}{ll}
&i=0,\dots,m,\\
&\alpha=1,\dots,n_i;
\end{array}\\
&\mathbf{v}^i(\omega_0):=\underset{\infty^i}{{\rm res}\ }\lambda(P)\omega_0(P),\quad \quad\quad  \quad i=1,\dots,m;\\
&\mathbf{s}^i(\omega_0):= p.v.\int_{\infty^0}^{\infty^i}\omega_0(P),\quad \quad \quad \quad i=1,\dots,m;\\
&\mathbf{r}(\omega_0):=\frac{1}{2{\rm{i}}\pi}\oint_{b}\omega_0(P);\\
&\mathbf{u}(\omega_0):=\oint_{a}\lambda(P)\omega_0(P)
\end{split}
\end{equation}
give a system of flat coordinates of the metric $\eta(\omega_0)$ and the nonzero entries of the constant Gram matrix of the  metric $\eta(\omega_0)$ are as follows:
\begin{equation}\label{Entries eta}
\begin{split}
&\eta(\omega_0)\big(\partial_{\mathbf{t}^{i,\alpha}(\omega_0)},\partial_{\mathbf{t}^{j,\beta}(\omega_0)}\big)=\delta_{ij}\delta_{\alpha+\beta,n_j+1};\\
&\eta(\omega_0)\big(\partial_{\mathbf{v}^{i}(\omega_0)},\partial_{\mathbf{s}^{j}(\omega_0)}\big)=\delta_{ij};\\
&\eta(\omega_0)\big(\partial_{\mathbf{r}(\omega_0)},\partial_{\mathbf{u}(\omega_0)}\big)=1.
\end{split}
\end{equation}
From \cite{Dubrovin2D}, it is known that the principal value in the flat function $\mathbf{s}^i$ is defined by omitting the divergent part of the integral as a function of the local parameters $\z_0(P):=\lambda(P)^{-\frac{1}{n_0+1}}$, $\z_i(P):=\lambda(P)^{-\frac{1}{n_i+1}}$ near the points $\infty^0$ and $\infty^i$, respectively.

\medskip

\noindent $\bullet$ \textbf{Duality relations}:   As it has been observed in \cite{Rejeb23},  from (\ref{Entries eta}), it follows   that all the rows  of the  constant matrix of the Darboux-Egoroff metric $\eta(\omega_0)$ contain exactly one nonzero entry which is 1, bringing to  duality relations  between   flat coordinates (\ref{Flat basis})
$$
\mathcal{S}(\omega_0):=\big\{\mathbf{t}^{i,\alpha}(\omega_0), \mathbf{v}^i(\omega_0), \mathbf{s}^i(\omega_0), \mathbf{r}(\omega_0), \mathbf{u}(\omega_0)\big\}.
$$
More precisely, two flat coordinates $\mathbf{t}^A, \mathbf{t}^{A^{\prime}}\in \mathcal{S}(\omega_0)$ are called dual to each other with respect to the metric $\eta(\omega_0)$ if
\begin{equation}\label{duality}
\eta(\omega_0)\big(\partial_{\mathbf{t}^A(\omega_0)}, \partial_{\mathbf{t}^{A^{\prime}}(\omega_0)}\big)=1.
\end{equation}
Due to (\ref{Entries eta}), the   operations $\mathbf{t}^{i,\alpha}, \mathbf{v}^i, \mathbf{s}^i, \mathbf{r}, \mathbf{u}$ and the corresponding primary differentials (\ref{Primary-e}) are connected by:
\begin{equation}\label{duality-picture}
\begin{split}
&\mathbf{t}^{({i,\alpha})^{\prime}}(\omega_0)=\mathbf{t}^{i,n_i+1-\alpha}(\omega_0),\quad \quad \quad \quad \mathbf{v}^{{i}^{\prime}}(\omega_0)=\mathbf{s}^i(\omega_0),\quad \quad \ \quad \mathbf{r}^{\prime}(\omega_0)=\mathbf{u}(\omega_0);\\
&\phi_{\mathbf{t}^{({i,\alpha})^{\prime}}}(P)=\phi_{\mathbf{t}^{i,n_i+1-\alpha}}(P),\quad \quad \quad\quad  \phi_{\mathbf{v}^{i^{\prime}}}(P)=\phi_{\mathbf{s}^i}(P),\quad \quad \quad \phi_{\mathbf{r}^{\prime}}(P)=\phi_{\mathbf{u}}(P).
\end{split}
\end{equation}
Moreover, relations (\ref{degrees}) together with (\ref{duality-picture}) show that the degrees of  primary differentials (\ref{Primary-e}) enjoy the following duality relation:
\begin{equation}\label{Duality-degree}
deg\big(\phi_{\mathbf{t}^A}\big)+deg\big(\phi_{\mathbf{t}^{A^{\prime}}}\big)=1,\quad \quad \quad \text{for all A}.
\end{equation}

\medskip

\noindent $\bullet$ \textbf{Prepotential:}  The following formula, proved in \cite{Rejeb23}, offers  a new method to calculate the WDVV prepotential of the semi-simple Hurwitz-Frobenius manifold structure on the open set $\mathcal{M}_{\omega_0}$ (\ref{open})  induced by the quasi-homogeneous differential $\omega_0$ (\ref{Int-rep-W}):
\begin{equation}\label{Prepotential}
\begin{split}
&\mathbf{F}_{\omega_0}=\frac{1}{2(1+d_0)}\sum_{A,B}\frac{\big((d_0+d_A)\mathbf{t}^A(\omega_0)+\rho_{\omega_0,\phi_{\mathbf{t}^A}}\big)
\big((d_0+d_B)\mathbf{t}^B(\omega_0)+\rho_{\omega_0,\phi_{\mathbf{t}^B}}\big)}{1+d_0+d_{A^{\prime}}}\mathbf{t}^{A^{\prime}}(\phi_{\mathbf{t}^{B^{\prime}}})\\
&\quad +\frac{\log(-1)}{4}\sum_{i,j=1, i\neq j}^m\mathbf{v}^i(\omega_0)\mathbf{v}^j(\omega_0)
-\frac{3}{4(1+d_0)}\sum_{i,j=1}^m\bigg(\frac{1}{n_0+1}+\delta_{ij}\frac{1}{n_i+1}\bigg)\mathbf{v}^i(\omega_0)\mathbf{v}^j(\omega_0),
\end{split}
\end{equation}
where $d_0, d_A, d_{A^{\prime}}$ are respectively the degrees of the quasi-homogeneous  differentials $\omega_0$, $\phi_{\mathbf{t}^A}$, $\phi_{\mathbf{t}^{A^{\prime}}}$ and $\mathbf{t}^{A^{\prime}}$ is the dual coordinate of $t^A$ (\ref{duality})-(\ref{duality-picture}), while $\rho_{\omega_0,\phi_{\mathbf{t}^A}}$ is the constant determined by:
\begin{equation}\label{r-AB}
\rho_{\omega_0,\phi_{\mathbf{t}^A}}=\sum_{i,j=1}^m\delta_{\omega_0,\phi_{\mathbf{s}^i}}\delta_{\phi_{\mathbf{t}^A},\phi_{\mathbf{s}^j}}
\Big(\frac{1}{n_0+1}+\delta_{ij}\frac{1}{n_i+1}\Big)=\rho_{\phi_{\mathbf{t}^A},\omega_0}.
\end{equation}
Because of the added quadratic terms in formula (\ref{Prepotential}), the functions $\big\{\mathbf{F}_{\omega_0}\big\}$ share the Hessian matrix which is as follows
\begin{equation}\label{Hessian}
\begin{split}
\partial_{\mathbf{t}^A(\omega_0)}\partial_{\mathbf{t}^B(\omega_0)}\mathbf{F}_{\omega_0}
&=\frac{1}{2}\left(\mathbf{t}^{A^{\prime}}\left(\phi_{\mathbf{t}^{B^{\prime}}}\right)+\mathbf{t}^{B^{\prime}}\left(\phi_{\mathbf{t}^{A^{\prime}}}\right)\right)\\
&=\mathbf{t}^{A^{\prime}}\left(\phi_{\mathbf{t}^{B^{\prime}}}\right)
+\frac{\log(-1)}{2}\sum_{i,j=1}^m\big(1-\delta_{ij}\big)\delta_{\mathbf{t}^A,\mathbf{v}^i}\delta_{\mathbf{t}^B,\mathbf{v}^j}.
\end{split}
\end{equation}
Lastly, let us point out that with respect to flat coordinates (\ref{Flat basis}), the ingredients of the Hurwitz-Frobenius manifold associated with the differential $\omega_0$ are as follows \cite{Dubrovin2D, Vasilisa, Rejeb23} (using notation of \cite{Rejeb23}):
\begin{description}
\item[i)] the unit vector field is $e=\partial_{\textbf{t}^{A_0^{\prime}}}$, whenever $\omega_0=\phi_{\mathbf{t}^{A_0}}$;
\item[ii)] the Euler vector field is given by
\begin{equation}\label{E-FC}
E=\sum_{A}\big((d_0+d_A)\mathbf{t}^A(\omega_0)+\rho_{\omega_0,\phi_{\mathbf{t}^A}}\big)\partial_{\mathbf{t}^A(\omega_0)},
\end{equation}
where $d_0=deg(\omega_0), d_A=deg(\phi_{\mathbf{t}^A})$ and $\rho_{\omega_0,\phi_{\mathbf{t}^A}}$ is the constant (\ref{r-AB});
\item[iii)] the   symmetric 3-tensor $\mathbf{c}_{\omega_0}$  has the form:
\begin{equation}\label{c-FC-eta}
\begin{split}
\mathbf{c}_{\omega_0}\big(\partial_{\mathbf{t}^A(\omega_0)},\partial_{\mathbf{t}^B(\omega_0)},\partial_{\mathbf{t}^C(\omega_0)}\big)
&=\partial_{\mathbf{t}^A(\omega_0)}\partial_{\mathbf{t}^B(\omega_0)}\partial_{\mathbf{t}^C(\omega_0)}\mathbf{F}_{\omega_0}\\
&=\frac{1}{2}
\sum_{j=1}^N\frac{\phi_{\mathbf{t}^{A^{\prime}}}(P_j)\phi_{\mathbf{t}^{B^{\prime}}}(P_j)\phi_{\mathbf{t}^{C^{\prime}}}(P_j)}{\omega_0(P_j)}.
\end{split}
\end{equation}
\end{description}
We also emphasize that the WDVV equations (\ref{WDVV}) for the function $\mathbf{F}_{\omega_0}$ (\ref{Prepotential}) take the following form
\begin{equation}\label{WDVV-2}
\mathbf{F}_{\omega_0, \alpha{AC}}\mathbf{F}_{\omega_0,\beta{BC^{\prime}}}=\mathbf{F}_{\omega_0, \beta{AC}}\mathbf{F}_{\omega_0,\alpha{BC^{\prime}}},\quad \quad A,B,C,\alpha,\beta=1,\dots,N,
\end{equation}
where  $N=2+2m+\sum_{j=0}^mn_j$ and
$$
\mathbf{F}_{\omega_0, ABC}:=\partial_{\mathbf{t}^A(\omega_0)}\partial_{\mathbf{t}^B(\omega_0)}\partial_{\mathbf{t}^C(\omega_0)}\mathbf{F}_{\omega_0}.
$$
Equations (\ref{WDVV-2}) follows from the fact that the constant Gram matrix $\eta$ of the metric $\eta(\omega_0)$ satisfies $\eta^{AB}=\delta_{B,A^{\prime}}$, by (\ref{Entries eta}).



\section{Abelian differentials and the corresponding Laurent series}
Throughout this section $(\C/{\mathbb{L}},\lambda)$ denotes  a branched covering of the Riemann sphere $\P^1$, with $\lambda$ being  a meromorphic function  having $m+1$ poles $\infty^0,\dots,\infty^m \in \C/{\mathbb{L}}$ of order $n_0+1,\dots,n_m+1$, respectively.
This section delves into the explicit form of  Abelian differentials of the second and third kind whose poles are among the points $\big\{\infty^0,\dots,\infty^m\big\}$, along with their Laurent expansions near these points. The obtained results involve some variables related to the normalized holomorphic differential and will serve as a crucial tool within the next sections.
\subsection{Normalized holomorphic differential}
The normalized holomorphic differential on the complex torus $\C/{\mathbb{L}}$ is defined by
\begin{equation}\label{phi-def}
\phi(P)=\frac{1}{2{\rm i}\pi}\oint_bW(P,Q)=dz_P
\end{equation}
and its periods are given by
$$
\oint_a\phi=1,\quad \quad \quad \oint_b\phi=\tau.
$$
For $j=0,\dots,m$, consider the (genus one) Abel map $\mathcal{A}(P;\infty^j)$ defined by
\begin{equation}\label{Abel-def}
\mathcal{A}(P;\infty^j)=\int_{\infty^j}^P\phi=z_P-\infty^j.
\end{equation}
Then, $\mathcal{A}(P;\infty^j)$ vanishes at $P=\infty^j$ and  its Taylor expansion near the point $P=\infty^j$ is of the following form:
\begin{equation}\label{Abel-Taylor}
\mathcal{A}(P;\infty^j)\underset{P\sim \infty^j}{=}\sum_{r=1}^{\infty}x_r(j)\z_j(P)^r,\quad \quad \quad \quad \lambda(P)\underset{P\sim \infty^j}{=}\z_j(P)^{-n_j-1}.
\end{equation}
The coefficients $x_r(j)$ in (\ref{Abel-Taylor}) occur also in the Taylor expansion for the holomorphic differential $\phi$ near the point $P=\infty^j$:
\begin{equation}\label{phi-Taylor}
\phi(P)\underset{P\sim \infty^j}{=}\bigg(\sum_{r=1}^{\infty}rx_r(j)\mathbf{z}_j(P)^{r-1}\bigg)d\z_j(P),
\quad \quad \quad \quad \lambda(P)\underset{P\sim \infty^j}{=}\z_j(P)^{-n_j-1}.
\end{equation}
This implies that
\begin{equation}\label{Evaluation}
x_r(j)=\frac{1}{r}\underset{\infty^j} {\rm res}\  \lambda(P)^{\frac{r}{n_j+1}}\phi(P),\quad \quad \quad  \forall\ r\geq 1
\end{equation}
and, since the genus one normalized holomorphic differential has no zero, we have
$$
x_1(j)\neq 0, \quad \quad \forall\ j=0,\dots,m.
$$
In addition, in view of  (\ref{Evaluation}) and the residue theorem applied to the differential $\lambda(P)\phi(P)=\lambda(P)dz_P$, we obtain
\begin{equation}\label{x-r-sum}
\sum_{j=0}^m(n_j+1)x_{n_j+1}(j)=\sum_{j=0}^m\underset{\infty^j}{{\rm res}}\ \lambda(P)\phi(P)=0.
\end{equation}

\medskip

We now introduce  the following operators related to holomorphic differential $\phi$ which are of particular importance in this paper.
\begin{Def} Let $\alpha$ be a positive integer and $\mathcal{B}_{\alpha,k}$  be  the partial ordinary  Bell polynomials defined  by (\ref{Bell-partial})-(\ref{Bell2}).
For $j=0,1,\dots,m$, we define the $(j,\alpha)$-Bell operator associated with the holomorphic differential $\phi$ by
\begin{equation}\label{Bell operator}
\mathbf{L}^{\phi}_{j,\alpha}:=\sum_{k=1}^{\alpha}\frac{1}{k!}\mathcal{B}_{\alpha,k}\big(x_1(j),\dots,x_{\alpha-k+1}(j)\big)\partial_{v}^k,
\end{equation}
where $\infty^j$ is one of  poles of the covering $\lambda$ of order $n_j+1$, the coefficients $x_k(j)$  are defined by (\ref{Evaluation}) and $v$ is a variables independent of $x_r(j)$.
\end{Def}

\bigskip

Let us mention that the Bell operator $\mathbf{L}^{\phi}_{j,\alpha}$  acts naturally  on functions as follows:
\begin{equation}\label{Bell-f}
\mathbf{L}^{\phi}_{j,\alpha}[f](v):=\sum_{k=1}^{\alpha}\frac{1}{k!}\mathcal{B}_{\alpha,k}\big(x_1(j),\dots,x_{\alpha-k+1}(j)\big)f^{(k)}(v).
\end{equation}
Moreover, using  the homogeneity property (\ref{Bell-homog}) of partial exponential Bell polynomials, we see that the Bell operator (\ref{Bell operator}) can also be expressed by means of  partial and complete exponential Bell polynomials (\ref{Bell-OE})-(\ref{Bell-complete}):
\begin{align*}
\mathbf{L}^{\phi}_{j,\alpha}&=\frac{1}{\alpha!}\sum_{k=1}^{\alpha}\mathbf{B}_{\alpha,k}\Big(1!x_1(j),\dots,(\alpha-k+1)!x_{\alpha-k+1}(j)\Big)\partial_{v}^k
=\frac{1}{\alpha!}\mathbf{B}_{\alpha}\Big(1!x_1(j)\partial_{v},\dots,\alpha!x_{\alpha}(j)\partial_{v}\Big).
\end{align*}

\medskip

For future use, we will take into account the following Laurent expansions near the prescribed poles $\infty^0,\dots,\infty^m$ of the covering $(\C/\mathbb{L},\lambda)$.
\begin{Prop} For $j=0,\dots,m$, let $\infty^j$ be a pole of $\lambda$ of order $n_j+1$ and  $\mathcal{A}(P;\infty^j)$ be the function defined by (\ref{Abel-def}). Then for any positive integer $k$ we have
\begin{align}
&\label{Abel-Laurent1}\big(\mathcal{A}(P;\infty^j)\big)^{k}\underset{P\sim \infty^j}{=}
\sum_{n=k}^{\infty}\mathcal{B}_{n,k}\big(x_1(j),\dots,x_{n-k+1}(j)\big)\z_j(P)^n;\\
&\label{Abel-Laurent2}\big(\mathcal{A}(P;\infty^j)\big)^{-k}\underset{P\sim \infty^j}{=}
\sum_{n=0}^{\infty}\left\{\sum_{\ell=0}^{n}\binom{-k}{\ell}\frac{1}{\big(x_1(j)\big)^{k+\ell}}
\mathcal{B}_{n,\ell}\big(x_2(j),\dots,x_{n-\ell+2}(j)\big)\right\}\z_j(P)^{n-k}.
\end{align}
Here $\z_j(P):=\lambda(P)^{-\frac{1}{n_j+1}}$ is the local parameter  induced by the covering $(\C/{\mathbb{L}},\lambda)$, $\mathcal{B}_{n,k}$ denotes the partial exponential Bell polynomials defined by (\ref{Bell-partial}) and (\ref{Bell2}) and $x_{r}(j)$ are the coefficients in the Taylor expansion (\ref{phi-Taylor}) of the normalized holomorphic differential $\phi(P)=dz_P$ defined by (\ref{Evaluation}).
\end{Prop}
\emph{Proof:} Formula (\ref{Abel-Laurent1}) follows from (\ref{Abel-Taylor}) and  generating function (\ref{Bell2}) for  partial Bell polynomials. Using again Taylor's expansion (\ref{Abel-Taylor}) for the Abel map and  (\ref{Bell2}), we obtain
\begin{align*}
&\big(\mathcal{A}(P;\infty^j)\big)^{-k}\underset{P\sim \infty^j}{=}\frac{1}{\big(x_1(j)\z_j(P)\big)^k}\bigg(1+\frac{1}{x_1(j)}\sum_{r=1}^{\infty}x_{r+1}(j)\z_j(P)^{r}\bigg)^{-k}\\
&=\frac{1}{\big(x_1(j)\z_j(P)\big)^k}+\frac{1}{\big(x_1(j)\z_j(P)\big)^k}\sum_{\ell=1}^{\infty}\left\{\binom{-k}{\ell}\frac{1}{\big(x_1(j)\big)^{\ell}}
\Bigg(\sum_{r=1}^{\infty}x_{r+1}(j)\z_j(P)^{r}\bigg)^{\ell}\right\}\\
&=\frac{1}{\big(x_1(j)\z_j(P)\big)^k}+\sum_{n=1}^{\infty}\left\{\sum_{\ell=1}^{n}\binom{-k}{\ell}\frac{1}{\big(x_1(j)\big)^{k+\ell}}
\mathcal{B}_{n,\ell}\big(x_2(j),\dots,x_{n-\ell+2}(j)\big)\right\}\z_j(P)^{n-k}.
\end{align*}
Therefore, since $\mathcal{B}_{n,0}=\delta_{n,0}$ we arrive at (\ref{Abel-Laurent2}).

\fd

\begin{Cor} With the notation of the preceding proposition, the following two equalities hold true for any  positive integers $\alpha,k$:
\begin{equation}\label{Bell-B-formula}
\frac{\alpha}{k+1}\mathcal{B}_{\alpha,k+1}\big(x_1(j),\dots,x_{\alpha-k}(j)\big)={\underset{\infty^j}{\rm res}}
\Big(\lambda(P)^{\frac{\alpha}{n_j+1}}\big(\mathcal{A}(P;\infty^j)\big)^{k}\phi(P)\Big)
\end{equation}
and
\begin{equation}\label{Bell-R-formula}
\alpha\mathcal{R}_{\alpha+k-1,k}\big(x_1(j),\dots,x_{\alpha+k}(j)\big)={\underset{\infty^j}{\rm res}}
\Big(\lambda(P)^{\frac{\alpha}{n_j+1}}\big(\mathcal{A}(P;\infty^j)\big)^{-k}\phi(P)\Big),
\end{equation}
where $\mathcal{R}_{\alpha+k-1,k}$ is the rational function given  by (\ref{R-function}) and (\ref{Bell-rec4}).
\end{Cor}
\emph{Proof:} Using the Laurent expansions (\ref{phi-Taylor}) and (\ref{Abel-Laurent1}) and applying recurrence relation (\ref{Bell-rec1}) , we deduce that
\begin{align*}
&{\underset{\infty^j}{\rm res}}\Big(\lambda(P)^{\frac{\alpha}{n_j+1}}(z_P-\infty^j)^{k}\phi(P)\Big)\\
&={\underset{0}{\rm res}}\left\{\sum_{n=k}^{\infty}\sum_{r=1}^{\infty}rx_r(j)\mathcal{B}_{n,k}\big(x_1(j),\dots,x_{n-k+1}(j)\big)\z_j(P)^{n+r-\alpha-1}d\z_j(P)\right\}\\
&=\sum_{n=k}^{\alpha-1}(\alpha-n)x_{\alpha-n}(j)\mathcal{B}_{n,k}\big(x_1(j),\dots,x_{n-k+1}(j)\big)\\
&=\frac{\alpha}{k+1}\mathcal{B}_{\alpha,k+1}\big(x_1(j),\dots,x_{\alpha-k}(j)\big).
\end{align*}
Similarly, by (\ref{phi-Taylor}), (\ref{Abel-Laurent2}) and (\ref{R-function}), we obtain
\begin{align*}
&{\underset{\infty^j}{\rm res}}\Big(\lambda(P)^{\frac{\alpha}{n_j+1}}(z_P-\infty^j)^{-k}\phi(P)\Big)\\
&=\sum_{n=0}^{\alpha+k-1}\left\{(\alpha+k-n)x_{\alpha+k-n}(j)\sum_{\ell=0}^{n}\binom{-k}{\ell}\frac{1}{\big(x_1(j)\big)^{k+\ell}}
\mathbf{B}_{n,\ell}\big(x_2(j),\dots,x_{n-\ell+2}(j)\big)\right\}\\
&=\alpha\mathcal{R}_{\alpha+k-1,k}\big(x_1(j),\dots,x_{\alpha+k}(j)\big).
\end{align*}

\fd
\medskip

At the end of this subsection let us emphasize that  the genus one  bidifferential (\ref{W-g1}) can be rewritten in terms of the normalized holomorphic differential (\ref{phi-def}) as follows:
\begin{equation}\label{W-g1-phi}
W(P,Q)=\Big(\wp(z_P-z_Q|\tau)+2\zeta(1/2|\tau)\Big)\phi(P)\phi(Q).
\end{equation}
This formula and   expansions (\ref{Abel-Taylor}),  (\ref{phi-Taylor}), (\ref{Abel-Laurent1}) and (\ref{Abel-Laurent2}) serve as key ingredients   for deriving   explicit forms for Abelian differentials of the second and third kind, as well as for  their Laurent series expansions  near the prescribed  poles $\infty^0,\dots,\infty^m$ of the covering $(\C/{\mathbb{L}},\lambda)$. This will be discussed in the upcoming two subsections.



\subsection{Abelian differentials of the third kind}
In this subsection we are interested  in  the Laurent expansions for the Abelian differentials of the third kind $\Omega_{\infty^0\infty^j}:=\int_{\infty^0}^{\infty^j}W(P,Q)$, $j=1,\dots,m$, near the poles $\infty^0,\dots,\infty^m$ of the given covering $(\C/{\mathbb{L}}, \lambda)$.  We start with the following explicit formula for the differential $\Omega_{\infty^0\infty^j}(P)$ involving  the Weierstrass zeta function (\ref{zeta-function}).
\begin{Prop} For $j=1,\dots,m$, the normalized Abelian differential of the third kind $\Omega_{\infty^0\infty^j}(P)$ is given by
\begin{equation}\label{Third kind}
\Omega_{\infty^0\infty^j}(P)=\bigg(2\zeta(1/2)(\infty^j-\infty^0)+\zeta(z_P-\infty^j)-\zeta(z_P-\infty^0)\bigg)\phi(P),
\end{equation}
where $\phi$ is the normalized holomorphic differential (\ref{phi-def}). The periods of the differential $\Omega_{\infty^0\infty^j}$ are such that
\begin{equation}\label{Third kind-periods}
\oint_a\Omega_{\infty^0\infty^j}=0\quad \quad \text{and}\quad \quad \oint_b\Omega_{\infty^0\infty^j}=2{\rm i}\pi(\infty^j-\infty^0).
\end{equation}
\end{Prop}
\emph{Proof:} Due to relations (\ref{zeta-derivative}), (\ref{W-g1}) and (\ref{phi-def}) we have
\begin{align*}
\Omega_{\infty^0\infty^j}(P)&:=\int_{\infty^0}^{\infty^j}W(P,Q)=\left(\int_{\infty^0}^{\infty^j}\Big(\wp(z_P-z_Q)+2\zeta(1/2)\Big)dz_Q\right)dz_P\\
&=\bigg(2\zeta(1/2)(\infty^j-\infty^0)+\zeta(z_P-\infty^j)-\zeta(z_P-\infty^0)\bigg)\phi(P).
\end{align*}
We obtain the periods (\ref{Third kind-periods}) using successively formula (\ref{Third kind}),  equality (\ref{zeta-sigma}) which states  that  $\zeta(u)=\partial_{u}\log\big(\sigma(u)\big)$, the quasi-periodicity property (\ref{zeta-sigma periods})  of the Weierstrass function $\sigma$ and the Legendre identity (\ref{Legendre}).

\fd

\medskip

The purpose of the next theorem is to calculate the  quantities:
\begin{equation}\label{rho-ij}
\mathcal{I}_{i,\alpha}[\Omega_{\infty^0\infty^j}]:=
\underset{\infty^i}{{\rm res}}\ \lambda(P)^{\frac{\alpha}{n_i+1}}\Omega_{\infty^0\infty^j}(P), \quad \quad \forall\ \alpha\geq 1.
\end{equation}
To simplify, in what follows, we shall frequently use the following convention: for any positive integers $k_0,\mu$, we understand  the sum $\sum_{\nu=k_0}^{\mu}$ in the
sense that
\begin{equation}\label{convention}
\sum_{\nu=k_0}^{\mu}:=\bigg(1-\sum_{j=0}^{k_0-1}\delta_{\mu,j}\bigg) \sum_{\nu=k_0}^{\mu},
\end{equation}
(with $\delta$ being the Kronecker delta), that is the sum $\sum_{\nu=k_0}^{\mu}$ is 0 when $\mu<k_0$.
\begin{Thm}\label{Third kind-FCij} Let $i\in \{0,1,\dots,m\}$,  $j\in \{1,\dots,m\}$ and $\alpha$ be a positive integer. For $n\geq k\geq 1$, let $\mathcal{B}_{n,k}(i):=\mathcal{B}_{n,k}\big(x_1(i),\dots,x_{n-k+1}(i))$ be the  partial  Bell polynomials (\ref{Bell-partial}), where  the variables
$\big\{x_n(i)\big\}_{n\geq 1}$ are defined by (\ref{Evaluation}).
\begin{description}
\item[1)] If $i\neq j$ and $i\neq 0$, then the quantity (\ref{rho-ij}) is as follows:
\begin{equation}\label{Third-alpha-ij}
\begin{split}
&\mathcal{I}_{i,\alpha}[\Omega_{\infty^0\infty^j}]=2{\alpha}x_{\alpha}(i)(\infty^j-\infty^0)\zeta(1/2)\\
&+\alpha\sum_{k=1}^{\alpha}\frac{1}{k!}\mathcal{B}_{\alpha,k}(i)\Big(\zeta^{(k-1)}(\infty^i-\infty^j)-\zeta^{(k-1)}(\infty^i-\infty^0)\Big)\\
&=2{\alpha}x_{\alpha}(i)(\infty^j-\infty^0)\zeta(1/2)
+\alpha\mathbf{L}_{i,\alpha}^{\phi}[\K](\infty^i-\infty^j)-\alpha\mathbf{L}_{i,\alpha}^{\phi}[\K](\infty^i-\infty^0),
\end{split}
\end{equation}
where $\K(u)=\log\sigma(u)$ and $\mathbf{L}_{i,\alpha}^{\phi}[\K]$ is the function defined by (\ref{Bell-f}).
\item[2)] When $j=i$, we have
\begin{equation}\label{Third-alpha-jj}
\begin{split}
\mathcal{I}_{j,\alpha}[\Omega_{\infty^0\infty^j}]&=2{\alpha}x_{\alpha}(j)(\infty^j-\infty^0)\zeta(1/2)+\alpha\mathcal{R}_{\alpha,1}\big(x_1(j),\dots,x_{\alpha+1}(j)\big)\\
&\quad -\alpha\sum_{\ell=4}^{\alpha}\frac{\mathcal{B}_{\alpha,\ell}(j)}{\ell}G_{\ell}
-\alpha\mathbf{L}_{j,\alpha}^{\phi}\big[\K\big](\infty^j-\infty^0),
\end{split}
\end{equation}
where $\mathcal{R}_{\alpha,1}$ is the rational function given by (\ref{R-function}) and (\ref{Bell-rec4}) and $G_{\ell}$ are the Eisenstein series (\ref{Eisenstein G}).
\item[3)] Assume that  $i=0$. Then the quantity  $\mathcal{I}_{0,\alpha}[\Omega_{\infty^0\infty^j}]$ is of the following form:
\begin{equation}\label{Third-alpha-0j}
\begin{split}
\mathcal{I}_{0,\alpha}[\Omega_{\infty^0\infty^j}]&=2{\alpha}x_{\alpha}(0)(\infty^j-\infty^0)\zeta(1/2)-\alpha\mathcal{R}_{\alpha,1}\big(x_1(0),\dots,x_{\alpha+1}(0)\big)\\
&\quad +\alpha \sum_{\ell=4}^{\alpha}\frac{\mathcal{B}_{\alpha,\ell}(0)}{\ell}G_{\ell}+\alpha\mathbf{L}_{0,\alpha}^{\phi}\big[\K\big](\infty^0-\infty^j).
\end{split}
\end{equation}
\end{description}
Using (\ref{convention}), the sum $\sum_{\ell=4}^{\alpha}$ in formulas (\ref{Third-alpha-jj}) and (\ref{Third-alpha-0j}) takes the value 0 for $\alpha<4$.
\end{Thm}
\emph{Proof:} Consider  the meromorphic  function $h$ on the complex torus $\C/{\mathbb{L}}$ defined by
\begin{equation}\label{h-zeta}
h_{0j}(P):=\zeta(z_P-\infty^j)-\zeta(z_P-\infty^0), \quad \quad j=1,\dots,m
\end{equation}
which appears in expression (\ref{Third kind}) for the differential $\Omega_{\infty^0\infty^j}(P)$. The proof of the theorem relies essentially on the asymptotic behavior of the meromorphic $h_{0j}$ near the point $\infty^i$ with respect to the standard local parameter $\z_i(P)$ defined by $\lambda(P)=\z_i(P)^{-n_i-1}$.

\medskip

\noindent \textbf{1)}  Assume that  $i\neq 0$ and $i\neq j$. We observe that the function $h_{0j}$ (\ref{h-zeta}) is holomorphic on a small neighborhood of the point $\infty^i$ and its Taylor series at $P=\infty^i$ is given by
\begin{align*}
h_{0j}(P)&:=\zeta(z_P-\infty^j)-\zeta(z_P-\infty^0)\\
&=\zeta\big(\infty^i-\infty^j+z_P-\infty^i\big)-\zeta\big(\infty^i-\infty^0+z_P-\infty^i\big)\\
&{\underset{P\sim\infty^i}{=}}h_{0j}(\infty^i)+\sum_{k=1}^{\infty}\Big(\zeta^{(k)}(\infty^i-\infty^j)-\zeta^{(k)}(\infty^i-\infty^0)\Big)\frac{(z_P-\infty^i)^k}{k!}\\
&=h_{0j}(\infty^i)+\sum_{k=1}^{\infty}\Bigg(\frac{1}{k!}\Big[\zeta^{(k)}(\infty^i-\infty^j)-\zeta^{(k)}(\infty^i-\infty^0)\Big]
\bigg(\sum_{r=1}^{\infty}x_r(i)\z_i(P)^r\bigg)^k\bigg)\\
&=h_{0j}(\infty^i)+\sum_{k=1}^{\infty}\left\{\frac{1}{k!}\Big(\zeta^{(k)}(\infty^i-\infty^j)-\zeta^{(k)}(\infty^i-\infty^0)\Big)
\sum_{n=k}^{\infty}\mathcal{B}_{n,k}(i)\z_i(P)^n\right\},
\end{align*}
where we used (\ref{Abel-Taylor}) in the forth equality and (\ref{Abel-Laurent1}) in the last one.\\
This and (\ref{phi-Taylor}) imply that the Taylor expansion for the Abelian differential (\ref{Third kind}) near the point $P=\infty^i$ takes the following form:
\begin{align*}
&\Omega_{\infty^0\infty^j}(P):=\Big(2\zeta(1/2)\big(\infty^j-\infty^0\big)+h_{0j}(P)\Big)\phi(P)\\
&\underset{P\sim\infty^i}{=}\bigg(2\zeta(1/2)\big(\infty^j-\infty^0\big)+
\zeta(\infty^i-\infty^j)-\zeta(\infty^i-\infty^0)\bigg)\bigg(\sum_{r=1}^{\infty}rx_r(i)\z_i(P)^{r-1}\bigg)d\z_i(P)\\
&\quad +\Bigg(\sum_{k=1}^{\infty}\left\{\frac{1}{k!}\Big(\zeta^{(k)}(\infty^i-\infty^j)-\zeta^{(k)}(\infty^i-\infty^0)\Big)
\sum_{n=k}^{\infty}\sum_{r=1}^{\infty}rx_r(i)\mathcal{B}_{n,k}(i)\z_i(P)^{n+r-1}\right\}\Bigg)d\z_i(P).
\end{align*}
Therefore
\begin{align*}
&\mathcal{I}_{i,\alpha}[\Omega_{\infty^0\infty^j}]:=\underset{\infty^i}{{\rm res}}\ \lambda(P)^{\frac{\alpha}{n_i+1}}\Omega_{\infty^0\infty^j}(P)\\
&={\alpha}x_{\alpha}(i)\bigg(2\zeta(1/2)\big(\infty^j-\infty^0\big)+\zeta(\infty^i-\infty^j)-\zeta(\infty^i-\infty^0)\bigg)\\
&\quad+\big(1-\delta_{\alpha,1}\big)\sum_{k=1}^{\alpha-1}\left\{\frac{1}{k!}\Big(\zeta^{(k)}(\infty^i-\infty^j)-\zeta^{(k)}(\infty^i-\infty^0)\Big)
\sum_{n=k}^{\alpha-1}(\alpha-n)x_{\alpha-n}(i)\mathcal{B}_{n,k}(i)\right\}\\
&=2{\alpha}x_{\alpha}(i)(\infty^j-\infty^0)\zeta(1/2)+{\alpha}x_{\alpha}(i)\Big(\zeta(\infty^i-\infty^j)-\zeta(\infty^i-\infty^0)\Big)\\
&\quad+\big(1-\delta_{\alpha,1}\big)\alpha\sum_{k=1}^{\alpha-1}\frac{1}{k!}\mathcal{B}_{\alpha,k+1}(i)\Big(\zeta^{(k)}(\infty^i-\infty^j)-\zeta^{(k)}(\infty^i-\infty^0)\Big)\\
&=2{\alpha}x_{\alpha}(i)(\infty^j-\infty^0)\zeta(1/2)
+\alpha\sum_{k=0}^{\alpha-1}\frac{1}{(k+1)!}\mathcal{B}_{\alpha,k+1}(i)\Big(\zeta^{(k)}(\infty^i-\infty^j)-\zeta^{(k)}(\infty^i-\infty^0)\Big)\\
&=2{\alpha}x_{\alpha}(i)(\infty^j-\infty^0)\zeta(1/2)
+\alpha\mathbf{L}_{i,\alpha}^{\phi}[\K](\infty^i-\infty^j)-\alpha\mathbf{L}_{i,\alpha}^{\phi}[\K](\infty^i-\infty^0),
\end{align*}
where we applied recurrence formula (\ref{Bell-rec1}) for  partial   Bell polynomials in the third equality and the relation
$\mathcal{B}_{\alpha,1}(i)=x_{\alpha}(i)$ in the forth  equality. In the last equality, we have used  expressions (\ref{Bell operator}) and (\ref{Bell-f}) for the Bell operator $\mathbf{L}_{i,\alpha}^{\phi}$ acting on the function $\K(u)=\log\sigma(u)$. This  establishes (\ref{Third-alpha-ij}).

\medskip

\noindent \textbf{2)} Making use of (\ref{Abel-Laurent1}), (\ref{Abel-Laurent2}) with $k=1$  and   the Laurent series (\ref{Laurent zeta}) for the Weierstrass zeta function near 0,  we see that  the meromorphic function $h_{0j}$ (\ref{h-zeta}) behaves as follows near the point $\infty^j$:
\begin{align*}
h_{0j}(P)&:=\zeta(z_P-\infty^j)-\zeta(z_P-\infty^0)\\
&\underset{P\sim\infty^j}{=}\frac{1}{z_P-\infty^j}-\sum_{\ell=3}^{\infty}G_{\ell+1}(z_P-\infty^j)^{\ell}
-\zeta\big(\infty^j-\infty^0)-\sum_{k=1}^{\infty}\frac{\zeta^{(k)}(\infty^j-\infty^0)}{k!}(z_P-\infty^j)^k\\
&=-\zeta\big(\infty^j-\infty^0)+\sum_{n=0}^{\infty}\left\{\sum_{\ell=0}^{n}\frac{(-1)^{\ell}}{\big(x_1(j)\big)^{\ell+1}}
\mathcal{B}_{n,\ell}\big(x_2(j),\dots,x_{n-\ell+2}(j)\big)\right\}\z_j(P)^{n-1}\\
&\quad -\sum_{\ell=3}^{\infty}\sum_{n=\ell}^{\infty}G_{\ell+1}\mathcal{B}_{n,\ell}(j)\z_j(P)^n
-\sum_{k=1}^{\infty}\left\{\frac{\zeta^{(k)}(\infty^j-\infty^0)}{k!}\sum_{n=k}^{\infty}\mathcal{B}_{n,k}(j)\z_j(P)^n\right\}.
\end{align*}
As consequence, the Laurent series of the differential $\Omega_{\infty^0\infty^j}(P)$ near $P=\infty^j$ has  the following form:
\begin{align*}
&\Omega_{\infty^0\infty^j}(P):=\Big(2\big(\infty^j-\infty^0\big)\zeta(1/2)+h_{0j}(P)\Big)\phi(P)\\
&\underset{P\sim\infty^j}{=}\bigg(2\big(\infty^j-\infty^0\big)\zeta(1/2)-\zeta\big(\infty^j-\infty^0)\bigg)
\bigg(\sum_{r=1}^{\infty}rx_r(j)\z_j(P)^{r-1}\bigg)d\z_j(P)\\
&\quad +\left(\sum_{n=0}^{\infty}\sum_{r=1}^{\infty}\left\{rx_r(j)\sum_{\ell=0}^{n}\frac{(-1)^{\ell}}{\big(x_1(j)\big)^{\ell+1}}
\mathcal{B}_{n,\ell}\big(x_2(j),\dots,x_{n-\ell+2}(j)\big)\right\}\z_j(P)^{n+r-2}\right)d\z_j(P)\\
&\quad -\left(\sum_{\ell=3}^{\infty}\sum_{n=\ell}^{\infty}\sum_{r=1}^{\infty}rx_r(j)G_{\ell+1}\mathcal{B}_{n,\ell}(j)\z_j(P)^{n+r-1}\right)d\z_j(P)\\
&\quad -\left(\sum_{k=1}^{\infty}\left\{\frac{\zeta^{(k)}(\infty^j-\infty^0)}{k!}\sum_{n=k}^{\infty}\sum_{r=1}^{\infty}
rx_r(j)\mathcal{B}_{n,k}(j)\z_j(P)^{n+r-1}\right\}\right)d\z_j(P).
\end{align*}
Thus we are now in position to calculate the residue $\mathcal{I}_{j,\alpha}[\Omega_{\infty^0\infty^j}]$:
\begin{align*}
&\mathcal{I}_{j,\alpha}[\Omega_{\infty^0\infty^j}]:=\underset{\infty^j}{{\rm res}}\ \lambda(P)^{\frac{\alpha}{n_j+1}}\Omega_{\infty^0\infty^j}(P)\\
&=2{\alpha}x_{\alpha}(j)(\infty^j-\infty^0)\zeta(1/2)-{\alpha}x_{\alpha}(j)\zeta\big(\infty^j-\infty^0)\\
&\quad +\sum_{n=0}^{\alpha}\left\{(\alpha+1-n)x_{\alpha+1-n}(j)\sum_{\ell=0}^{n}\frac{(-1)^{\ell}}{\big(x_1(j)\big)^{\ell+1}}
\mathcal{B}_{n,\ell}\big(x_2(j),\dots,x_{n-\ell+2}(j)\big)\right\}\\
&\quad -\big(1-\delta_{\alpha,1}-\delta_{\alpha,2}-\delta_{\alpha,3}\big)\sum_{\ell=3}^{\alpha-1}G_{\ell+1}\left\{
\sum_{n=\ell}^{\alpha-1}(\alpha-n)x_{\alpha-n}(j)\mathcal{B}_{n,\ell}(j)\right\}\\
&\quad -\big(1-\delta_{\alpha,1}\big)\sum_{k=1}^{\alpha-1}\left\{\frac{\zeta^{(k)}(\infty^j-\infty^0)}{k!}\sum_{n=k}^{\alpha-1}(\alpha-n)
x_{\alpha-n}(j)\mathcal{B}_{n,k}(j)\right\}\\
&=2{\alpha}x_{\alpha}(j)(\infty^j-\infty^0)\zeta(1/2)+\alpha\mathcal{R}_{\alpha,1}\big(x_1(j),\dots,x_{\alpha+1}(j)\big)\\
&\quad -\big(1-\delta_{\alpha,1}-\delta_{\alpha,2}-\delta_{\alpha,3}\big)\alpha
\sum_{\ell=4}^{\alpha}\frac{\mathcal{B}_{\alpha,\ell}(j)}{\ell}G_{\ell}-\alpha\sum_{k=1}^{\alpha}\frac{\mathcal{B}_{\alpha,k}(j)}{k!}\zeta^{(k-1)}(\infty^j-\infty^0)\\
&=2{\alpha}x_{\alpha}(j)(\infty^j-\infty^0)\zeta(1/2)+\alpha\mathcal{R}_{\alpha,1}\big(x_1(j),\dots,x_{\alpha+1}(j)\big)\\
&\quad -\alpha\sum_{\ell=4}^{\alpha}\frac{\mathcal{B}_{\alpha,\ell}(j)}{\ell}G_{\ell}-\alpha\mathbf{L}_{j,\alpha}^{\phi}\big[\K\big](\infty^j-\infty^0),
\end{align*}
where in the third equality we used expression (\ref{R-function}) for the rational function $\mathcal{R}_{\alpha,1}$ and applied twice recurrence relation (\ref{Bell-rec1}).

\medskip

\noindent \textbf{3)} We have
\begin{align*}
\mathcal{I}_{0,\alpha}[\Omega_{\infty^0\infty^j}]:=\underset{\infty^0}{{\rm res}}\ \lambda(P)^{\frac{\alpha}{n_0+1}}\Omega_{\infty^0\infty^j}(P)
=-\underset{\infty^0}{{\rm res}}\ \lambda(P)^{\frac{\alpha}{n_0+1}}\Omega_{\infty^j\infty^0}(P)=-\mathcal{I}_{0,\alpha}[\Omega_{\infty^j\infty^0}].
\end{align*}
Thus (\ref{Third-alpha-0j}) follows from (\ref{Third-alpha-jj}) by permuting the role of the indices $j$ and 0.

\fd

\begin{Remark}
\end{Remark}
$\bullet$ Let $i=0,1,\dots,m$ and $j=1,\dots,m$ be fixed. Then the above theorem yields that the Laurent expansions of the differential $\Omega_{\infty^0\infty^j}$ near the point $\infty^i$ takes the following explicit form:
$$
\Omega_{\infty^0\infty^j}(P)\underset{P\sim \infty^i}{=}
\bigg(\frac{\delta_{ij}}{z_j(P)}-\frac{\delta_{i0}}{z_i(P)}+\sum_{\alpha=1}^{\infty}\Big(\mathcal{I}_{i,\alpha}[\Omega_{\infty^0\infty^j}]\Big)\z_i(P)^{\alpha-1}\bigg)d\z_i(P),
$$
where the coefficients $\mathcal{I}_{i,\alpha}[\Omega_{\infty^0\infty^j}]$ are given by formulas  (\ref{Third-alpha-ij})-(\ref{Third-alpha-0j}) and $\z_i(P)$ is the standard  local parameter near $\infty^i$ defined by $\z_i(P)=\lambda(P)^{-\frac{1}{n_i+1}}$.

\medskip

\noindent $\bullet$ Due to relations (\ref{zeta-theta1}) and (\ref{Third kind}), the Abelian differential $\Omega_{\infty^0\infty^j}$ can be rewritten in terms of the Jacobi $\theta_1$-function as follows:
\begin{align*}
\Omega_{\infty^0\infty^j}(P)&=\left(\frac{\theta'_1(z_P-\infty^j)}{\theta_1(z_P-\infty^j)}-\frac{\theta'_1(z_P-\infty^0)}{\theta_1(z_P-\infty^0)}\right)dz_P.
\end{align*}



\subsection{Abelian differentials of the second kind}
This subsection focuses on providing an  explicit expression for the differential
\begin{equation}\label{Psi-ialpha-def}
\Psi_{j,\alpha}(P):=\underset{\infty^j} {\rm res}\  \lambda(Q)^{\frac{\alpha}{n_j+1}}W(P,Q),\quad \quad \quad j=0,\dots,m\quad \text{and}\quad \alpha\geq 1,
\end{equation}
as well as for the coefficients of the  corresponding Laurent series near the points $\infty^i$, $i=0,\dots,m$. Here, as above, $W(P,Q)$ denotes the genus one bidifferential given by (\ref{W-g1}) and (\ref{W-g1-phi}).

\begin{Prop} Assume that $j\in \{0,1,\dots,m\}$. Let $\mathcal{B}_{n,k}(j):=\mathcal{B}_{n,k}(x_1(j),\dots,x_{n-k+1}(j))$, $n\geq k\geq 1$, be the partial Bell polynomials (\ref{Bell-partial})-(\ref{Bell2}) and $x_n(j)$ be the coefficients (\ref{Evaluation}) appearing in Taylor's expansion (\ref{phi-Taylor}) of  the normalized holomorphic differential $\phi$ near the point $P=\infty^j$.
Then for any positive integer $\alpha$, the differential $\Psi_{j,\alpha}(P)$ (\ref{Psi-ialpha-def}) is given by:
\begin{equation}\label{Psi-i-alpha}
\begin{split}
\Psi_{j,\alpha}(P)&=\bigg(2{\alpha}x_{\alpha}(j)\zeta(1/2)-\alpha
\sum_{k=1}^{\alpha}\frac{(-1)^k}{k!}\mathcal{B}_{\alpha,k}(j)\wp^{(k-1)}\big(z_P-\infty^j\big)\bigg)\phi(P)\\
&=\bigg(2{\alpha}x_{\alpha}(j)\zeta(1/2)+\alpha\mathbf{L}_{j,\alpha}^{\phi}\big[\zeta\big](\infty^j-z_P)\bigg)\phi(P).
\end{split}
\end{equation}
Here $\mathbf{L}_{j,\alpha}^{\phi}\big[\zeta\big]$ is the function  (\ref{Bell-f}) obtained by the action of the Bell operator (\ref{Bell operator}) on the Weierstrass zeta function.
\end{Prop}
\emph{Proof:} Using the relation $\wp(u)=-\partial_u\zeta(u)$ (\ref{zeta-derivative}) and expression (\ref{Bell-f}) for $\mathbf{L}_{j,\alpha}^\phi[f]$, we deduce that the second equality in (\ref{Psi-i-alpha}) is an immediate consequence of the first one. \\
The Taylor series of the function $x\longmapsto \wp(u-x)$ at $x=0$ (with $u\neq 0$) is given by
$$
\wp(u-x){\underset{x\sim0}=}\sum_{n=0}^{\infty}(-1)^n\frac{\wp^{(n)}(u)}{n!}x^n.
$$
From this and (\ref{Abel-Laurent1}), we find
\begin{align*}
\wp\big(z_P-z_Q\big)&=\wp\big((z_P-\infty^j)-(z_Q-\infty^j)\big)\\
&{\underset{Q\sim\infty^j}=}\sum_{k=0}^{\infty}\frac{(-1)^k}{k!}\wp^{(k)}\big(z_P-\infty^j\big)\Big(\mathcal{A}(Q;\infty^j)\Big)^k\\
&=\wp\big(z_P-\infty^j\big)+\sum_{k=1}^{\infty}\frac{(-1)^k}{k!}\wp^{(k)}\big(z_P-\infty^j\big)\bigg(\sum_{n=1}^{\infty}x_n(j)\big(\z_j(Q)\big)^{j}\bigg)^k\\
&=\wp\big(z_P-\infty^j\big)+\sum_{k=1}^{\infty}\frac{(-1)^k}{k!}\wp^{(k)}\big(z_P-\infty^j\big)
\bigg(\sum_{n=k}^{\infty}\mathcal{B}_{n,k}\big(x_1(j),\dots,x_{n-k+1}(j)\big)\big(\z_j(Q)\big)^{n}\bigg)\\
&=\wp\big(z_P-\infty^j\big)+\sum_{k=1}^{\infty}\sum_{n=k}^{\infty}\bigg(\frac{(-1)^k}{k!}\mathcal{B}_{n,k}(j)\wp^{(k)}\big(z_P-\infty^j\big)\big(\z_j(Q)\big)^{n}\bigg).
\end{align*}
Accordingly, the differential $\frac{W(P,Q)}{\phi(P)}$ has  the following Laurent series at $Q=\infty^j$:
\begin{align*}
\frac{W(P,Q)}{\phi(P)}&=\Big(\wp\big(z_P-z_Q\big)+2\zeta(1/2)\Big)\phi(Q)\\
&{\underset{Q\sim\infty^j}=}\Big(2\zeta(1/2)+\wp\big(z_P-\infty^j\big)\Big)\bigg(\sum_{r=1}^{\infty}rx_{r}(j)\big(\z_j(Q)\big)^{r-1}\bigg)d\z_j(Q)\\
&\quad +\Bigg(\sum_{k=1}^{\infty}\sum_{n=k}^{\infty}\bigg\{\frac{(-1)^k}{k!}\mathcal{B}_{n,k}(j)\wp^{(k)}\big(z_P-\infty^j\big)\big(\z_j(Q)\big)^{n}\bigg\}\Bigg)
\bigg(\sum_{r=1}^{\infty} rx_{r}(j)\big(\z_j(Q)\big)^{r-1}\bigg)d\z_j(Q)\\
&=\Big(2\zeta(1/2)+\wp\big(z_P-\infty^i\big)\Big)\bigg(\sum_{r=1}^{\infty}rx_{r}(j)\big(\z_j(Q)\big)^{r-1}\bigg)d\z_j(Q)\\
&\quad+\Bigg(\sum_{k=1}^{\infty}\sum_{n=k}^{\infty}\sum_{r=1}^{\infty}\bigg\{\frac{(-1)^k}{k!}rx_r(j)\mathcal{B}_{n,k}(j)\wp^{(k)}\big(z_P-\infty^j\big)
\big(\z_j(Q)\big)^{n+r-1}\bigg\}\Bigg)d\z_j(Q),
\end{align*}
where we used formula (\ref{W-g1-phi}) in the first equality and  Taylor's expansion (\ref{phi-Taylor}) for the holomorphic differential $\phi$ in the second equality.\\
Now, the obtained behavior of the bidifferential $W(P,Q)$ near $Q=\infty^j$ and  recurrence formula (\ref{Bell-rec1}) for the partial exponential Bell polynomials imply that
\begin{align*}
&\Psi_{j,\alpha}(P):=\underset{\infty^j} {\rm res}\  \lambda(Q)^{\frac{\alpha}{n_j+1}}W(P,Q)\\
&=\Big(2{\alpha}x_{\alpha}(j)\zeta(1/2)+{\alpha}x_{\alpha}(j)\wp(z_P-\infty^j)\Big)\phi(P)\\
&+\left\{\big(1-\delta_{\alpha,1}\big)
\sum_{k=1}^{\alpha-1}\sum_{n=k}^{\alpha-1}\frac{(-1)^k}{k!}(\alpha-n)x_{\alpha-n}(j)\mathcal{B}_{n,k}(j)\wp^{(k)}\big(z_P-\infty^j\big)\right\}\phi(P)\\
&=\Big(2{\alpha}x_{\alpha}(j)\zeta(1/2)+{\alpha}x_{\alpha}(j)\wp(z_P-\infty^j)\Big)\phi(P)\\
&\quad +\left\{\big(1-\delta_{\alpha,1}\big)\alpha
\sum_{k=1}^{\alpha-1}\frac{(-1)^k}{(k+1)!}\mathcal{B}_{\alpha,k+1}(j)\wp^{(k)}\big(z_P-\infty^j\big)\right\}\phi(P).
\end{align*}
Finally, since $\mathcal{B}_{\alpha,1}(j)=x_{\alpha}(j)$ by (\ref{Bell-Examples}), we get the expression claimed in the proposition.

\fd

\medskip

In the next proposition we study the characteristic properties of the differential $\Psi_{j,\alpha}$ (\ref{Psi-i-alpha}).
\begin{Prop} Let $j=0,\dots,m$, $\alpha$ be a positive integer and $\Psi_{j,\alpha}$ be the differential defined by (\ref{Psi-i-alpha}). Then $\Psi_{j,\alpha}$ is a normalized  Abelian differential of the
second kind. It has the only pole at the point $\infty^j$ with the following principal part:
\begin{equation}\label{Psi-i-alpha-PP}
\Psi_{j,\alpha}(P)\underset{P\sim\infty^j}{=}\Big(\frac{\alpha}{\z_j(P)^{\alpha+1}}+\mathcal{O}(1)\Big)d\z_j(P),\quad \quad \quad \lambda(P)=\z_j(P)^{-n_j-1}.
\end{equation}
Moreover, the periods of the differential $\Psi_{j,\alpha}$ are as follows:
\begin{equation}\label{Psi-ialpha-periods}
\oint_a\Psi_{j,\alpha}=0\quad \quad \text{and}\quad \quad \oint_b\Psi_{j,\alpha}=2{\rm i}\pi{\alpha}x_{\alpha}(j).
\end{equation}
\end{Prop}
\emph{Proof:} This statement follows from definition (\ref{Psi-ialpha-def}) of $\Psi_{j,\alpha}$ together with the properties (\ref{W-asymptotic})-(\ref{W-periods}) of the bidifferential $W(P,Q)$ (\ref{W-g1}) (see \cite{Vasilisa, Rejeb23} for details), but let us give a direct proof using expression (\ref{Psi-i-alpha}).\\
Due to (\ref{Psi-i-alpha}), the periodicity (resp. the quasi-periodicity) property of the function $\wp$ (resp. $\zeta$) and Lendgre's identity (\ref{Legendre}) and the fact that $\mathcal{B}_{\alpha,1}(j)=x_{\alpha}(j)$, we have
\begin{align*}
&\oint_a\Psi_{j,\alpha}=2{\alpha}x_{\alpha}(j)\zeta(1/2)+\alpha\mathcal{B}_{\alpha,1}(j)\int_{x}^{x+1}\wp(z_P-\infty^j)dz_P=0;\\
&\oint_b\Psi_{j,\alpha}=2{\alpha}x_{\alpha}(j)\tau\zeta(1/2)+\alpha\mathcal{B}_{\alpha,1}(j)\int_{x}^{x+\tau}\wp(z_P-\infty^i)dz_P=2{\rm i}\pi{\alpha}x_{\alpha}(j).
\end{align*}
Now, in order to arrive at the asymptotic behavior (\ref{Psi-i-alpha-PP}), we shall calculate the residue  of the differentials $(z_P-\infty^j)^{k}\Psi_{j,\alpha}(P)$, $k=0,\dots,\alpha$, using the two local parameters $z_P-\infty^j$ and $\z_j(P)=\lambda(P)^{-\frac{1}{n_j+1}}$ near the point $P=\infty^j$ and then compare the results.\\
In  view of the following singular part of the function $\wp^{(k)}$ near $u=0$ which follows from (\ref{Laurend-p})
\begin{align*}
\forall\ k\geq 0,\quad \quad \wp^{(k)}(u)\underset{u\sim0}=(-1)^k\frac{(k+1)!}{u^{k+2}}+\mathcal{O}(1),
\end{align*}
we deduce that  the differential $\Psi_{j,\alpha}$ (\ref{Psi-i-alpha}) behaves as follows near the point $P=\infty^j$:
\begin{align*}
\Psi_{j,\alpha}(P)&\underset{P\sim\infty^j}{=}\left(\sum_{k=1}^{\alpha}\frac{\alpha\mathcal{B}_{\alpha,k}(j)}{(z_P-\infty^j)^{k+1}}+\mathcal{O}(1)\right)dz_P.
\end{align*}
As consequence, we obtain
\begin{equation}\label{Residue1}
\forall\ k=0,\dots,\alpha, \quad \quad \underset{\infty^j} {\rm res}\Big((z_P-\infty^j)^{k}\Psi_{j,\alpha}(P)\Big)=
\big(1-\delta_{k,0}\big)\alpha\mathcal{B}_{\alpha,k}(j).
\end{equation}
On the other hand, with respect to the local parameter $\z_j(P)=\lambda(P)^{-\frac{1}{n_j+1}}$ near  the pole $\infty^j$ of $\lambda$,  the singular part of the differential  $\Psi_{j,\alpha}(P)$  takes the following form:
$$
\Psi_{j,\alpha}(P)\underset{P\sim\infty^j}{=}\bigg(\sum_{\ell=0}^{\alpha}c_{\ell}\z_j(P)^{-\ell-1}+\mathcal{O}(1)\bigg)d\z_j(P).
$$
This and Laurent expansion   (\ref{Abel-Laurent1})  for $\Big(\mathcal{A}(P;\infty^j)\Big)^{k}$  imply that
\begin{align*}
(z_P-\infty^j)^{k}\Psi_{j,\alpha}(P)&=\Big(\mathcal{A}(P;\infty^j)\Big)^{k}\Psi_{j,\alpha}(P)
\underset{P\sim\infty^j}{=}\bigg(\sum_{\ell=0}^{\alpha}\sum_{n=k}^{\infty}c_{\ell}\mathcal{B}_{n,k}(j)\z_j(P)^{n-\ell-1}+\dots\bigg)d\z_j(P)
\end{align*}
and then
\begin{equation}\label{Residue2}
\begin{split}
&\underset{\infty^j} {\rm res}\Big((z_P-\infty^j)^{k}\Psi_{j,\alpha}(P)\Big)=\sum_{\ell=k}^{\alpha}c_{\ell}\mathcal{B}_{\ell,k}(j)\\
&=\delta_{k,0}c_0+\big(1-\delta_{k,0}\big)\sum_{\ell=k}^{\alpha}c_{\ell}\mathcal{B}_{\ell,k}(j),
\end{split}
\quad \quad \quad \quad \forall\ k=0,\dots,\alpha.
\end{equation}
Finally, the results in (\ref{Residue1}) and (\ref{Residue2}) prove  that $c_0=0$ and
$$
\forall\ k=1,\dots,\alpha, \quad \quad \sum_{\ell=k}^{\alpha}c_{\ell}\mathcal{B}_{\ell,k}(j)=\alpha\mathcal{B}_{\alpha,k}(j)
$$
and this leads to the following system:
\begin{align*}
&c_{\alpha}\mathcal{B}_{\alpha,\alpha}=\alpha\mathcal{B}_{\alpha,\alpha};\\
&c_{\alpha-1}\mathcal{B}_{\alpha-1,\alpha-1}+c_{\alpha}\mathcal{B}_{\alpha,\alpha-1}=\alpha\mathcal{B}_{\alpha,\alpha-1};\\
&\quad \quad \vdots\\
&c_1\mathcal{B}_{1,1}+c_2\mathcal{B}_{2,1}+\dots +c_{\alpha-1}\mathcal{B}_{\alpha-1,1}+c_{\alpha}\mathcal{B}_{\alpha,1}=\alpha\mathcal{B}_{\alpha,1}.
\end{align*}
Thus by taking into account the fact that
$$
\mathcal{B}_{n,n}(j)=\big(x_1(j)\big)^n\neq 0, \quad \quad \forall\ n,
$$
we get
$$
c_{\alpha}=\alpha,\quad \quad c_{\alpha-1}=\dots=c_1=0
$$
as desired.

\fd

\medskip

Now we move on  to calculate the following quantities:
\begin{equation}\label{chi-ij-def}
\mathcal{I}_{i,\beta}[\Psi_{j,\alpha}]:=\underset{\infty^i} {\rm res}\  \lambda(P)^{\frac{\beta}{n_i+1}}\Psi_{j,\alpha}(P)
\end{equation}
which appear as the coefficients of the Laurent expansion  of the Abelian differential $\Psi_{j,\alpha}$ (\ref{Psi-i-alpha}) near the prescribed pole $\infty^i$ of order $n_i+1$ of the covering $\lambda$.
\begin{Thm}\label{Second kind-thm} Let $i,j\in \{0,1,\dots,m\}$ and $\alpha, \beta$ be two positive integers. For $n\geq k\geq 1$, denote by $\mathcal{B}_{n,k}(i):=\mathcal{B}_{n,k}(x_1(i),\dots,x_{n-k+1}(i))$  the partial Bell polynomials
(\ref{Bell-partial}), with $\big\{x_n(i)\big\}_{n\geq 1}$ being the coefficients defined by (\ref{Evaluation}).
\begin{description}
\item[1)] If $i\neq j$, then the quantity $\mathcal{I}_{i,\beta}[\Psi_{j,\alpha}]$ (\ref{chi-ij-def}) is given by
\begin{equation}\label{chi-ij}
\begin{split}
\mathcal{I}_{i,\beta}[\Psi_{j,\alpha}]&=2{\alpha\beta}x_{\alpha}(j)x_{\beta}(i)\zeta(1/2)
-\alpha\beta\sum_{k=1}^{\alpha}\sum_{\ell=1}^{\beta}\frac{(-1)^k}{k!\ell!}\mathcal{B}_{\alpha,k}(j)\mathcal{B}_{\beta,\ell}(i)\wp^{(k+l-2)}(\infty^i-\infty^j)\\
&=2{\alpha\beta}x_{\alpha}(j)x_{\beta}(i)\zeta(1/2)+\alpha\beta\sum_{k=1}^{\alpha}\frac{(-1)^k}{k!}\mathcal{B}_{\alpha,k}(j)
\mathbf{L}_{i,\beta}^{\phi}\big[\K^{(k)}](\infty^i-\infty^j).
\end{split}
\end{equation}
Here, as above, $\mathbf{L}_{i,\beta}^{\phi}\big[\K]$ denotes the function  defined by (\ref{Bell-f}) and $\K(u)=\log\sigma(u)$, with $\sigma$ being the Weierstrass sigma function (\ref{sigma-function}).
\item[2)] When $i=j$, then by means of the function $\mathcal{R}_{\mu,k}$ (\ref{R-function})-(\ref{Bell-rec4}) and the non-normalized
Eisenstein series (\ref{Eisenstein G}), the quantity $\mathcal{I}_{j,\beta}[\Psi_{j,\alpha}]$ in (\ref{chi-ij-def}) is given by
\begin{equation}\label{chi-ii}
\begin{split}
\mathcal{I}_{j,\beta}[\Psi_{j,\alpha}]&=2{\alpha\beta}x_{\alpha}(j)x_{\beta}(i)\zeta(1/2)+\alpha\beta
\sum_{k=1}^{\alpha}\mathcal{B}_{\alpha,k}(j)\mathcal{R}_{\beta+k,k+1}\big(x_1(j),\dots,x_{\beta+k+1}(j)\big)\\
&\quad -\alpha\beta\sum_{k=1}^{\alpha}\left((-1)^k\mathcal{B}_{\alpha,k}(j)
\sum_{\ell=1, k+\ell\geq 4}^{\beta}\frac{1}{k+\ell}\binom{k+\ell}{\ell}\mathcal{B}_{\beta,\ell}(j)G_{k+\ell}\right),
\end{split}
\end{equation}
where, as in (\ref{convention}), the sum $\sum_{\ell=1, k+\ell\geq 4}^{\beta}$ takes the value 0 whenever $\beta+k<4$.
\end{description}

\end{Thm}
\emph{Proof:} Using (\ref{Psi-i-alpha}) and (\ref{Evaluation}), we see that
\begin{align*}
&\mathcal{I}_{i,\beta}[\Psi_{j,\alpha}]:=\underset{\infty^i} {\rm res}\  \lambda(P)^{\frac{\beta}{n_i+1}}\Psi_{j,\alpha}(P)\\
&=2{\alpha\beta}x_{\alpha}(j)x_{\beta}(i)\zeta(1/2)-\alpha\sum_{k=1}^{\alpha}\frac{(-1)^k}{k!}\mathcal{B}_{\alpha,k}(j)
\underset{\infty^i} {\rm res}\bigg(\lambda(P)^{\frac{\beta}{n_i+1}}\wp^{(k-1)}\big(z_P-\infty^j\big)\phi(P)\bigg).
\end{align*}
Therefore, it suffices to compute the following residue:
\begin{equation}\label{R-kij}
\mathcal{I}_{i,\beta}\big[\wp^{(k)}(z_P-\infty^j)\phi\big]:=\underset{\infty^i} {\rm res}
\bigg(\lambda(P)^{\frac{\beta}{n_i+1}}\wp^{(k)}\big(z_P-\infty^j\big)\phi(P)\bigg),\quad \quad \forall\ k\geq 0.
\end{equation}
\textbf{1)} Let $k$ be a nonnegative integer. According to  (\ref{Abel-Laurent1}), we can write the following Taylor series for the function $\wp^{(k)}(z_P-\infty^j)$ at $P=\infty^i$, with $i\neq j$:
\begin{align*}
\wp^{(k)}(z_P-\infty^j)&=\wp^{(k)}(z_P-\infty^i+\infty^i-\infty^j))=\wp^{(k)}\big(\infty^i-\infty^j+ \mathcal{A}(P,\infty^i)\big)\\
&\underset{P\sim \infty^i}{=}\wp^{(k)}(\infty^i-\infty^j)
+\sum_{\ell=1}^{\infty}\frac{\wp^{(k+\ell)}(\infty^i-\infty^j)}{\ell!}\bigg(\sum_{r=1}^{\infty}x_r(i)\z_j(P)^r\bigg)^{\ell}\\
&=\wp^{(k)}(\infty^i-\infty^j)+\sum_{\ell=1}^{\infty}\sum_{\varepsilon=\ell}^{\infty}
\bigg(\frac{1}{\ell!}\mathcal{B}_{\varepsilon,\ell}(i)\wp^{(k+l)}(\infty^i-\infty^j)\bigg)\z_i(P)^{\varepsilon}.
\end{align*}
From this and Taylor's series (\ref{phi-Taylor}) for the holomorphic differential $\phi$ we find
\begin{align*}
&\wp^{(k)}(z_P-\infty^j)\phi(P)\underset{P\sim \infty^i}{=}\wp^{(k)}(\infty^i-\infty^j)\bigg(\sum_{r=1}^{\infty}rx_r(i)\z_i(P)^{r-1}\bigg)d\z_i(P)\\
&+\left\{\sum_{\ell=1}^{\infty}\frac{1}{\ell!}\wp^{(k+l)}(\infty^i-\infty^j)\sum_{\varepsilon=\ell}^{\infty}\sum_{r=1}^{\infty}
rx_r(i)\mathcal{B}_{\varepsilon,\ell}(i)\z_i(P)^{\varepsilon+r-1}\right\}d\z_i(P).
\end{align*}
Thus, due to recurrence relation (\ref{Bell-rec1}) and the fact that $\mathcal{B}_{\beta,1}(i)=x_{\beta}(i)$, we get
\begin{align*}
&\mathcal{I}_{i,\beta}\big[\wp^{(k)}(z_P-\infty^j)\phi\big]:=\underset{\infty^i} {\rm res}
\bigg(\lambda(P)^{\frac{\beta}{n_i+1}}\wp^{(k)}\big(z_P-\infty^j\big)\phi(P)\bigg)\\
&={\beta}x_{\beta}(i)\wp^{(k)}(\infty^i-\infty^j)+\big(1-\delta_{\ell,1}\big)
\sum_{\ell=1}^{\beta-1}\frac{1}{\ell!}\wp^{(k+l)}(\infty^i-\infty^j)\left\{\sum_{\varepsilon=\ell}^{\beta-1}(\beta-\varepsilon)
x_{\beta-\varepsilon}(i)\mathcal{B}_{\varepsilon,\ell}(i)\right\}\\
&=\beta\sum_{\ell=0}^{\beta-1}\frac{1}{(\ell+1)!}\mathcal{B}_{\beta,\ell+1}(i)\wp^{(k+l)}(\infty^i-\infty^j)
=\beta\sum_{\ell=1}^{\beta}\frac{1}{\ell!}\mathcal{B}_{\beta,\ell}(i)\wp^{(k+l-1)}(\infty^i-\infty^j).
\end{align*}
The second equality in (\ref{chi-ij}) is an immediate consequence of the first equality,  the fact that $\wp(u)=-\K''(u)$ and expression (\ref{Bell-f}) for the  action of the Bell operator on the function $\K(u)=\log\sigma(u)$.

\medskip

\noindent \textbf{2)}   As above, in order to calculate  the residue $\mathcal{I}_{j,\beta}\big[\wp^{(k)}(z_P-\infty^j)\phi\big]$ (\ref{R-kij}),
we start with writing the Laurent series of the meromorphic differential  $\wp^{(k)}(z_P-\infty^j)\phi(P)$ near its pole $P=\infty^j$.\\
By (\ref{Laurend-p}), we know that the series expansion of $\wp^{(k)}$, $\forall\ k\geq 0$, near the origin  is given by
\begin{equation}\label{Laurent p-k}
\begin{split}
\wp^{(k)}(u)&=(-1)^{k}\frac{(k+1)!}{u^{k+2}}+\sum_{\ell=2, \ell\geq k}^{\infty}(\ell+1)!G_{\ell+2}\frac{u^{\ell-k}}{(\ell-k)!}\\
&=(-1)^{k}\frac{(k+1)!}{u^{k+2}}+(k+1)!\sum_{\ell=0,\ \ell+k\geq 2}^{\infty}\binom{\ell+k+1}{\ell}G_{\ell+k+2}u^{\ell}.
\end{split}
\end{equation}
Therefore, due to (\ref{Abel-Laurent1}) and (\ref{Abel-Laurent2}) we have
\begin{align*}
&\wp^{(k)}(z_P-\infty^j)-(-1)^k(k+1)!(z_P-\infty^j)^{-k-2}\\
&=\wp^{(k)}\Big(\mathcal{A}(P;\infty^j)\Big)-(-1)^k(k+1)!\Big(\mathcal{A}(P;\infty^j)\Big)^{-k-2}\\
&\underset{P\sim \infty^j}{=}(k+1)!\sum_{\ell=0,\ \ell+k\geq2}^{\infty}\binom{k+\ell+1}{\ell}G_{\ell+k+2}\Big(\mathcal{A}(P;\infty^j)\Big)^{\ell}\\
&=(k+1)!\sum_{\ell=0,\ \ell+k\geq2}^{\infty}\left\{\binom{k+\ell+1}{\ell}G_{\ell+k+2}\sum_{n=\ell}^{\infty}\mathcal{B}_{n,\ell}(j)\z_j(P)^n\right\}.
\end{align*}
In view of this  and (\ref{phi-Taylor}), we get  the following expansion for the differential $\wp^{(k)}(z_P-\infty^i)\phi(P)$:
\begin{align*}
&\quad \quad \wp^{(k)}(z_P-\infty^j)\phi(P)-(-1)^k(k+1)!\Big(\mathcal{A}(P;\infty^j)\Big)^{-k-2}\phi(P)\\
&\underset{P\sim \infty^j}{=}(k+1)!\sum_{\ell=0, k+\ell\geq 2}^{\infty}\left\{\binom{k+\ell+1}{\ell}G_{\ell+k+2}\sum_{n=\ell}^{\infty}\sum_{r=1}^{\infty}
rx_r(j)\mathcal{B}_{n,\ell}(j)\right\}\z_j(P)^{n+r-1}d\z_j(P).
\end{align*}
Thus, using (\ref{Bell-R-formula}) and recurrence relation (\ref{Bell-rec1}) for Bell polynomials, we arrive at
\begin{align*}
&\mathcal{I}_{j,\beta}\big[\wp^{(k)}(z_P-\infty^j)\phi\big]:=\underset{\infty^j} {\rm res}
\bigg(\lambda(P)^{\frac{\beta}{n_j+1}}\wp^{(k)}\big(z_P-\infty^j\big)\phi(P)\bigg)\\
&=(-1)^k(k+1)!\underset{\infty^j} {\rm res}\bigg(\lambda(P)^{\frac{\beta}{n_j+1}}\Big(\mathcal{A}(P;\infty^j)\Big)^{-k-2}\phi(P)\bigg)\\
&\quad+\big(1-\delta_{\beta+k,1}-\delta_{\beta+k,2}\big)(k+1)!\sum_{\ell=0, k+\ell\geq 2}^{\beta-1}
\left\{\binom{k+\ell+1}{\ell}G_{\ell+k+2}\sum_{n=\ell}^{\beta-1}(\beta-n)x_{\beta-n}(j)\mathcal{B}_{n,\ell}(j)\right\}\\
&=(-1)^k(k+1)!\beta\mathcal{R}_{\beta+k+1,k+2}\big(x_1(j),\dots,x_{\beta+k+2}(j)\big)\\
&\quad+\big(1-\delta_{\beta+k,1}-\delta_{\beta+k,2}\big)(k+1)!\beta\sum_{\ell=1, k+\ell\geq 3}^{\beta}\frac{1}{k+\ell+1}\binom{k+\ell+1}{\ell}\mathcal{B}_{\beta,\ell}(j)G_{k+\ell+1}.
\end{align*}

\fd

\begin{Cor} With respect to the local parameter $\z_i(P)$ defined by $\z_i(P):=\lambda(P)^{-\frac{1}{n_i+1}}$,  the Laurent expansion of the differential $\Psi_{j,\alpha}$ (\ref{Psi-i-alpha}) near the pole  $P=\infty^i$ of $\lambda$ is given by
$$
\Psi_{j,\alpha}(P)\underset{P\sim \infty^i}{=}\bigg(\delta_{ij}\frac{\alpha}{\z_i(P)^{\alpha+1}}+\sum_{\beta=1}^{\infty}\Big(\mathcal{I}_{i,\beta}[\Psi_{j,\alpha}]\Big)\z_i(P)^{\beta-1}\bigg)d\z_i(P),
$$
where the coefficients $\mathcal{I}_{i,\beta}[\Psi_{j,\alpha}]$ are  given by (\ref{chi-ij}) and (\ref{chi-ii}).

\end{Cor}

As an application of the results (\ref{Psi-i-alpha}) and (\ref{Psi-i-alpha-PP}), we can derive  the following (new) recurrence relation for partial ordinary Bell polynomials involving the rational function $\mathcal{R}_{\mu,k}$ (\ref{Bell-rec4}).
\begin{Prop} For any   positive integers $\alpha\leq n$, the partial ordinary Bell polynomial $\mathcal{B}_{n+1,\alpha}$ satisfies
\begin{equation}\label{Bell-rec5}
\begin{split}
\mathcal{B}_{n+1,\alpha}&=\alpha x_1^{\alpha}\sum_{k=\alpha+1}^{n+1}\mathcal{B}_{n+1,k}\mathcal{R}_{k-\alpha,k+1},
\end{split}
\end{equation}
where $\mathcal{R}_{k-\alpha,k+1}=\mathcal{R}_{k-\alpha,k+1}(x_1,\dots,x_{k-\alpha+1})$ is the function defined by (\ref{Bell-rec4}).
\end{Prop}
\emph{Proof:} Denote by $x_r:=x_r(0)$ the coefficients in the Taylor series (\ref{phi-Taylor}) of the holomorphic differential $\phi(P)=dz_P$ near the pole $\infty^0$.
Then, reasoning  as in the proof of (\ref{Bell-R-formula}), we get
\begin{align*}
{\underset{\infty^0}{\rm res}}
\Big(\lambda(P)^{\frac{-\alpha}{n_0+1}}\big(z_P-\infty^0\big)^{-k-1}\phi(P)\Big)
=\left\{
\begin{array}{llll}
0 & \hbox{if} \quad \quad 1\leq k<\alpha;\\
\\
x_1^{-\alpha}& \hbox{if} \quad \quad k=\alpha\geq 1;\\
\\
-\alpha\mathcal{R}_{k-\alpha,k+1}& \hbox{if} \quad \quad k>\alpha\geq1.
\end{array}
\right.
\end{align*}
Now, due to  expression (\ref{Psi-i-alpha}) for the particular  Abelian differential $\Psi_{0,n_0+1}(P)$ as well as its behavior (\ref{Psi-i-alpha-PP}) near the pole $\infty^0$, we can write for $\alpha=1,\dots,n_0$ the following:
\begin{align*}
0&={\underset{\infty^0}{\rm res}}\Big(\lambda(P)^{\frac{-\alpha}{n_0+1}}\Psi_{0,n_0+1}(P)\Big)\\
&=\sum_{k=1}^{n_0+1}\frac{(-1)^k}{k!}\mathcal{B}_{n_0+1,k}\big(x_1,\dots,x_{n_0+2-k}\big){\underset{\infty^0}{\rm res}}
\Big(\lambda(P)^{\frac{-\alpha}{n+1}}\wp^{(k-1)}\big(z_P-\infty^0\big)\phi(P)\Big)\\
&=\sum_{k=1}^{n_0+1}\mathcal{B}_{n_0+1,k}\big(x_1,\dots,x_{n_0+2-k}\big){\underset{\infty^0}{\rm res}}
\Big(\lambda(P)^{\frac{-\alpha}{n+1}}\big(z_P-\infty^0\big)^{-k-1}\phi(P)\Big)\\
&=x_1^{-\alpha}\mathcal{B}_{n_0+1,\alpha}\big(x_1,\dots,x_{n_0+2-\alpha}\big)
-\alpha\sum_{k=\alpha+1}^{n_0+1}\mathcal{B}_{n_0+1,k}\big(x_1,\dots,x_{n_0+2-k}\big)\mathcal{R}_{k-\alpha,k+1}\big(x_1,\dots,x_{k-\alpha+1}\big),
\end{align*}
where we used the Laurent series (\ref{Laurent p-k}) for $\wp^{(k)}$ in the third equality.

\fd


\subsection{Landau-Ginzburg superpotentials parametrization of the  Hurwitz space $\mathcal{H}_1(n_0,\dots,n_m)$}
This purpose of this section is  to construct  explicit Landau-Ginzburg (LG) superpotential of the Hurwitz space $\mathcal{H}_1(n_0,\dots,n_m)$ in terms of the variables
$$
\Big\{x_1(j),\dots,x_{n_j+1}(j):\quad j=0,\dots,m\Big\}
$$
related to the holomorphic differential $\phi$ via (\ref{phi-Taylor})-(\ref{Evaluation}). The main result here is the following.

\begin{Thm}\label{LG}
Let $\lambda$  be a meromorphic function on the torus $\C/{\mathbb{L}}$, with $\mathbb{L}$ being  the lattice $\mathbb{L}=\Z+\tau\Z$ and $\Im\tau>0$, and $\infty^0,\dots,\infty^m$ be the poles of $\lambda$ of order $n_0+1,\dots,n_m+1$, respectively. Let  $x_{r}(j)$, $j=0,\dots,m$,  be the coefficients  in  Taylor's expansion (\ref{phi-Taylor}) of the normalized holomorphic differential $\phi$ near the point $\infty^j$. Then  $\lambda$ admits the following representation:
\begin{equation}\label{lambda-zeta}
\lambda(P)=c-\sum_{j=0}^m(n_j+1)\left(\sum_{k=1}^{n_j+1}\frac{(-1)^k}{k!}\mathcal{B}_{n_j+1,k}(j)\zeta^{(k-1)}\big(z_P-\infty^j\big)\right), \quad P\in \C/{\mathbb{L}}
\end{equation}
where $c$ is a constant (with respect to $z_P\in \C$), $\mathcal{B}_{n_j+1,k}(j):=\mathcal{B}_{n_j+1,k}\big(x_1(j),\dots,x_{n_j+1-k+1}(j)\big)$ is the partial ordinary  Bell polynomial (\ref{Bell-partial})-(\ref{Bell2})
and $\zeta$ is the Weierstrass function (\ref{zeta-function}).
\end{Thm}
\emph{Proof:} Using the relation $\zeta^{(k-1)}=-\wp^{(k-2)}$ for all $k\geq 2$ and the fact that  $\mathcal{B}_{n_j+1,1}(j)=x_{n_j+1}(j)$ and taking into account equality (\ref{x-r-sum}), we easily see that the right hand side of (\ref{lambda-zeta}) defines an $\mathbb{L}$-elliptic function. \\
Let $\Psi_{j,n_j+1}$ be the Abelian  differential of the second kind given by (\ref{Psi-i-alpha}).
Then, in view of the principal part (\ref{Psi-i-alpha-PP}) of the differential $\Psi_{j,n_j+1}$ near $\infty^j$ and the fact that
$$
\forall\ j=0,\dots,m,\quad \quad d\lambda(P)\underset{P\sim \infty^j}{=}-\frac{n_j+1}{\z_j(P)^{n_j+2}}d\z_j(P),\quad \quad \lambda(P)=\z_j(P)^{-n_j-1},
$$
we deduce that  the differential
$$
\Psi(P):=d\lambda(P)+\sum_{j=0}^m\Psi_{j,n_j+1}(P)
$$
has no pole and its $a$-period is zero. Thus, $\Psi$ must vanishes identically. Therefore, due to (\ref{Psi-i-alpha}), we have
\begin{align*}
d\lambda(P)&=-\sum_{j=0}^m\Psi_{j,n_j+1}(P)\\
&=-\sum_{j=0}^m(n_j+1)\left\{2x_{n_j+1}(j)\zeta(1/2)-\sum_{k=1}^{n_j+1}\frac{(-1)^k}{k!}\mathcal{B}_{n_j+1,k}(j)\wp^{(k-1)}\big(z_P-\infty^j\big)\right\}dz_P\\
&=\partial_{z_P}\sum_{j=0}^m(n_j+1)\left\{-2x_{n_j+1}(j)\zeta(1/2)z_P-\sum_{k=1}^{n_j+1}\frac{(-1)^k}{k!}\mathcal{B}_{n_j+1,k}(j)\zeta^{(k-1)}\big(z_P-\infty^j\big)\right\}dz_P.
\end{align*}
This implies that there is a constant $c$ (with respect to $z_P$) such that
\begin{align*}
\lambda(P)&=c-2z_P\zeta(1/2)\sum_{j=0}^m(n_j+1)x_{n_j+1}(j)-\sum_{j=0}^m(n_j+1)\left(
\sum_{k=1}^{n_j+1}\frac{(-1)^k}{k!}\mathcal{B}_{n_j+1,k}(j)\zeta^{(k-1)}\big(z_P-\infty^j\big)\right)\\
&=c-\sum_{j=0}^m(n_j+1)\left(\sum_{k=1}^{n_j+1}\frac{(-1)^k}{k!}\mathcal{B}_{n_j+1,k}(j)\zeta^{(k-1)}\big(z_P-\infty^j\big)\right),
\end{align*}
where the second equality follows from (\ref{x-r-sum}).

\fd

As a corollary, we investigate further equivalent expressions for the LG superpotential (\ref{lambda-zeta}).
\begin{Cor} Let $\lambda$ be the meromorphic function given by (\ref{lambda-zeta}). Then $\lambda$ takes also the following forms:
\begin{align}
\begin{split}\label{lambda-p}
\lambda(P)&=c+(1-\delta_{m,0})\sum_{j=1}^m(n_j+1)x_{n_j+1}(j)\Big(\zeta(z_P-\infty^j)-\zeta(z_P-\infty^0)\Big)\\
&\quad +\sum_{j=0}^m\big(1-\delta_{n_j,0}\big)(n_j+1)\left(\sum_{k=2}^{n_j+1}\frac{(-1)^k}{k!}\mathcal{B}_{n_j+1,k}(j)\wp^{(k-2)}\big(z_P-\infty^j\big)\right)
\end{split}\\
\begin{split}\label{lambda-sigma}
&=c-\sum_{j=0}^m(n_j+1)\left(\sum_{k=1}^{n_j+1}\frac{(-1)^k}{k!}\mathcal{B}_{n_j+1,k}(j)\partial_{z_P}^k\log\sigma\big(z_P-\infty^j\big)\right)
\end{split}\\
\begin{split}\label{lambda-Bell L}
&=c-\sum_{j=0}^m(n_j+1)\mathbf{L}_{j,n_j+1}^{\phi}\big[\K\big](\infty^j-z_P)
\end{split}\\
\begin{split}\label{lambda-theta}
&=c-\zeta(1/2)\sum_{j=0}^m(n_j+1)\left(2x_{n_j+1}(j)\infty^j+\big(1-\delta_{n_j,0}\big)\sum_{k=1}^{n_j}x_k(j)x_{n_j+1-k}(j)\right)\\
&\quad -\sum_{j=0}^m(n_j+1)\left(\sum_{k=1}^{n_j+1}\frac{(-1)^k}{k!}\mathcal{B}_{n_j+1,k}(j)\partial_{z_P}^k\log\big(\theta_1(z_P-\infty^i)\big)\right).
\end{split}
\end{align}
Here  the functions $\wp,\zeta,\sigma$ and $\K=\log\sigma$ are the Weierstrass functions (\ref{p-function})-(\ref{sigma-function}), $\theta_1$ is the Jacobi function (\ref{Jacobi}), $\mathcal{B}_{n_j+1,k}(j)=\mathcal{B}_{n_j+1,k}\big(x_1(j),\dots,x_{n_j+1-k+1}(j)\big)$ is the partial ordinary  Bell polynomial (\ref{Bell-partial})-(\ref{Bell2}) and $\mathbf{L}_{j,n_j+1}^{\phi}$ is the  Bell operator given by (\ref{Bell operator}).
\end{Cor}
\emph{Proof:} Formula (\ref{lambda-p}) is a direct consequence of (\ref{lambda-zeta}), (\ref{x-r-sum}) and the equality $\wp(u)=-\zeta'(u)$. Using (\ref{lambda-zeta}) and the fact that the function $\zeta$ is the logarithmic derivative of the function $\sigma$ (\ref{zeta-sigma}), we see that (\ref{lambda-sigma}) holds true. The expression defining  the function $\K$ and definition (\ref{Bell operator}) of the Bell operators imply that formulas (\ref{lambda-sigma}) and  (\ref{lambda-Bell L}) are equal.  Furthermore, relation (\ref{sigma-theta1}) implies that
\begin{equation}\label{sigma-theta-log}
\log\big(\sigma(u)\big)=\zeta(1/2)u^2+\log\big(\theta_1(u)\big)-\log\big(\theta_1'(0)\big).
\end{equation}
Therefore, recalling the expressions for the Bell polynomials $\mathcal{B}_{n_j+1,1}(j)$ and $\mathcal{B}_{n_j+1,2}(j)$ presented  in  (\ref{Bell-Examples}), we deduce that  expression   (\ref{lambda-theta}) follows  from (\ref{lambda-sigma}), (\ref{sigma-theta-log}) and  (\ref{x-r-sum}).

\fd

\begin{Remark}\label{Rk-Hurwitz}
\end{Remark}
Let $\lambda$ be the meromorphic function (\ref{lambda-zeta}). Then the divisor of $\lambda'(P)$ is of the form
$$
\big(\lambda'(P)\big)=\sum_{r=1}^NP_r-\sum_{j=0}^{m}(n_j+2).\infty^j=\big(d\lambda(P)\big), \quad \quad \text{with}\quad N=\sum_{j=0}^m(n_j+2).
$$
Thus, because of the LG superpotentials (\ref{lambda-zeta}), the Hurwitz space $\mathcal{H}_1(n_0,\dots,n_m)$ in genus one (see Section 2.3.2)  can be identified with the open set of elements
$$
\Big(c, \infty^0,\dots,\infty^m, x_1(0), \dots, x_{n_0+1}(0),\ \dots, x_1(m),\ \dots,x_{n_m+1}(m)\Big)\in \C\times\C^{N}\times\mathbb{H}
$$
(with $\mathbb{H}$ being the upper half-plane) such that the following conditions are satisfied:
\begin{description}
\item[i)] $\lambda''(P_j)\neq 0$ for all $j=1,\dots,N$ (by the assumption of simple ramification at the ramification points $P_j$);
\item[ii)] the branch  points $\lambda_j=\lambda(P_j)$ are distinct;
\item[iii)] $\sum_{j=0}^m(n_j+1)x_{n_j+1}(j)=0$, by (\ref{x-r-sum});
\item[iv)] the divisor $(\lambda')$ is equal to zero in the quotient $\C/(\Z+\tau\Z)$.
\end{description}

\begin{Remark}
\end{Remark} It is known that the elliptic function from Theorem \ref{LG} can also be written as rational function in the Weierstrass sigma function (\ref{sigma-function}) (or in the Jacobi $\theta_1$-function (\ref{Jacobi})) as follows \cite{Akhiezer, Watson}
\begin{equation}\label{lambda-zeros}
\lambda(P)=C_1\prod_{j=0}^m\prod_{k=1}^{n_j+1}\frac{\sigma\big(z_P-\mathfrak{z}_{j,k}\big)}{\sigma\big(z_P-\infty^j)}
=C_2\prod_{j=0}^m\prod_{k=1}^{n_j+1}\frac{\theta_1\big(z_P-\mathfrak{z}_{j,k}\big)}{\theta_1\big(z_P-\infty^j)},
\end{equation}
where $C_1$, $C_2$ are two nonzero constants and $\mathfrak{z}_{j,k}$ are the zeros of $\lambda$. In our context, the aforementioned  expressions (\ref{lambda-zeta})-(\ref{lambda-p}) are more appropriate  than the product versions in (\ref{lambda-zeros}), as we shall  see below.

\medskip

The coming result presents a first useful consequence stemming from  expressions (\ref{lambda-zeta})-(\ref{lambda-p}) for $\lambda$.
\begin{Cor} The periods of the Abelian  differential $\lambda(P)dz_P=\lambda(P)\phi(P)$ are given by
\begin{align}
\begin{split}\label{u-a}
u:=\oint_a\lambda(P)\phi(P)&=c-2\big(1-\delta_{m,0}\big)\zeta(1/2)\sum_{j=1}^m(n_j+1)x_{n_j+1}(j)(\infty^j-\infty^0)\\
&\quad -\zeta(1/2)\sum_{j=0}^m\big(1-\delta_{n_j,0}\big)(n_j+1)\mathcal{B}_{n_j+1,2}(j);
\end{split}\\
\begin{split}\label{u-b}
\widehat{u}:=\oint_b\lambda(P)\phi(P)&=c\tau-2\big(1-\delta_{m,0}\big)\big(\tau\zeta(1/2)-{\rm i}\pi\big)\sum_{j=1}^m(n_j+1)x_{n_j+1}(j)(\infty^j-\infty^0)\\
&\quad -\big(\tau\zeta(1/2)-{\rm i}\pi\big)\sum_{j=0}^m\big(1-\delta_{n_j,0}\big)(n_j+1)\mathcal{B}_{n_j+1,2}(j).
\end{split}
\end{align}
\end{Cor}
\emph{Proof:} Let $\omega_1=1/2$ and $\omega_2=\tau/2$ be the half-period.
Then, making use of expression (\ref{lambda-p}) for the meromorphic function $\lambda$ and the properties (\ref{zeta-derivative}), (\ref{zeta-sigma}) and (\ref{zeta-sigma periods}) of the Weierstrass functions, we find for $r=1,2$:
\begin{align*}
&\int_x^{x+2\omega_r}\lambda(P)dz_P\\
&=2\omega_rc+(1-\delta_{m,0})\sum_{j=1}^m(n_j+1)x_{n_j+1}(j)
\int_x^{x+2\omega_r}\Big(\zeta(z_P-\infty^j)-\zeta(z_P-\infty^0)\Big)dz_P\\
&\quad +\frac{1}{2}\sum_{j=0}^m\big(1-\delta_{n_j,0}\big)(n_j+1)\mathcal{B}_{n_j+1,2}(j)\int_x^{x+2\omega_r}\wp\big(z_P-\infty^j\big)dz_P\\
&=2\omega_rc-2(1-\delta_{m,0})\zeta(\omega_r)\sum_{j=1}^m(n_j+1)x_{n_j+1}(j)(\infty^j-\infty^0)\\
&\quad -\zeta(\omega_r)\sum_{j=0}^m\big(1-\delta_{n_j,0}\big)(n_j+1)\mathcal{B}_{n_j+1,2}(j).
\end{align*}
This and Legendre's identity (\ref{Legendre}) give the desired result.

\fd

\begin{Prop} Let $j=0,\dots,m$. Then, with respect to the local parameter $z_P-\infty^j$, the Laurent expansion for the meromorphic function $\lambda$ (\ref{lambda-zeta}) near its pole $\infty^j$ of order $n_j+1$ is given by
\begin{equation}\label{lambda-Laurent}
\begin{split}
\lambda(P)\underset{P\sim \infty^j}{=} (n_j+1)\sum_{k=1}^{n_j+1}\frac{\mathcal{B}_{n_j+1,k}(j)}{k(z_P-\infty^j)^k}
+\sum_{\ell=0}^{\infty}\mathbf{f}_{\ell}(j)(z_P-\infty^j)^{\ell},
\end{split}
\end{equation}
where for $\ell\geq 0$, the coefficient $\mathbf{f}_{\ell}(j)$ is defined by
\begin{equation}\label{f-function}
\forall\ \ell\geq 0,\quad \quad \mathbf{f}_{\ell}(j):=\frac{\Lambda_j^{(\ell)}(\infty^j)}{\ell!}+(n_j+1)\sum_{k=1, k+\ell\geq 4}^{n_j+1}
\frac{(-1)^k}{k+\ell}\binom{k+\ell}{\ell}\mathcal{B}_{n_j+1,k}(j)G_{k+\ell}
\end{equation}
and  $\Lambda_j^{(\ell)}(\infty^j)$ is  given by
\begin{equation}\label{Lambdai-inftyi}
\begin{split}
\Lambda_j^{(\ell)}(\infty^j)
&=\delta_{\ell,0}c-\sum_{i=0,i\neq j}^m(n_i+1)\left(\sum_{k=1}^{n_i+1}\frac{(-1)^k}{k!}\mathcal{B}_{n_i+1,k}(i)\zeta^{(k+\ell-1)}\big(\infty^j-\infty^i\big)\right).
\end{split}
\end{equation}
\end{Prop}
\emph{Proof:} Consider the function
\begin{align*}
\Lambda_j(P)&:=c-\sum_{i=0,i\neq j}^m(n_i+1)\left(\sum_{k=1}^{n_i+1}\frac{(-1)^k}{k!}\mathcal{B}_{n_i+1,k}(i)\zeta^{(k-1)}\big(z_P-\infty^i\big)\right)
\end{align*}
which is holomorphic near the point $P=\infty^j$.
Then, expression (\ref{lambda-zeta}) for $\lambda(P)$ and the Laurent expansions (\ref{Laurent zeta}) and (\ref{Laurent p-k}) of the functions $\zeta$ and  $\zeta^{(k-1)}=-\wp^{(k-2)}$, $k\geq 2$,  near the origin, permit us to write
\begin{align*}
&\lambda(P)=\Lambda_j(P)-(n_j+1)\sum_{k=1}^{n_j+1}\frac{(-1)^k}{k!}\mathcal{B}_{n_j+1,k}(j)\zeta^{(k-1)}\big(z_P-\infty^j\big)\\
&\underset{P\sim \infty^j}{=}\sum_{\ell=0}^{\infty}\frac{\Lambda_j^{(\ell)}(\infty^j)}{\ell!}(z_P-\infty^j)^{\ell}
-(n_j+1)\sum_{k=1}^{n_j+1}\frac{(-1)^k}{k!}\mathcal{B}_{n_j+1,k}(j)\Bigg\{(-1)^{k-1}\frac{(k-1)!}{(z_P-\infty^j)^k}\\
&-(k-1)!\sum_{\ell=0, \ell+k\geq 4}^{\infty}\binom{\ell+k-1}{\ell}G_{\ell+k}(z_P-\infty^j)^{\ell}\Bigg\}\\
&=(n_j+1)\sum_{k=1}^{n_j+1}\frac{\mathcal{B}_{n_j+1,k}(j)}{k(z_P-\infty^j)^k}+\Lambda_j(\infty^j)+
\big(1-\delta_{n_j,0}-\delta_{n_j,1}-\delta_{n_j,2}\big)(n_j+1)\sum_{k=4}^{n_j+1}\frac{(-1)^k}{k}\mathcal{B}_{n_j+1,k}(j)G_{k}\\
&+\sum_{\ell=1}^{\infty}\left\{\frac{\Lambda_j^{(\ell)}(\infty^j)}{\ell!}+\big(1-\delta_{n_j,0}-\delta_{n_j,1}\big)(n_j+1)
\sum_{k=1, k+\ell\geq 4}^{n_j+1}\frac{(-1)^k}{k+\ell}\binom{k+\ell}{\ell}\mathcal{B}_{n_j+1,k}(j)G_{k+\ell}\right\}(z_P-\infty^j)^{\ell}.
\end{align*}
Hence we arrive at (\ref{lambda-Laurent}) by taking into account the convention (\ref{convention}).

\fd

\medskip

The upcoming result demonstrates that the  coefficients $x_{n_j+2}(j)$, $x_{n_j+3}(j)$, $\dots$  in  formulas (\ref{phi-Taylor})-(\ref{Evaluation}) can be obtained recursively as explicit function (although non trivial) of the  parameters $\tau$, $c$, $x_1(i)$, $\dots$, $x_{n_i+1}(i)$, $\infty^i$ of the Hurwitz space $\mathcal{H}_1(n_0,\dots,n_m)$. Note that formula (\ref{x-alpha2}) is useful since  the coefficients $x_{n_j+2}(j)$, $x_{n_j+3}(j)$, $\dots$  occur in  expression (\ref{chi-ii}) for the functions $\mathcal{I}_{j,\beta}[\Psi_{j,\alpha}]$ which are among the required terms  for calculating  the desired WDVV prepotential associated with the holomorphic differential (see  formula (\ref{Prep-phi1}) below). The examples of WDVV solutions listed in  Corollaries \ref{A123} and \ref{AA123}  illustrate  the usefulness  of formula (\ref{x-alpha2}).
\begin{Prop} Let $x_r(j)$ be defined by (\ref{Evaluation}) and (\ref{phi-Taylor}), with $j=0,1,\dots,m$ and $r\geq 1$. Then for any positive integer $\alpha$, the following equality holds:
\begin{equation}\label{x-alpha2}
\begin{split}
x_{n_j+1+\alpha}(j)&=\frac{\alpha}{n_j+1}\sum_{k=1}^{n_j+\alpha-1}\frac{1}{k+1}\binom{-n_j-1}{k}\big(x_1(j)\big)^{-k}
\mathcal{B}_{n_j+\alpha,k+1}\big(x_2(j),\dots,x_{n_j+\alpha-k+1}(j)\big)\\
&\quad +\alpha\sum_{k=1}^{n_j}\frac{1}{k}\mathcal{B}_{n_j+1,k}(j)\mathcal{R}_{\alpha+k-1,k}(j)
+\frac{\alpha}{n_j+1}\sum_{\ell=0}^{\alpha-1}\frac{1}{\ell+1}\mathbf{f}_{\ell}(j)\mathcal{B}_{\alpha,\ell+1}(j),
\end{split}
\end{equation}
where , as above, $\mathcal{B}_{n,k}(j)$ and $\mathcal{R}_{\mu,k}(j)$ are respectively the Bell polynomial (\ref{Bell-partial})-(\ref{Bell2})and the rational function (\ref{Bell-rec4}), while $\mathbf{f}_{\ell}$  is defined  by (\ref{f-function}).
\end{Prop}
\noindent \emph{Proof:} Due to (\ref{Evaluation}) and the behavior  (\ref{lambda-Laurent}) of $\lambda(P)$ near $\infty^j$, we can write
\begin{align*}
&(n_j+1+\alpha)x_{n_j+1+\alpha}(j)=\underset{\infty^j} {\rm res}\ \lambda(P)^{\frac{n_j+1+\alpha}{n_j+1}}\phi(P)\\
&= (n_j+1)\sum_{k=1}^{n_j+1}\frac{1}{k}\mathcal{B}_{n_j+1,k}(j)\underset{\infty^j} {\rm res}\Big((z_P-\infty^j)^{-k}\lambda(P)^{\frac{\alpha}{n_j+1}}\phi(P)\Big)
+\sum_{\ell=0}^{\infty}\mathbf{f}_{\ell}(j)\underset{\infty^j} {\rm res}\Big((z_P-\infty^j)^{\ell}\lambda(P)^{\frac{\alpha}{n_j+1}}\phi(P)\Big)\\
&=\alpha(n_j+1)\sum_{k=1}^{n_j+1}\frac{1}{k}\mathcal{B}_{n_j+1,k}(j)\mathcal{R}_{\alpha+k-1,k}(j)
+\alpha\sum_{\ell=0}^{\alpha-1}\frac{1}{\ell+1}\mathbf{f}_{\ell}(j)\mathcal{B}_{\alpha,\ell+1}(j)\\
&=\alpha\big(x_1(j)\big)^{n_j+1}\mathcal{R}_{\alpha+n_j,n_j+1}(j)+\alpha(n_j+1)\sum_{k=1}^{n_j}\frac{1}{k}\mathcal{B}_{n_j+1,k}(j)\mathcal{R}_{\alpha+k-1,k}(j)
+\alpha\sum_{\ell=0}^{\alpha-1}\frac{1}{\ell+1}\mathbf{f}_{\ell}(j)\mathcal{B}_{\alpha,\ell+1}(j),
\end{align*}
with  the third equality being a direct consequence of (\ref{Bell-B-formula}) and (\ref{Bell-R-formula}).\\
On the other hand, (\ref{Bell-rec4}) implies that
\begin{align*}
&\alpha\big(x_1(j)\big)^{n_j+1}\mathcal{R}_{\alpha+n_j,n_j+1}(j)={\alpha}x_{n_j+1+\alpha}(j)\\
&+\alpha\sum_{k=1}^{n_j+\alpha-1}\frac{1}{k+1}\binom{-n_j-1}{k}\big(x_1(j)\big)^{-k}
\mathcal{B}_{n_j+\alpha,k+1}\big(x_2(j),\dots,x_{n_j+\alpha-k+1}(j)\big).
\end{align*}
Thus we get
\begin{align*}
(n_j+1)x_{n_j+1+\alpha}(j)&=\alpha\sum_{k=1}^{n_j+\alpha-1}\frac{1}{k+1}\binom{-n_j-1}{k}\big(x_1(j)\big)^{-k}
\mathcal{B}_{n_j+\alpha,k+1}\big(x_2(j),\dots,x_{n_j+\alpha-k+1}(j)\big)\\
&\quad +\alpha(n_j+1)\sum_{k=1}^{n_j}\frac{1}{k}\mathcal{B}_{n_j+1,k}(j)\mathcal{R}_{\alpha+k-1,k}(j)
+\alpha\sum_{\ell=0}^{\alpha-1}\frac{1}{\ell+1}\mathbf{f}_{\ell}(j)\mathcal{B}_{\alpha,\ell+1}(j)
\end{align*}
as desired.

\fd

\begin{Example}
\end{Example}
Consider the even elliptic  function defined by
$$
\lambda(v)=c+x_1^2\wp(v|\tau),\quad \quad \text{with}\quad x_1\neq 0\quad \text{and}\quad  c\in \C.
$$
By the Laurent expansion (\ref{Laurend-p}), we have
$$
\lambda(v)\underset{v\sim0}{=}c+\frac{x_1^2}{v^2}+\sum_{\ell=2}^{\infty}(\ell+1)G_{\ell+2}v^{\ell}.
$$
Therefore formulas (\ref{f-function}) and (\ref{x-alpha2}) imply that $\mathbf{f}_{\ell}=\delta_{\ell,0}c$ for all $\ell\geq 0$ and
\begin{align*}
&x_2:=\frac{1}{2}\underset{0}{{\rm res}\ } \lambda(v)dv=0;\\
&x_{\alpha+2}:=\frac{1}{\alpha+2}\underset{0}{{\rm res}\ } \lambda(v)^{\frac{\alpha}{2}+1}dv
=\frac{c}{2}{\alpha}x_{\alpha}+\frac{\alpha}{2}\sum_{k=1}^{\alpha}\frac{(-1)^k}{x_1^k}
\mathcal{B}_{\alpha+1,k+1}(x_2,\dots,x_{\alpha-k+2}),\quad \quad \forall\ \alpha\geq 1.
\end{align*}


\section{WDVV prepotential associated with the normalized holomorphic differential}
\subsection{The main result}
In this section, we look for an explicit form of the prepotential $\mathbf{F}_{\phi}$ of the semi-simple Frobenius manifold structure associated with the primary  differential $\phi(P)=dz_P=\frac{1}{2{\rm i}\pi}\oint_bW(P,Q)$ which is quasi-homogeneous of degree 0 (in the sense of Definition \ref{degree-diff}).\\
Let $\eta(\phi)$ be the Darboux-Egoroff metric (\ref{eta-def}) on the Hurwitz space $\widehat{H}_1(n_0,\dots,n_m)$ induced by $\phi(P)=dz_P$.
From Section 2.3 (see also \cite{Dubrovin2D} and \cite{Rejeb23, Vasilisa}), we already  know that the following $N$ functions, with $N=2+2m+\sum_{i=0}^mn_i$,
provide a system of flat coordinates of the flat metric $\eta(\phi)$:
\begin{equation}\label{FC-phi}
\begin{array}{lllll}
\mathbf{t}^{i,\alpha}(\phi):=\displaystyle\frac{\sqrt{n_i+1}}{\alpha} \underset{\infty^i}{{\rm res}\ } \lambda(P)^{\frac{\alpha}{n_i+1}}\phi(P),&\quad i=0,\dots,m,\quad \alpha=1,\dots,n_i;\\
\\
\mathbf{v}^i(\phi)=\displaystyle \underset{\infty^i}{{\rm res}\ }\lambda(P)\phi(P),&\quad i=1,\dots,m;\\
\\
s^i:=\mathbf{s}^i(\phi):=\displaystyle \int_{\infty^0}^{\infty^i}\phi(P)=\infty^i-\infty^0,&\quad i=1,\dots,m;\\
\\
t:=\mathbf{r}(\phi):=\displaystyle \frac{1}{2{\rm{i}}\pi}\oint_{b}\phi(P)=\frac{\tau}{2{\rm i}\pi};\\
\\
u:=\mathbf{u}(\phi):=\displaystyle \oint_{a}\lambda(P)\phi(P).
\end{array}
\end{equation}

\begin{Lemma}\label{Lemma1} Let $j\in \{0,\dots,m\}$,  $\alpha\in \{1,\dots,n_j+1\}$ and $x_{\alpha}(j)$ be the coefficients given by (\ref{Evaluation}) and  appearing in the Taylor series (\ref{phi-Taylor}) for the holomorphic differential $\phi$. Then, with the notation (\ref{FC-phi}), we have
\begin{equation}\label{x-t-v}
\begin{split}
&x_{\alpha}(j)=
\left\{
\begin{array}{ll}
\displaystyle \frac{\mathbf{t}^{j,\alpha}(\phi)}{\sqrt{n_j+1}},\quad \quad  & \hbox{if}\quad  n_j\geq1\quad \text{and}\quad  \alpha=1,\dots,n_j;\\
\\
\displaystyle \frac{\mathbf{v}^j(\phi)}{n_j+1},\quad \quad  & \hbox{if}\quad  j\neq 0\quad \text{and}\quad \alpha=n_j+1;\\
\\
\displaystyle-\frac{1}{n_0+1}\sum_{r=1}^m\mathbf{v}^r(\phi),\quad \quad  & \hbox{if}\quad   j=0\quad \text{and}\quad \alpha=n_0+1.
\end{array}
\right.
\end{split}
\end{equation}
\end{Lemma}
\emph{Proof:} The   cases $1\leq \alpha\leq n_j$ and $\alpha=\big(1-\delta_{j,0}\big)(n_j+1)$  can be easily checked using (\ref{Evaluation}) and  (\ref{FC-phi}). The expression for $x_{n_0+1}(0)$ in (\ref{x-t-v}) is nothing but (\ref{x-r-sum}) which follows from the residue theorem applied to the differential $\lambda(P)\phi(P)$.

\fd

\medskip

Relationship (\ref{x-t-v}) implies that the variables
\begin{equation}\label{FC-phi2}
\bigg\{x_1(0),\dots,x_{n_0}(0),x_1(j),\dots,x_{n_j+1}(j), s^j, \tau={2{\rm i}\pi}t, u: \quad \text{with}\quad  j=1,\dots,m\bigg\}
\end{equation}
serve also as a system of flat coordinates for the metric $\eta(\phi)$. Moreover, because of (\ref{Entries eta}) and  (\ref{x-t-v}), we deduce that the nonzero entries of the constant matrix of $\eta(\phi)$ with respect to coordinates (\ref{FC-phi2})  are given by
\begin{equation}\label{Entries eta-phi}
\begin{split}
&\eta(\phi)\big(\partial_{x_{\alpha}(i)},\partial_{x_{\beta}(j)}\big)=(n_j+1)\delta_{ij}\delta_{\alpha+\beta,n_j+1};\\
&\eta(\phi)\big(\partial_{x_{n_i+1}(i)},\partial_{s^{j}}\big)=(n_j+1)\delta_{ij};\\
&\eta(\phi)\big(\partial_{\tau},\partial_u\big)=\frac{1}{2{\rm i}\pi},
\end{split}
\end{equation}
with $\alpha=1,\dots,n_i$, $\beta=1,\dots,n_j$.

\medskip

For the sake of simplicity and readability, we choose to express  the WDVV prepotential $\mathbf{F}_{\phi}$ associated with the holomorphic differential $\phi$ (\ref{phi-def}) as a function of the variables listed in (\ref{FC-phi2}) and (\ref{x-t-v}) instead of those in (\ref{FC-phi}). We  start with describing a preliminary expression  for $\mathbf{F}_{\phi}$ by means of  flat coordinates introduced in (\ref{FC-phi2}) and (\ref{x-t-v}). To achieve this goal we shall apply formula  (\ref{Prepotential}) giving  the WDVV solution induced by any primary and quasi-homogeneous differential $\omega_0$.

\begin{Prop}\label{Preliminary prep} Let $\mathbf{F}_{\phi}$ be the WDVV prepotential associated with the Frobenius manifold structure on the Hurwitz space $\widehat{H}_1(n_0,\dots,n_m)$ and induced by the normalized holomorphic differential $\phi$ (\ref{phi-def}). Let  $\big\{x_{\alpha}(j):\quad j=0,\dots,m \big\}$ be the coefficients appearing in the Taylor series (\ref{phi-Taylor}), $\mathbf{s}^i(\phi_{\mathbf{s}^j})$ be the operation defined by the principal value in the family (\ref{Flat basis}) and  $\mathcal{I}_{i,\alpha}[\Omega_{\infty^0\infty^j}]$ and $\mathcal{I}_{i,\alpha}\big[\Psi_{j,\beta}\big]$ be respectively the quantities  defined by (\ref{rho-ij}) and (\ref{chi-ij-def}). Then $\mathbf{F}_{\phi}$ takes the following form:
\begin{equation}\label{Prep-phi1}
\begin{split}
&\mathbf{F}_{\phi}=\frac{u^2}{4{\rm i}\pi}\tau+u\sum_{j=1}^{m}(n_j+1)x_{n_j+1}(j)s^j+\frac{u}{2}\sum_{i=0}^{m}\sum_{\alpha=1}^{n_i}\frac{(n_i+1-\alpha)(2n_i+2+\alpha)}{(n_i+1+\alpha)}
x_{\alpha}(i)x_{n_i+1-\alpha}(i)\\
&+\frac{1}{2}\sum_{i=1}^{m}\sum_{j=1}^m(n_i+1)(n_j+1)x_{n_i+1}(i)x_{n_j+1}(j)\mathbf{s}^i(\phi_{\mathbf{s}^j})\\
&+\frac{1}{2}\sum_{j=1}^m\sum_{i=0}^{m}\sum_{\alpha=1}^{n_i}\left(\frac{n_i+1-\alpha}{n_i+1+\alpha}+\frac{n_i+1-\alpha}{n_i+1}\right)
(n_i+1)(n_j+1)x_{n_i+1-\alpha}(i)x_{n_j+1}(j)\Big(\frac{1}{\alpha}\mathcal{I}_{i,\alpha}[\Omega_{\infty^0\infty^j}]\Big)\\
&+\frac{1}{2}\sum_{i,j=0}^{m}\sum_{\alpha=1}^{n_i}\sum_{\beta=1}^{n_j}\frac{(n_i+1-\alpha)(n_j+1-\beta)}{(n_i+1+\alpha)}
(n_i+1)x_{n_i+1-\alpha}(i)x_{n_j+1-\beta}(j)\Big(\frac{1}{\alpha\beta}\mathcal{I}_{i,\alpha}\big[\Psi_{j,\beta}\big]\Big)\\
&+\frac{\log(-1)}{4}\sum_{i,j=1, i\neq j}^m(n_i+1)(n_j+1)x_{n_i+1}(i)x_{n_j+1}(j)\\
&-\frac{3}{4}\sum_{i,j=1}^m\bigg(\frac{(n_i+1)(n_j+1)}{n_0+1}+\delta_{ij}(n_j+1)\bigg)x_{n_i+1}(i)x_{n_j+1}(j).
\end{split}
\end{equation}
\end{Prop}
\emph{Proof:} Since the normalized holomorphic differential $\phi$ is quasi-homogeneous of  degree $d_0=0$ and by (\ref{r-AB}) the quantities $\rho_{\phi,\phi_{\mathbf{t}^A}}$ appearing in (\ref{Prepotential}) vanish, it follows that formula  (\ref{Prepotential}) becomes:
\begin{align*}
&\mathbf{F}_{\phi}-\frac{\log(-1)}{4}\sum_{i,j=1, i\neq j}^m\mathbf{v}^i(\phi)\mathbf{v}^j(\phi)
+\frac{3}{4}\sum_{i,j=1}^m\bigg(\frac{1}{n_0+1}+\delta_{ij}\frac{1}{n_i+1}\bigg)\mathbf{v}^i(\phi)\mathbf{v}^j(\phi)\\
&=\frac{1}{2}\sum_{A,B}\frac{d_{A^{\prime}}d_{B^{\prime}}}{1+d_A}\mathbf{t}^{A^{\prime}}(\phi)\mathbf{t}^{B^{\prime}}(\phi)\mathbf{t}^{A}\big(\phi_{\mathbf{t}^B}\big)\\
&=\frac{1}{2}\sum_{A,B}\frac{\big(1-d_A\big)\big(1-d_B\big)}{1+d_A}\mathbf{t}^{A^{\prime}}(\phi)\mathbf{t}^{B^{\prime}}(\phi)\mathbf{t}^{A}\big(\phi_{\mathbf{t}^B}\big)
\end{align*}
where we have used  duality relation (\ref{Duality-degree}) which states that $d_{A^{\prime}}=1-d_A$  in the second equality. \\
Thus, observing that, due to (\ref{degrees}), $1-d_A\neq 0$ only when $\mathbf{t}^A\in \big\{\mathbf{t}^{i,\alpha}, \mathbf{s}^i,\mathbf{r}\}$ and taking into account notation (\ref{FC-phi}) as well as the precise duality relations listed in (\ref{duality-picture}),  we arrive at the following expression in terms of flat coordinates given by (\ref{FC-phi}):
\begin{equation}\label{Prep-phi0}
\begin{split}
\mathbf{F}_{\phi}&=\frac{u^2}{2}t+u\sum_{j=1}^{m}\mathbf{v}^i(\phi)s^j+\frac{u}{2}\sum_{i=0}^{m}\sum_{\alpha=1}^{n_i}
\frac{(n_i+1-\alpha)(2n_i+2+\alpha)}{(n_i+1+\alpha)(n_i+1)}\mathbf{t}^{i,\alpha}(\phi)\mathbf{t}^{i,n_i+1-\alpha}(\phi)\\
&\quad +\frac{1}{2}\sum_{i=1}^{m}\sum_{j=1}^m\mathbf{v}^i(\phi)\mathbf{v}^j(\phi)\mathbf{s}^i(\phi_{\mathbf{s}^j})\\
&\quad +\frac{1}{2}\sum_{j=1}^m\sum_{i=0}^{m}\sum_{\alpha=1}^{n_i}\left(\frac{n_i+1-\alpha}{n_i+1+\alpha}+\frac{n_i+1-\alpha}{n_i+1}\right)
\mathbf{t}^{i,n_i+1-\alpha}(\phi)\mathbf{v}^j(\phi)\mathbf{t}^{i,\alpha}\big(\phi_{\mathbf{s}^j}\big)\\
&\quad +\frac{1}{2}\sum_{i=0}^{m}\sum_{j=0}^m\sum_{\alpha=1}^{n_i}\sum_{\beta=1}^{n_j}\frac{(n_i+1-\alpha)(n_j+1-\beta)}{(n_i+1+\alpha)(n_j+1)}\mathbf{t}^{i,n_i+1-\alpha}(\phi)
\mathbf{t}^{j,n_j+1-\beta}(\phi)\mathbf{t}^{i,\alpha}\big(\phi_{\mathbf{t}^{j,\beta}}\big)\\
&\quad +\frac{\log(-1)}{4}\sum_{i,j=1, i\neq j}^m\mathbf{v}^i(\phi)\mathbf{v}^j(\phi)
-\frac{3}{4}\sum_{i,j=1}^m\bigg(\frac{1}{n_0+1}+\delta_{ij}\frac{1}{n_i+1}\bigg)\mathbf{v}^i(\phi)\mathbf{v}^j(\phi).
\end{split}
\end{equation}
On the other hand, because of (\ref{Primary-e}), (\ref{Flat basis}), (\ref{rho-ij}) and (\ref{chi-ij-def}), we have
$$
\phi_{\mathbf{t}^{j,\beta}}(P)=\frac{\sqrt{n_j+1}}{\beta}\Psi_{j,\beta}(P)\quad \quad \text{and}\quad \quad \mathbf{t}^{i,\alpha}(\omega_0)=\frac{\sqrt{n_i+1}}{\alpha}\mathcal{I}_{i,\alpha}[\omega_0].
$$
This brings us to write
\begin{align*}
&\mathbf{t}^{i,\alpha}\big(\phi_{\mathbf{s}^j}\big)=\mathbf{t}^{i,\alpha}\big(\Omega_{\infty^0\infty^j}\big)
=\frac{\sqrt{n_i+1}}{\alpha}\mathcal{I}_{i,\alpha}[\Omega_{\infty^0\infty^j}];\\
&\mathbf{t}^{i,\alpha}\big(\phi_{\mathbf{t}^{j,\beta}}\big)=\frac{\sqrt{(n_i+1)(n_j+1)}}{\alpha\beta}\mathcal{I}_{i,\alpha}\big[\Psi_{j,\beta}\big].
\end{align*}
Finally, the claimed result (\ref{Prep-phi1}) follows  from (\ref{Prep-phi0}), (\ref{x-t-v}), (\ref{FC-phi2}) and the two previous equalities:

\fd

\medskip

It remains to write  expressions for lines two, three and four by means of variables (\ref{FC-phi2}). We begin with the second line.
\begin{description}
\item[$\bullet$] \textbf{Expression for the second line in (\ref{Prep-phi1})}
\end{description}
The next  result deals with  the expression  for the following  functions given  by the principal value
$$
\mathbf{s}^i\big(\phi_{\mathbf{s}^j}\big)=\mathbf{s}^i(\Omega_{\infty^0\infty^j}):= p.v.\int_{\infty^0}^{\infty^i}\Omega_{\infty^0\infty^j},\quad \quad i,j=1,\dots,m.
$$
Here we recall that the principal value is defined by subtracting  the divergent part of the integral as a function of the local parameters $\z_0(P):=\lambda(P)^{-\frac{1}{n_0+1}}$ and  $\z_j(P):=\lambda(P)^{-\frac{1}{n_j+1}}$ near the points $\infty^0$ and $\infty^j$, respectively.\\
A general formula for $\mathbf{s}^i\big(\phi_{\mathbf{s}^j}\big)$  in terms of the Riemann theta function with (odd and non-singular) half integer characteristics  was recently  obtained in  \cite{Rejeb23} for an arbitrary genus $g\geq 1$ of the underlying Riemann surfaces. Here we prefer to rewrite the genus one case by means of the Weierstrass function $\K(u):=\log\sigma(u)$, with $\sigma$ being the sigma function (\ref{sigma-function}).
\begin{Lemma} Let $i,j=1,\dots,m$,  $\K(u):=\log\sigma(u)$ and $s^j:=\int_{\infty^0}^{\infty^j}\phi=\infty^j-\infty^0$, where $\phi=dz_P$ is the holomorphic differential (\ref{phi-def}) and $\infty^0,\dots,\infty^m$ are the prescribed poles of a covering $\big(\C/{\mathbb{L}},\lambda\big)$.
\begin{description}
\item[i)] If $i\neq j$, then
\begin{equation}\label{sij}
\begin{split}
\mathbf{s}^i\big(\phi_{\mathbf{s}^j}\big)&=\K(s^i-s^j)-\K(s^i)-\K(s^j)+2s^is^j\zeta(1/2)+\log\big(-x_1(0)\big).
\end{split}
\end{equation}
\item[ii)] When $i=j$, we have
\begin{equation}\label{sjj}
\begin{split}
\mathbf{s}^j\big(\phi_{\mathbf{s}^j}\big)&=\log\big(x_1(j)\big)+\log\big(-x_1(0)\big)-2\K(s^j)+2(s^j)^2\zeta(1/2).
\end{split}
\end{equation}
\end{description}
\end{Lemma}
\emph{Proof:} Let  $\theta_1(\cdot|\tau)=-\Theta_{[\frac{1}{2},\frac{1}{2}]}(\cdot|\tau)$  be the  Jacobi function defined by (\ref{Jacobi}). Using the genus one case of Theorem 3.3 in \cite{Rejeb23}, we can  deduce  that the functions $\mathbf{s}^i\big(\phi_{\mathbf{s}^j}\big)$ are explicitly expressed by means of the Jacobi $\theta_1$-function as follows:
\begin{align*}
\mathbf{s}^i\big(\phi_{\mathbf{s}^j}\big)=
\left\{
\begin{array}{ll}
\displaystyle \log\left(\frac{\theta_1'(0)\theta_1(\infty^i-\infty^j)}{\theta_1(\infty^i-\infty^0)\theta_1(\infty^j-\infty^0)}\right)
+\log\big(\phi(\infty^0)\big)-\log(-1), & \hbox{if}\  i\neq j\\
\\
\displaystyle\log\big(\phi(\infty^j)\big)+\log\big(\phi(\infty^0)\big)
-2\log\left(\frac{\theta_1(\infty^j-\infty^0)}{\theta_1'(0)}\right)-\log(-1), & \hbox{if}\  i=j.
\end{array}
\right.
\end{align*}
On the other hand, due to relation  (\ref{sigma-theta1}) which gives the link between the functions $\sigma$ and $\theta_1$, we conclude that $\mathbf{s}^i\big(\phi_{\mathbf{s}^j}\big)$ can be rewritten in terms  of the Weierstrass function $\K(u):=\log\sigma(u)$. More precisely,  when $i\neq j$, we have
\begin{align*}
\mathbf{s}^i\big(\phi_{\mathbf{s}^j}\big)&=\log\left(\frac{\sigma(\infty^i-\infty^j)}{\sigma(\infty^i-\infty^0)\sigma(\infty^j-\infty^0)}\right)
+2(\infty^i-\infty^0)(\infty^j-\infty^0)\zeta(1/2)+\log\big(-\phi(\infty^0)\big)\\
&=\log\left(\frac{\sigma(s^i-s^j)}{\sigma(s^i)\sigma(s^j)}\right)+2s^is^j\zeta(1/2)+\log\big(-\phi(\infty^0)\big)\\
&=\K(s^i-s^j)-\K(s^i)-\K(s^j)+2s^is^j\zeta(1/2)+\log\big(-x_1(0)\big)
\end{align*}
and when $i=j$,
\begin{align*}
\mathbf{s}^j\big(\phi_{\mathbf{s}^j}\big)&=\log\big(\phi(\infty^j)\big)+\log\big(-\phi(\infty^0)\big)-2\log\left(\sigma(\infty^j-\infty^0)\right)+2(\infty^j-\infty^0)^2\zeta(1/2)\\
&=\log\big(\phi(\infty^j)\big)+\log\big(-\phi(\infty^0)\big)-2\K(s^j)+2(s^j)^2\zeta(1/2)\\
&=\log\big(x_1(j)\big)+\log\big(-x_1(0)\big)-2\K(s^j)+2(s^j)^2\zeta(1/2).
\end{align*}
This establishes  the desired formulas (\ref{sij}) and (\ref{sjj}).

\fd

\medskip

As a direct consequence of formulas (\ref{sij}) and (\ref{sjj}) and the following link between $\zeta(1/2|\tau)$ and  the quasi-modular Eisenstein series $G_2(\tau)$:
\begin{equation}\label{G2-zeta}
\zeta(1/2|\tau)=\frac{1}{2}G_2(\tau),\quad\quad \quad  \forall\ \tau, \quad \Im\tau>0,
\end{equation}
we obtain the expression for the second  line of the prepotential $\mathbf{F}_{\phi}$  (\ref{Prep-phi1}). Relation (\ref{G2-zeta}) can be found in \cite{Akhiezer} (Section 20) and \cite{Dubrovin2D} (formulas (C.72)-(C.73)) and can be seen from the Fourier expansion  (\ref{Eisentein-Fourier}) for the Eisenstein series $G_2$.

\begin{Prop}\label{Line2} Let $\K(u)=\log\sigma(u)$ and $G_2$ be the quasi-modular Eisenstein series defined by (\ref{G2-def}) and (\ref{Eisentein-Fourier}).  Then the second line in expression (\ref{Prep-phi1}) for the WDVV prepotential  $\mathbf{F}_{\phi}$ is given by
\begin{align*}
\Sigma_2&:=\frac{1}{2}\sum_{i,j=1}^{m}(n_i+1)(n_j+1)x_{n_i+1}(i)x_{n_j+1}(j)\mathbf{s}^i(\phi_{\mathbf{s}^j})\\
&=\frac{1}{2}G_2(\tau)\bigg(\sum_{j=1}^{m}(n_j+1)x_{n_j+1}(j)s^j\bigg)^2+\frac{1}{2}\sum_{i,j=1, i\neq j}^{m}(n_i+1)(n_j+1)x_{n_i+1}(i)x_{n_j+1}(j)\K(s^i-s^j)\\
&\quad -\bigg(\sum_{i=1}^{m}(n_i+1)x_{n_i+1}(i)\bigg)\bigg(\sum_{j=1}^{m}(n_j+1)x_{n_j+1}(j)\K(s^j)\bigg)
 +\frac{1}{2}\sum_{j=1}^{m}\Big((n_j+1)x_{n_j+1}(j)\Big)^2\log\big(x_1(j)\big)\\
&\quad +\frac{1}{2}\bigg(\sum_{j=1}^m(n_j+1)x_{n_j+1}(j)\bigg)^2\log\big(x_1(0)\big)+\frac{\log(-1)}{2}\bigg(\sum_{j=1}^m(n_j+1)x_{n_j+1}(j)\bigg)^2.
\end{align*}
\end{Prop}

\begin{description}
\item[$\bullet$] \textbf{Expression for the third line in (\ref{Prep-phi1})}
\end{description}
 For $0\leq i\leq m$, $1\leq j\leq m$ and $1\leq \alpha\leq n_i$, let us introduce the following quantities
\begin{equation}\label{rho-ij2}
\begin{split}
&\varrho(i,j,\alpha):=\mathcal{I}_{i,\alpha}[\Omega_{\infty^0\infty^j}]-2{\alpha}x_{\alpha}(i)s^j\zeta(1/2);\\
&A_{ij}:=\sum_{\alpha=1}^{n_i}\frac{(n_i+1-\alpha)(n_i+1)(n_j+1)}{n_i+1+\alpha}
x_{n_i+1-\alpha}(i)x_{n_j+1}(j)\Big(\frac{1}{\alpha}\varrho(i,j,\alpha)\Big);\\
&B_{ij}:=\sum_{\alpha=1}^{n_i}(n_i+1-\alpha)(n_j+1)x_{n_i+1-\alpha}(i)x_{n_j+1}(j)\Big(\frac{1}{\alpha}\varrho(i,j,\alpha)\Big).
\end{split}
\end{equation}
Then  in terms  of $\varrho(i,j,\alpha)$, $A_{ij}$ and $B_{ij}$, we can write the third line as follows:
\begin{align*}
\Sigma_3&:=\frac{1}{2}\sum_{j=1}^m\sum_{i=0}^{m}\sum_{\alpha=1}^{n_i}\left(\frac{n_i+1-\alpha}{n_i+1+\alpha}+\frac{n_i+1-\alpha}{n_i+1}\right)
(n_i+1)(n_j+1)x_{n_i+1-\alpha}(i)x_{n_j+1}(j)\Big(\frac{1}{\alpha}\mathcal{I}_{i,\alpha}[\Omega_{\infty^0\infty^j}]\Big)\\
&=\zeta(1/2)\sum_{j=1}^m\sum_{i=0}^{m}\sum_{\alpha=1}^{n_i}\left(\frac{n_i+1-\alpha}{n_i+1+\alpha}+\frac{n_i+1-\alpha}{n_i+1}\right)
(n_i+1)(n_j+1)x_{\alpha}(i)x_{n_i+1-\alpha}(i)x_{n_j+1}(j)s^j\\
&\quad +\frac{1}{2}\sum_{j=1}^m\sum_{i=0}^{m}\sum_{\alpha=1}^{n_i}\left(\frac{n_i+1-\alpha}{n_i+1+\alpha}+\frac{n_i+1-\alpha}{n_i+1}\right)
(n_i+1)(n_j+1)x_{n_i+1-\alpha}(i)x_{n_j+1}(j)\Big(\frac{1}{\alpha}\varrho(i,j,\alpha)\Big)\\
&=\zeta(1/2)\sum_{j=1}^m\sum_{i=0}^{m}\sum_{\alpha=1}^{n_i}\left(\frac{n_i+1-\alpha}{n_i+1+\alpha}+\frac{n_i+1-\alpha}{n_i+1}\right)
(n_i+1)(n_j+1)x_{\alpha}(i)x_{n_i+1-\alpha}(i)x_{n_j+1}(j)s^j\\
&\quad +\frac{1}{2}\sum_{j=1}^m\sum_{i=0}^{m}A_{ij}+\frac{1}{2}\sum_{j=1}^m\sum_{i=0}^{m}B_{ij}.
\end{align*}
Therefore, it suffices to express  the functions $A_{ij}$ and $B_{ij}$ in terms of flat coordinates (\ref{x-t-v})-(\ref{FC-phi2}). We distinguish three cases.\\
\textbf{First case: $i\neq j$.} Because of (\ref{Third-alpha-ij})  we have
\begin{align*}
A_{ij}
&=\sum_{\alpha=1}^{n_i}\frac{(n_i+1-\alpha)(n_i+1)(n_j+1)}{n_i+1+\alpha}
x_{n_i+1-\alpha}(i)x_{n_j+1}(j)\mathbf{L}_{i,\alpha}^{\phi}[\K](s^i-s^j)\\
&\quad -\sum_{\alpha=1}^{n_i}\frac{(n_i+1-\alpha)(n_i+1)(n_j+1)}{n_i+1+\alpha}
x_{n_i+1-\alpha}(i)x_{n_j+1}(j)\mathbf{L}_{i,\alpha}^{\phi}[\K](s^i).
\end{align*}
Moreover, (\ref{Third-alpha-ij})  and recurrence relation (\ref{Bell-rec1}) for partial ordinary Bell polynomials imply that
\begin{align*}
B_{ij}
&=\sum_{\alpha=1}^{n_i}(n_i+1-\alpha)(n_j+1)x_{n_i+1-\alpha}(i)x_{n_j+1}(j)\bigg(\sum_{k=1}^{\alpha}\frac{1}{k!}
\mathcal{B}_{\alpha,k}(i)\Big(\K^{(k)}(s^i-s^j)-\K^{(k)}(s^i)\Big)\bigg)\\
&=\sum_{k=1}^{n_i}\frac{(n_i+1)(n_j+1)}{(k+1)!}x_{n_j+1}(j)\mathcal{B}_{n_i+1,k+1}(i)\Big(\K^{(k)}(s^i-s^j)-\K^{(k)}(s^i)\Big)\\
&=\sum_{k=2}^{n_i+1}\frac{(n_i+1)(n_j+1)}{k!}x_{n_j+1}(j)\mathcal{B}_{n_i+1,k}(i)\Big(\K^{(k-1)}(s^i-s^j)-\K^{(k-1)}(s^i)\Big).
\end{align*}
\textbf{Second case: $i=j\geq 1$.} Expressions (\ref{Third-alpha-jj}) and (\ref{rho-ij2}) for $\mathcal{I}_{i,\alpha}[\Omega_{\infty^0\infty^j}]$
and $\varrho(i,j,\alpha)$ imply that
\begin{align*}
A_{jj}&:=\sum_{\alpha=1}^{n_j}\frac{(n_j+1-\alpha)(n_j+1)^2}{n_j+1+\alpha}
x_{n_j+1-\alpha}(j)x_{n_j+1}(j)\Big(\frac{1}{\alpha}\varrho(j,j,\alpha)\Big)\\
&=\sum_{\alpha=1}^{n_j}\frac{(n_j+1-\alpha)(n_j+1)^2}{n_j+1+\alpha}x_{n_j+1-\alpha}(j)x_{n_j+1}(j)\mathcal{R}_{\alpha,1}(j)\\
&\quad-\sum_{\alpha=1}^{n_j}\frac{(n_j+1-\alpha)(n_j+1)^2}{n_j+1+\alpha}x_{n_j+1-\alpha}(j)x_{n_j+1}(j)
\bigg(\sum_{\ell=4}^{\alpha}\frac{\mathcal{B}_{\alpha,\ell}(j)}{\ell}G_{\ell}\bigg)\\
&\quad -\sum_{\alpha=1}^{n_j}\frac{(n_j+1-\alpha)(n_j+1)^2}{n_j+1+\alpha}x_{n_j+1-\alpha}(j)x_{n_j+1}(j)\mathbf{L}_{j,\alpha}^{\phi}\big[\K\big](s^j)
\end{align*}
and
\begin{align*}
B_{jj}&:=\sum_{\alpha=1}^{n_j}(n_j+1-\alpha)(n_j+1)x_{n_j+1-\alpha}(j)x_{n_j+1}(j)
\Big(\frac{1}{\alpha}\varrho(j,j,\alpha)\Big)\\
&=\sum_{\alpha=1}^{n_j}(n_j+1-\alpha)(n_j+1)x_{n_j+1-\alpha}(j)x_{n_j+1}(j)\mathcal{R}_{\alpha,1}(j)\\
&\quad -\sum_{\alpha=1}^{n_j}(n_j+1-\alpha)(n_j+1)x_{n_j+1-\alpha}(j)x_{n_j+1}(j)
\bigg(\sum_{\ell=4}^{\alpha}\frac{\mathcal{B}_{\alpha,\ell}(j)}{\ell}G_{\ell}\bigg)\\
&\quad -\sum_{\alpha=1}^{n_j}(n_j+1-\alpha)(n_j+1)x_{n_j+1-\alpha}(j)x_{n_j+1}(j)
\bigg(\sum_{k=1}^{\alpha}\frac{1}{k!}\mathcal{B}_{\alpha,k}(j)\K^{(k)}(s^j) \bigg)\\
&=\sum_{\alpha=1}^{n_j}(n_j+1-\alpha)(n_j+1)x_{n_j+1-\alpha}(j)x_{n_j+1}(j)\mathcal{R}_{\alpha,1}(j)\\
&\quad -\sum_{\ell=4}^{n_j}\frac{(n_j+1)^2}{\ell(\ell+1)}x_{n_j+1}(j)\mathcal{B}_{n_j+1,\ell+1}(j)G_{\ell}
-\sum_{k=2}^{n_j+1}\frac{(n_j+1)^2}{k!}x_{n_j+1}(j)\mathcal{B}_{n_j+1,k}(j)\K^{(k-1)}(s^j),
\end{align*}
where, as in the first case, the last equality follows from   recurrence relation (\ref{Bell-rec1}) for Bell polynomials. \\
\textbf{Third case: $i=0$ and $j\geq 1$.} As above, using now (\ref{Third-alpha-0j}) we have
\begin{align*}
A_{0j}&:=\sum_{\alpha=1}^{n_0}\frac{(n_0+1-\alpha)(n_0+1)(n_j+1)}{n_0+1+\alpha}x_{n_0+1-\alpha}(0)x_{n_j+1}(j)
\Big(\frac{1}{\alpha}\varrho(0,j,\alpha)\Big)\\
&=-(n_j+1)x_{n_j+1}(j)\sum_{\alpha=1}^{n_0}\frac{(n_0+1-\alpha)(n_0+1)}{n_0+1+\alpha}x_{n_0+1-\alpha}(0)\mathcal{R}_{\alpha,1}(0)\\
&\quad +(n_j+1)x_{n_j+1}(j)\sum_{\alpha=1}^{n_0}\Bigg\{\frac{(n_0+1-\alpha)(n_0+1)}{n_0+1+\alpha}x_{n_0+1-\alpha}(0)
\bigg(\sum_{\ell=4}^{\alpha}\frac{\mathcal{B}_{\alpha,\ell}(0)}{\ell}G_{\ell}\bigg)\Bigg\}\\
&\quad +(n_j+1)x_{n_j+1}(j)\sum_{\alpha=1}^{n_0}\frac{(n_0+1-\alpha)(n_0+1)}{n_0+1+\alpha}x_{n_0+1-\alpha}(0) \mathbf{L}_{0,\alpha}^{\phi}\big[\K\big](-s^j)
\end{align*}
and
\begin{align*}
B_{0j}&:=\sum_{\alpha=1}^{n_0}(n_0+1-\alpha)(n_j+1)x_{n_0+1-\alpha}(0)x_{n_j+1}(j)\Big(\frac{1}{\alpha}\varrho(0,j,\alpha)\Big)\\
&=-\sum_{\alpha=1}^{n_0}(n_0+1-\alpha)(n_j+1)x_{n_0+1-\alpha}(0)x_{n_j+1}(j)\mathcal{R}_{\alpha,1}(0)\\
&\quad +\sum_{\alpha=1}^{n_0}(n_0+1-\alpha)(n_j+1)x_{n_0+1-\alpha}(0)x_{n_j+1}(j)\bigg(\sum_{\ell=4}^{\alpha}\frac{\mathcal{B}_{\alpha,\ell}(0)}{\ell}G_{\ell}\bigg)\\
&\quad +\sum_{\alpha=1}^{n_0}(n_0+1-\alpha)(n_j+1)x_{n_0+1-\alpha}(0)x_{n_j+1}(j)\bigg(\sum_{k=1}^{\alpha}\frac{1}{k!}\mathcal{B}_{\alpha,k}(0)\K^{(k)}(-s^j)\bigg)\\
&=-(n_j+1)x_{n_j+1}(j)\Bigg\{\sum_{\alpha=1}^{n_0}(n_0+1-\alpha)x_{n_0+1-\alpha}(0)\mathcal{R}_{\alpha,1}(0)
-\sum_{\ell=4}^{n_0}\frac{(n_0+1)}{\ell(\ell+1)}\mathcal{B}_{n_0+1,\ell+1}(0)G_{\ell}\Bigg\}\\
&\quad +(n_j+1)x_{n_j+1}(j)\sum_{k=2}^{n_0+1}\frac{(n_0+1)}{k!}\mathcal{B}_{n_0+1,k}(0)\K^{(k-1)}(-s^j)\\
&=-(n_j+1)x_{n_j+1}(j)\Bigg\{\sum_{\alpha=1}^{n_0}(n_0+1-\alpha)x_{n_0+1-\alpha}(0)\mathcal{R}_{\alpha,1}(0)
-\sum_{\ell=4}^{n_0}\frac{(n_0+1)}{\ell(\ell+1)}\mathcal{B}_{n_0+1,\ell+1}(0)G_{\ell}\Bigg\}\\
&\quad +(n_j+1)x_{n_j+1}(j)\Bigg\{\sum_{k=1}^{n_0+1}\frac{(n_0+1)}{k!}\mathcal{B}_{n_0+1,k}(0)\K^{(k-1)}(-s^j)-(n_0+1)x_{n_0+1}(0)\K(s^j)\Bigg\}\\
&\quad -\big(\log(-1)\big)(n_0+1)(n_j+1)x_{n_0+1}(0)x_{n_j+1}(j).
\end{align*}
Here, in the last equality we have taken into account the fact that
$$
\mathcal{B}_{n_0+1,1}(0)=x_{n_0+1}(0)\quad \quad \text{and}\quad \quad \K(-u)=\log\sigma(-u)=\log(-1)+\K(u).
$$

Summarizing the above computations and bearing in mind the relation $G_2(\tau)=2\zeta(1/2|\tau)$, we arrive at the following explicit form for the third line.
\begin{Prop}\label{Line3} Using notation of Theorem \ref{Third kind-FCij}, the third line in (\ref{Prep-phi1}) has the following form in terms of flat coordinates listed in (\ref{FC-phi2}) and (\ref{x-t-v}):
\begin{align*}
\Sigma_3&:=\frac{1}{2}\sum_{j=1}^m\sum_{i=0}^{m}\sum_{\alpha=1}^{n_i}\left(\frac{n_i+1-\alpha}{n_i+1+\alpha}+\frac{n_i+1-\alpha}{n_i+1}\right)
(n_i+1)(n_j+1)x_{n_i+1-\alpha}(i)x_{n_j+1}(j)\Big(\frac{1}{\alpha}\mathcal{I}_{i,\alpha}[\Omega_{\infty^0\infty^j}]\Big)\\
&=\frac{1}{2}G_2\sum_{j=1}^m\sum_{i=0}^{m}\sum_{\alpha=1}^{n_i}\frac{(n_i+1-\alpha)(2n_i+2+\alpha)}{n_i+1+\alpha}
(n_j+1)x_{\alpha}(i)x_{n_i+1-\alpha}(i)x_{n_j+1}(j)s^j\\
&\quad +\frac{1}{2}\sum_{i,j=1, i\neq j}^m\Bigg\{\sum_{\alpha=1}^{n_i}\frac{(n_i+1-\alpha)(n_i+1)(n_j+1)}{n_i+1+\alpha}
x_{n_i+1-\alpha}(i)x_{n_j+1}(j)\mathbf{L}_{i,\alpha}^{\phi}[\K](s^i-s^j)\Bigg\}\\
&\quad -\frac{1}{2}(n_0+1)x_{n_0+1}(0)\Bigg(\sum_{i=1}^m\Bigg\{\sum_{\alpha=1}^{n_i}\frac{(n_i+1-\alpha)(n_i+1)}{n_i+1+\alpha}
x_{n_i+1-\alpha}(i)\mathbf{L}_{i,\alpha}^{\phi}[\K](s^i)\Bigg\}\Bigg)\\
&\quad +\frac{1}{2}\sum_{i,j=1, i\neq j}^m\Bigg\{\sum_{k=2}^{n_i+1}\frac{(n_i+1)(n_j+1)}{k!}x_{n_j+1}(j)\mathcal{B}_{n_i+1,k}(i)\K^{(k-1)}(s^i-s^j)
\Bigg\}\\
&\quad -\frac{1}{2}\sum_{i,j=1}^m\Bigg\{\sum_{k=2}^{n_i+1}\frac{(n_i+1)(n_j+1)}{k!}x_{n_j+1}(j)\mathcal{B}_{n_i+1,k}(i)\K^{(k-1)}(s^i)
\Bigg\}\\
&\quad +\frac{1}{2}\sum_{j=0}^m\Bigg\{\sum_{\alpha=1}^{n_j}\frac{(n_j+1-\alpha)(n_j+1)^2}{n_j+1+\alpha}x_{n_j+1-\alpha}(j)x_{n_j+1}(j)\mathcal{R}_{\alpha,1}(j)\Bigg\}\\
&\quad-\frac{1}{2}\sum_{j=0}^m\Bigg\{\sum_{\alpha=1}^{n_j}\frac{(n_j+1-\alpha)(n_j+1)^2}{n_j+1+\alpha}x_{n_j+1-\alpha}(j)x_{n_j+1}(j)
\bigg(\sum_{\ell=4}^{\alpha}\frac{\mathcal{B}_{\alpha,\ell}(j)}{\ell}G_{\ell}\bigg)\Bigg\}\\
&\quad +\frac{1}{2}\sum_{j=0}^m\Bigg\{\sum_{\alpha=1}^{n_j}(n_j+1-\alpha)(n_j+1)x_{n_j+1-\alpha}(j)x_{n_j+1}(j)\mathcal{R}_{\alpha,1}(j)
 -\sum_{\ell=4}^{n_j}\frac{(n_j+1)^2}{\ell(\ell+1)}x_{n_j+1}(j)\mathcal{B}_{n_j+1,\ell+1}(j)G_{\ell}\Bigg\}\\
&\quad +\frac{1}{2}\sum_{j=1}^m\sum_{\alpha=1}^{n_0}\frac{(n_0+1-\alpha)(n_0+1)(n_j+1)}{n_0+1+\alpha}x_{n_0+1-\alpha}(0)x_{n_j+1}(j) \mathbf{L}_{0,\alpha}^{\phi}\big[\K\big](-s^j)\bigg)\\
&\quad +\frac{1}{2}\sum_{j=1}^m\left\{(n_j+1)x_{n_j+1}(j)\sum_{k=1}^{n_0+1}\frac{(n_0+1)}{k!}\mathcal{B}_{n_0+1,k}(0)\K^{(k-1)}(-s^j)\right\}\\
&\quad -\frac{1}{2}(n_0+1)x_{n_0+1}(0)\sum_{j=1}^m(n_j+1)x_{n_j+1}(j)\K(s^j)+\frac{\log(-1)}{2}\bigg(\sum_{j=1}^m(n_j+1)x_{n_j+1}(j)\bigg)^2,
\end{align*}
where, by (\ref{x-r-sum}), we have
$$
(n_0+1)x_{n_0+1}(0)=-\sum_{j=1}^m(n_j+1)x_{n_j+1}(j).
$$
\end{Prop}

\begin{description}
\item[$\bullet$] \textbf{Expression for the fourth line in (\ref{Prep-phi1})}
\end{description}
Lastly, we move on to find the explicit form of the fourth line involving the functions $\mathcal{I}_{i,\alpha}\big[\Psi_{j,\beta}\big]$ (\ref{chi-ij-def}):
\begin{align*}
\Sigma_4&:=\frac{1}{2}\sum_{i,j=0}^{m}\sum_{\alpha=1}^{n_i}\sum_{\beta=1}^{n_j}\frac{(n_i+1-\alpha)(n_j+1-\beta)}{(n_i+1+\alpha)}
(n_i+1)x_{n_i+1-\alpha}(i)x_{n_j+1-\beta}(j)\Big(\frac{1}{\alpha\beta}\mathcal{I}_{i,\alpha}\big[\Psi_{j,\beta}\big]\Big)\\
&=\frac{1}{2}G_2(\tau)\sum_{i,j=0}^{m}\sum_{\alpha=1}^{n_i}\sum_{\beta=1}^{n_j}\frac{(n_i+1-\alpha)(n_j+1-\beta)}{(n_i+1+\alpha)}
(n_i+1)x_{\alpha}(i)x_{\beta}(j)x_{n_i+1-\alpha}(i)x_{n_j+1-\beta}(j)\\
&\quad +\frac{1}{2}\sum_{i,j=0}^{m}\sum_{\alpha=1}^{n_i}\sum_{\beta=1}^{n_j}\frac{(n_i+1-\alpha)(n_j+1-\beta)(n_i+1)}{(n_i+1+\alpha)}x_{n_i+1-\alpha}(i)
x_{n_j+1-\beta}(j)\chi(i,j;\alpha,\beta),
\end{align*}
where
\begin{equation}\label{chi2-ij}
\chi(i,j;\alpha,\beta):=\frac{1}{\alpha\beta}\Big(\mathcal{I}_{i,\alpha}\big[\Psi_{j,\beta}\big]-2\alpha\beta x_{\alpha}(i)x_{\beta}(j)\zeta(1/2)\Big).
\end{equation}
\begin{Lemma} For $i,j=0,\dots,m$, define the functions
\begin{equation}\label{Q-ij-def}
\mathbf{Q}_{ij}:=\sum_{\alpha=1}^{n_i}\sum_{\beta=1}^{n_j}\frac{(n_i+1-\alpha)(n_j+1-\beta)(n_i+1)}{(n_i+1+\alpha)}x_{n_i+1-\alpha}(i)
x_{n_j+1-\beta}(j)\chi(i,j;\alpha,\beta),
\end{equation}
where $\chi(i,j;\alpha,\beta)$ is given by (\ref{chi2-ij}).
\begin{description}
\item[1)] If $i\neq j$, then
\begin{equation}\label{Q-ij}
\begin{split}
\mathbf{Q}_{ij}&=-\sum_{\alpha=1}^{n_i}\Bigg\{\left(\frac{(n_i+1-\alpha)(n_i+1)(n_j+1)}{(n_i+1+\alpha)}x_{n_i+1-\alpha}(i)\right)\\
&\quad \quad \times \bigg(\sum_{k=2}^{n_j+1}\frac{(-1)^k}{k!}\mathcal{B}_{n_j+1,k}(j)\mathbf{L}_{i,\alpha}^{\phi}\big[\K^{(k-1)}\big](\infty^i-\infty^j)\bigg)\Bigg\}.
\end{split}
\end{equation}
\item[2)] When $i=j$ we have
\begin{equation}\label{Q-jj}
\begin{split}
\mathbf{Q}_{jj}&=\sum_{\alpha=1}^{n_j}\sum_{k=2}^{n_j+1}\frac{(n_j+1-\alpha)(n_j+1)^2}{k(n_i+1+\alpha)}x_{n_j+1-\alpha}(j)
\mathcal{B}_{n_j+1,k}(j)\mathcal{R}_{\alpha+k-1,k}(j)\\
&\quad +\sum_{\alpha=1}^{n_j}\sum_{k=2, k+\alpha\geq 5}^{n_j+1}\Bigg\{
\left[(-1)^k\frac{(n_j+1-\alpha)(n_j+1)^2}{k(n_i+1+\alpha)}x_{n_j+1-\alpha}(j)\mathcal{B}_{n_j+1,k}(j)\right]\\
&\quad \quad \times \bigg[\sum_{\ell=1, k+\ell\geq 5}^{\alpha}\frac{1}{k+\ell-1}\binom{k+\ell-1}{\ell}\mathcal{B}_{\alpha,\ell}(j)G_{k+\ell-1}\bigg]\Bigg\},
\end{split}
\end{equation}
where $\mathcal{R}_{\alpha+k-1,k}(j)=\mathcal{R}_{\alpha+k-1,k}\big(x_1(j),\dots,x_{\alpha+k}(j)\big)$ denotes the rational function (\ref{Bell-rec4}).
\end{description}
\end{Lemma}
\emph{Proof:} Assume first that  $i\neq j$. Then expression (\ref{chi-ij}) for $\mathcal{I}_{i,\alpha}\big[\Psi_{j,\beta}\big]$ and (\ref{chi2-ij})  imply that
\begin{align*}
\chi(i,j;\alpha,\beta)
&=-\sum_{k=1}^{\beta}\sum_{\ell=1}^{\alpha}\frac{(-1)^k}{k!\ell!}\mathcal{B}_{\beta,k}(j)\mathcal{B}_{\alpha,\ell}(i)\wp^{(k+l-2)}(\infty^i-\infty^j).
\end{align*}
Now, using (\ref{Q-ij-def}) and applying recurrence relation (\ref{Bell-rec1}) for partial ordinary Bell polynomials and observing that $\K''(u)=-\wp(u)$, we get
\begin{align*}
\mathbf{Q}_{ij}
&=-\sum_{\alpha=1}^{n_i}\sum_{\beta=1}^{n_j}\left(\frac{(n_i+1-\alpha)(n_j+1-\beta)(n_i+1)}{(n_i+1+\alpha)}x_{n_i+1-\alpha}(i)x_{n_j+1-\beta}(j)\right)\\
&\quad \quad \times
\bigg(\sum_{k=1}^{\beta}\sum_{\ell=1}^{\alpha}\frac{(-1)^k}{k!\ell!}\mathcal{B}_{\beta,k}(j)\mathcal{B}_{\alpha,\ell}(i)\wp^{(k+l-2)}(\infty^i-\infty^j)\bigg)\\
&=-\sum_{\alpha=1}^{n_i}\sum_{k=1}^{n_j}\left((-1)^k\frac{(n_i+1-\alpha)(n_i+1)(n_j+1)}{(n_i+1+\alpha)(k+1)!}x_{n_i+1-\alpha}(i)
\mathcal{B}_{n_j+1,k+1}(j)\bigg(\sum_{\ell=1}^{\alpha}\frac{1}{\ell!}\mathcal{B}_{\alpha,\ell}(i)\wp^{(k+l-2)}(\infty^i-\infty^j)\bigg)\right)\\
&=-\sum_{\alpha=1}^{n_i}\sum_{k=2}^{n_j+1}\left((-1)^k\frac{(n_i+1-\alpha)(n_i+1)(n_j+1)}{(n_i+1+\alpha)k!}x_{n_i+1-\alpha}(i)
\mathcal{B}_{n_j+1,k}(j)\mathbf{L}_{i,\alpha}^{\phi}\big[\K^{(k-1)}\big](\infty^i-\infty^j)\right).
\end{align*}
Let us now  deal with the case where $i=j$. As above, by employing  (\ref{chi-ii}) and (\ref{chi2-ij}) we deduce
\begin{align*}
&\chi(j,j;\alpha,\beta):=\frac{1}{\alpha\beta}\Big(\mathcal{I}_{j,\alpha}\big[\Psi_{j,\beta}\big]-2\alpha\beta x_{\alpha}(j)x_{\beta}(j)\zeta(1/2)\Big)\\
&=\sum_{k=1}^{\beta}\mathcal{B}_{\beta,k}(j)\mathcal{R}_{\alpha+k,k+1}(j)-\sum_{k=1}^{\beta}\bigg((-1)^k\mathcal{B}_{\beta,k}(j)
\sum_{\ell=1, k+\ell\geq 4}^{\alpha}\frac{1}{k+\ell}\binom{k+\ell}{\ell}\mathcal{B}_{\alpha,\ell}(j)G_{k+\ell}\bigg).
\end{align*}
Therefore
\begin{align*}
&\mathbf{Q}_{jj}:=\sum_{\alpha=1}^{n_j}\sum_{\beta=1}^{n_j}\frac{(n_j+1-\alpha)(n_j+1-\beta)(n_j+1)}{(n_j+1+\alpha)}x_{n_j+1-\alpha}(j)
x_{n_j+1-\beta}(j)\chi(j,j;\alpha,\beta)\\
&=\sum_{\alpha=1}^{n_j}\sum_{\beta=1}^{n_j}\frac{(n_j+1-\alpha)(n_j+1-\beta)(n_j+1)}{(n_j+1+\alpha)}x_{n_j+1-\alpha}(j)
x_{n_j+1-\beta}(j)\left(\sum_{k=1}^{\beta}\mathcal{B}_{\beta,k}(j)\mathcal{R}_{\alpha+k,k+1}(j)\right)\\
&\quad -\sum_{\alpha=1}^{n_j}\sum_{\beta=1}^{n_j}\Bigg\{\left(\frac{(n_j+1-\alpha)(n_j+1-\beta)(n_j+1)}{(n_i+1+\alpha)}x_{n_j+1-\alpha}(j)x_{n_j+1-\beta}(j)\right)\\
&\quad \quad \times \left[\sum_{k=1}^{\beta}\bigg[(-1)^k\mathcal{B}_{\beta,k}(j)
\sum_{\ell=1, k+\ell\geq 4}^{\alpha}\frac{1}{k+\ell}\binom{k+\ell}{\ell}\mathcal{B}_{\alpha,\ell}(j)G_{k+\ell}\bigg]\right]\Bigg\}\\
&=\sum_{\alpha=1}^{n_j}\sum_{k=2}^{n_j+1}\frac{(n_j+1-\alpha)(n_j+1)^2}{k(n_j+1+\alpha)}x_{n_j+1-\alpha}(j)
\mathcal{B}_{n_j+1,k}(j)\mathcal{R}_{\alpha+k-1,k}(j)\\
&\quad +\sum_{\alpha=1}^{n_j}\sum_{k=2, k+\alpha\geq 5}^{n_j+1}\Bigg\{\left((-1)^k\frac{(n_j+1-\alpha)(n_j+1)^2}{k(n_i+1+\alpha)}x_{n_j+1-\alpha}(j)
\mathcal{B}_{n_j+1,k}(j)\right)\\
&\quad \quad \times \bigg[\sum_{\ell=1, k+\ell\geq 5}^{\alpha}\frac{1}{k+\ell-1}\binom{k+\ell-1}{\ell}\mathcal{B}_{\alpha,\ell}(j)G_{k+\ell-1}\bigg]\Bigg\}.
\end{align*}

\fd

\begin{Prop}\label{Line4} Let $\mathcal{R}_{\alpha+k-1,k}(j)=\mathcal{R}_{\alpha+k-1,k}\big(x_1(j),\dots,x_{\alpha+k}(j)\big)$ be the rational function (\ref{Bell-rec4}). Then with the notation of Theorem \ref{Second kind-thm}, the expression for the fourth line in (\ref{Prep-phi1}) is given by
\begin{align*}
\Sigma_4&:=\frac{1}{2}\sum_{i,j=0}^{m}\sum_{\alpha=1}^{n_i}\sum_{\beta=1}^{n_j}\frac{(n_i+1-\alpha)(n_j+1-\beta)}{(n_i+1+\alpha)}
(n_i+1)x_{n_i+1-\alpha}(i)x_{n_j+1-\beta}(j)\Big(\frac{1}{\alpha\beta}\mathcal{I}_{i,\alpha}\big[\Psi_{j,\beta}\big]\Big)\\
&=\frac{1}{2}G_2\sum_{i,j=0}^{m}\sum_{\alpha=1}^{n_i}\sum_{\beta=1}^{n_j}\frac{(n_i+1-\alpha)(n_j+1-\beta)(n_i+1)}{(n_i+1+\alpha)}x_{\alpha}(i)x_{\beta}(j)x_{n_i+1-\alpha}(i)
x_{n_j+1-\beta}(j)\\
&\quad -\frac{1}{2}\sum_{i,j=1, i\neq j}^{m}\sum_{\alpha=1}^{n_i}\left\{\frac{(n_i+1-\alpha)(n_i+1)(n_j+1)}{(n_i+1+\alpha)}x_{n_i+1-\alpha}(i)
\bigg(\sum_{k=2}^{n_j+1}\frac{(-1)^k}{k!}\mathcal{B}_{n_j+1,k}(j)\mathbf{L}_{i,\alpha}^{\phi}\big[\K^{(k-1)}\big](s^i-s^j)\bigg)\right\}\\
&\quad -\frac{1}{2}\sum_{i=1}^{m}\sum_{\alpha=1}^{n_i}\left\{\frac{(n_i+1-\alpha)(n_i+1)(n_0+1)}{(n_i+1+\alpha)}x_{n_i+1-\alpha}(i)
\bigg(\sum_{k=2}^{n_0+1}\frac{(-1)^k}{k!}\mathcal{B}_{n_0+1,k}(0)\mathbf{L}_{i,\alpha}^{\phi}\big[\K^{(k-1)}\big](s^i)\bigg)\right\}\\
&\quad -\frac{1}{2}\sum_{j=1}^{m}\sum_{\alpha=1}^{n_0}\left\{\frac{(n_0+1-\alpha)(n_0+1)(n_j+1)}{(n_0+1+\alpha)}x_{n_0+1-\alpha}(0)
\bigg(\sum_{k=2}^{n_j+1}\frac{(-1)^k}{k!}\mathcal{B}_{n_j+1,k}(j)\mathbf{L}_{0,\alpha}^{\phi}\big[\K^{(k-1)}\big](-s^j)\bigg)\right\}\\
&\quad +\frac{1}{2}\sum_{j=0}^{m}\sum_{\alpha=1}^{n_j}\sum_{k=2}^{n_j+1}\frac{(n_j+1-\alpha)(n_j+1)^2}{k(n_j+1+\alpha)}x_{n_j+1-\alpha}(j)
\mathcal{B}_{n_j+1,k}(j)\mathcal{R}_{\alpha+k-1,k}(j)\\
&\quad +\frac{1}{2}\sum_{j=0}^{m}\sum_{\alpha=1}^{n_j}\sum_{k=2, k+\alpha\geq 5}^{n_j+1}\Bigg\{
\left[(-1)^k\frac{(n_j+1-\alpha)(n_j+1)^2}{k(n_j+1+\alpha)}x_{n_j+1-\alpha}(j)\mathcal{B}_{n_j+1,k}(j)\right]\\
&\quad \quad \times \bigg[\sum_{\ell=1, k+\ell\geq 5}^{\alpha}\binom{k+\ell-1}{\ell}\mathcal{B}_{\alpha,\ell}(j)\frac{G_{k+\ell-1}}{k+\ell-1}\bigg]\Bigg\}.
\end{align*}
\end{Prop}
\emph{Proof:} Since
$$
\infty^i-\infty^j=\delta_{j,0}s^i-\delta_{i,0}s^j+(1-\delta_{i,0}-\delta_{j,0})(s^i-s^j), \quad \quad \forall\ i,j=0,\dots,m,
$$
and
\begin{align*}
\Sigma_4&=\frac{1}{2}G_2(\tau)\sum_{i,j=0}^{m}\sum_{\alpha=1}^{n_i}\sum_{\beta=1}^{n_j}\frac{(n_i+1-\alpha)(n_j+1-\beta)(n_i+1)}{(n_i+1+\alpha)}
x_{\alpha}(i)x_{\beta}(j)x_{n_i+1-\alpha}(i)x_{n_j+1-\beta}(j)\\
&+\frac{1}{2}\sum_{i,j=1, i\neq j}^{m}\mathbf{Q}_{ij}+\frac{1}{2}\sum_{i=1}^{m}\mathbf{Q}_{i0}+\frac{1}{2}\sum_{j=1}^{m}\mathbf{Q}_{0j}+\frac{1}{2}\sum_{j=0}^{m}\mathbf{Q}_{jj},
\end{align*}
it follows that the desired result holds by employing expressions (\ref{Q-ij}) and (\ref{Q-jj}).

\fd

\medskip

We are  in position to finally state our main theorem which is an immediate consequence of (\ref{Prep-phi1}), Propositions \ref{Line2}, \ref{Line3} and \ref{Line4} and simples computations.

\begin{Thm}\label{Main result} Consider the semi-simple Frobenius-Hurwitz manifold $\mathcal{H}_1(n_0,\dots,n_m)$  induced by the normalized holomorphic differential $\phi(P)=dz_P$ and denote by $\mathbf{F}_{\phi}$  the corresponding WDVV prepotential.
Let
\begin{enumerate}
\item $\K$  be the function $\K(v)=\K(v|\tau):=\log\sigma(v|\tau)$, with $\sigma$ being the Weierstrass sigma function and $G_{2\ell}$ be the Eisenstein series (\ref{Eisentein-Fourier});
\item  $\Big\{x_{\alpha}(j):\quad\text{with}\quad j=0,\dots,m,\ \alpha=1,\dots, n_j+1\Big\}$ be the variables  defined by (\ref{phi-Taylor})-(\ref{Evaluation}) and satisfying (\ref{x-r-sum}) and $\tau,u,s^j$ be such that
$$
\tau:=\oint_b\phi(P),\quad \quad  u:=\oint_a\lambda(P)\phi(P)\quad\quad \text{and}\quad s^j:=\int_{\infty^0}^{\infty^j}\phi(P)=\infty^j-\infty^0;
$$
\item $\mathcal{B}_{n_j+1,k}(j)=\mathcal{B}_{n_j+1,k}\big(x_1(j),\dots,x_{n_j+2-k}(j)\big)$ be the partial ordinary Bell polynomials (\ref{Bell-partial})- (\ref{Bell2});
\item $\mathbf{L}_{j,\alpha}^{\phi}$ be the Bell  operator defined by (\ref{Bell operator});
\item $\mathcal{R}_{\alpha+k-1,k}(j)=\mathcal{R}_{\alpha+k-1,k}\big(x_1(j),\dots, x_{\alpha+k}(j)\big)$ be the rational function given by (\ref{Bell-rec4}) which  depends (when $\alpha+k>n_j+1)$ on the quantities $x_{n_j+1+\beta}(j)$, $\beta=1,\dots, n_j$, determined by formula (\ref{x-alpha2}).
\end{enumerate}
Then, with the above notation, the WDVV prepotential $\mathbf{F}_{\phi}$  is given by
\begin{align*}
&\mathbf{F}_{\phi}=\frac{u^2}{4{\rm i}\pi}\tau+u\sum_{j=1}^{m}(n_j+1)x_{n_j+1}(j)s^j
+\frac{u}{2}\sum_{i=0}^{m}\sum_{\alpha=1}^{n_i}\frac{(n_i+1-\alpha)(2n_i+2+\alpha)}{(n_i+1+\alpha)}x_{\alpha}(i)x_{n_i+1-\alpha}(i)\\
&+\frac{1}{2}G_2(\tau)\bigg(\sum_{j=1}^{m}(n_j+1)x_{n_j+1}(j)s^j\bigg)^2\\
&+\frac{1}{2}G_2(\tau)\bigg(\sum_{j=1}^{m}(n_j+1)x_{n_j+1}(j)s^j\bigg)
\bigg(\sum_{i=0}^{m}\sum_{\alpha=1}^{n_i}\frac{(n_i+1-\alpha)(2n_i+2+\alpha)}{n_i+1+\alpha}x_{\alpha}(i)x_{n_i+1-\alpha}(i)\bigg)\\
&+\frac{1}{2}G_2(\tau)\sum_{i,j=0}^{m}\sum_{\alpha=1}^{n_i}\sum_{\beta=1}^{n_j}\frac{(n_i+1-\alpha)(n_j+1-\beta)(n_i+1)}{(n_i+1+\alpha)}
x_{\alpha}(i)x_{\beta}(j)x_{n_i+1-\alpha}(i)x_{n_j+1-\beta}(j)\\
&+\frac{1}{2}\sum_{i,j=1, i\neq j}^m\Bigg\{\sum_{k=1}^{n_i+1}\frac{(n_i+1)(n_j+1)}{k!}x_{n_j+1}(j)\mathcal{B}_{n_i+1,k}(i)\K^{(k-1)}\big(s^i-s^j|\tau\big)\Bigg\}\\
&-\frac{1}{2}\sum_{i,j=1}^m\Bigg\{\sum_{k=1}^{n_i+1}\frac{(n_i+1)(n_j+1)}{k!}x_{n_j+1}(j)\mathcal{B}_{n_i+1,k}(i)\K^{(k-1)}\big(s^i|\tau\big)\Bigg\}\\
&+\frac{1}{2}\sum_{j=1}^m\left\{(n_j+1)x_{n_j+1}(j)\sum_{k=1}^{n_0+1}\frac{(n_0+1)}{k!}\mathcal{B}_{n_0+1,k}(0)\K^{(k-1)}\big(-s^j|\tau\big)\right\}\\
&-\frac{1}{2}\sum_{i,j=1, i\neq j}^{m}\sum_{\alpha=1}^{n_i}\left\{\frac{(n_i+1-\alpha)(n_i+1)(n_j+1)}{(n_i+1+\alpha)}x_{n_i+1-\alpha}(i)
\bigg(\sum_{k=1}^{n_j+1}\frac{(-1)^k}{k!}\mathcal{B}_{n_j+1,k}(j)\mathbf{L}_{i,\alpha}^{\phi}\big[\K^{(k-1)}(\cdot|\tau)\big](s^i-s^j)\bigg)\right\}\\
&-\frac{1}{2}\sum_{i=1}^{m}\sum_{\alpha=1}^{n_i}\left\{\frac{(n_i+1-\alpha)(n_i+1)(n_0+1)}{(n_i+1+\alpha)}x_{n_i+1-\alpha}(i)
\bigg(\sum_{k=1}^{n_0+1}\frac{(-1)^k}{k!}\mathcal{B}_{n_0+1,k}(0)\mathbf{L}_{i,\alpha}^{\phi}\big[\K^{(k-1)}(\cdot|\tau)\big](s^i)\bigg)\right\}\\
&-\frac{1}{2}\sum_{j=1}^{m}\sum_{\alpha=1}^{n_0}\left\{\frac{(n_0+1-\alpha)(n_0+1)(n_j+1)}{(n_0+1+\alpha)}x_{n_0+1-\alpha}(0)
\bigg(\sum_{k=1}^{n_j+1}\frac{(-1)^k}{k!}\mathcal{B}_{n_j+1,k}(j)\mathbf{L}_{0,\alpha}^{\phi}\big[\K^{(k-1)}(\cdot|\tau)\big](-s^j)\bigg)\right\}\\
&+\frac{1}{2}\sum_{j=0}^{m}\sum_{\alpha=1}^{n_j}\sum_{k=1}^{n_j+1}\frac{(n_j+1-\alpha)(n_j+1)^2}{k(n_j+1+\alpha)}x_{n_j+1-\alpha}(j)
\mathcal{B}_{n_j+1,k}(j)\mathcal{R}_{\alpha+k-1,k}(j)\\
& +\frac{1}{2}\sum_{j=0}^{m}\sum_{\alpha=1}^{n_j}\sum_{k=1, k+\alpha\geq 5}^{n_j+1}\Bigg\{
\left[(-1)^k\frac{(n_j+1-\alpha)(n_j+1)^2}{k(n_j+1+\alpha)}x_{n_j+1-\alpha}(j)\mathcal{B}_{n_j+1,k}(j)\right]\\
&\quad \quad \times \bigg[\sum_{\ell=1, k+\ell\geq 5}^{\alpha}\binom{k+\ell-1}{\ell}\mathcal{B}_{\alpha,\ell}(j)\frac{G_{k+\ell-1}(\tau)}{k+\ell-1}\bigg]\Bigg\}\\
&+\frac{1}{2}\sum_{j=0}^m\Bigg\{\sum_{\alpha=1}^{n_j}(n_j+1-\alpha)(n_j+1)x_{n_j+1-\alpha}(j)x_{n_j+1}(j)\mathcal{R}_{\alpha,1}(j)
 -\sum_{\ell=4}^{n_j}\frac{(n_j+1)^2}{\ell(\ell+1)}x_{n_j+1}(j)\mathcal{B}_{n_j+1,\ell+1}(j)G_{\ell}(\tau)\Bigg\}\\
&+\frac{1}{2}\sum_{j=1}^{m}\Big((n_j+1)x_{n_j+1}(j)\Big)^2\log\big(x_1(j)\big)
+\frac{1}{2}\bigg(\sum_{j=1}^m(n_j+1)x_{n_j+1}(j)\bigg)^2\log\big(x_1(0)\big)\\
&+\log(-1)\bigg(\sum_{j=1}^m(n_j+1)x_{n_j+1}(j)\bigg)^2+\frac{\log(-1)}{4}\sum_{i,j=1, i\neq j}^m(n_i+1)(n_j+1)x_{n_i+1}(i)x_{n_j+1}(j)\\
&-\frac{3}{4}\sum_{i,j=1}^m\bigg(\frac{(n_i+1)(n_j+1)}{n_0+1}+\delta_{ij}(n_j+1)\bigg)x_{n_i+1}(i)x_{n_j+1}(j).
\end{align*}
\end{Thm}
Let us point out that, by (\ref{E-FC}) and (\ref{x-t-v}), the unit and Euler vector fields of the $\phi$-Frobenius manifold structure on $\mathcal{H}_1(n_0,\dots,n_m)$ are such that
$e=\partial_{u}$ and
\begin{equation}\label{E-phi}
E=u\partial_{u}+\sum_{j=1}^mx_{n_j+1}(j)\partial_{x_{n_j+1}(j)}+\sum_{j=0}^m\sum_{\alpha=1}^{n_j}\frac{\alpha}{n_j+1}x_{\alpha}(j)\partial_{x_{\alpha}(j)}.
\end{equation}
In addition, the  prepotential $\mathbf{F}_{\phi}$ form Theorem \ref{Main result} is a quasi-homogeneous function of degree $2$ with respect to the Euler vector field (\ref{E-phi}):
$$
E.\mathbf{F}_{\phi}=2\mathbf{F}_{\phi}
$$
and solves  the WDVV equations (\ref{WDVV}) with respect to the $N=2+2m+\sum_{k=0}^mn_k$ variables
$$
u,\tau,s^1,\dots,s^m,x_1(0),\dots,x_{n_0}(0),x_1(j),\dots,x_{n_j+1}(j), \quad\quad \text{with}\quad  j=1,\dots,m
$$
where the distinguished variable corresponding to $F$ is $u$, i.e. the constant Gram matrix (\ref{Entries eta-phi}) of the flat metric $\eta(\phi)$ coincides with the Hessian matrix of $\partial_{u}\mathbf{F}_{\phi}$.\\
On the other hand, as we have already highlighted, a crucial aspect in  constructing  of the explicit form of the specific WDVV prepotential $\mathbf{F}_{\phi}$ relies on the fact that the procedure  requires only the computation  of some flat coordinates  among  the list $\big\{\mathbf{t}^A\big(\phi_{\mathbf{t}^B}\big)\big\}$ related to the  $N$ primary differentials listed in (\ref{Primary-e}). In fact, the remaining operations $\big\{\mathbf{t}^A\big(\phi_{\mathbf{t}^B}\big)\big\}$, whose calculations are in general rather difficult, can be directly obtained from the Hessian matrix of the function $\mathbf{F}_{\phi}$ using formula (\ref{Hessian}). For example,  equality  (\ref{Hessian}), duality relations (\ref{duality-picture}) and (\ref{x-t-v}) lead to  the following results:
\begin{equation}\label{a-periods-hessian}
\begin{split}
&\oint_a\lambda(P)\Omega_{\infty^0\infty^j}(P)=\mathbf{u}\big(\phi_{\mathbf{s}^{j}}\big)=
\partial_{\mathbf{u}^{\prime}(\phi)}\partial_{{\mathbf{s}^j}^{\prime}(\phi)}\mathbf{F}_{\phi}
=\frac{2{\rm i}\pi}{n_j+1}\partial_{\tau}\partial_{x_{n_j+1}(j)}\mathbf{F}_{\phi};\\
&\oint_a\lambda(P)\Psi_{j,n_j+1}(P)=\mathbf{u}\big(\phi_{\mathbf{v}^{j}}\big)=\partial_{\mathbf{u}^{\prime}(\phi)}\partial_{{\mathbf{v}^j}^{\prime}(\phi)}\mathbf{F}_{\phi}
=2{\rm i}\pi\partial_{\tau}\partial_{s^j}\mathbf{F}_{\phi};\\
&\oint_a\lambda(P)\Psi_{j,\alpha}(P)=\frac{\alpha}{\sqrt{n_j+1}}\mathbf{u}\big(\phi_{\mathbf{t}^{j,\alpha}}\big)
=\frac{2{\rm i}\pi\alpha}{n_j+1}\partial_{\tau}\partial_{x_{n_j+1-\alpha}(j)}\mathbf{F}_{\phi};\\
&\oint_a\lambda(P)\bigg(\oint_a\lambda(Q)W(P,Q)\bigg)=\mathbf{u}\big(\phi_{\mathbf{u}}\big)=\big(2{\rm i}\pi\big)^2\partial_{\tau}^2\mathbf{F}_{\phi},
\end{split}
\end{equation}
where $\lambda$ is the meromorphic function given by (\ref{lambda-p}), $\Omega_{\infty^0\infty^j}$ is the Abelian  differential of the third d given by (\ref{Third kind}),  $\Psi_{j,\alpha}$ is the Abelian  differentials of the second kind  (\ref{Psi-i-alpha}), $W(P,Q)$ is the genus one bidifferential (\ref{W-g1}) and $\mathbf{F}_{\phi}$ is given by Theorem \ref{Main result}.

\subsection{Some special cases}
By considering particular choices of the assigned combinatorial parameters $(n_0,\dots,n_m)$ of the Hurwitz-Frobenius manifold $\mathcal{H}_1(n_0,\dots,n_m)$ associated with $\phi$, this section is devoted  to discuss and  highlight some interesting specific cases of the WDVV solution $\mathbf{F}_{\phi}$ presented
in Theorem \ref{Main result}.\\
To accomplish this goal, we  need the following special cases of the rational function $\mathcal{R}_{\mu,k}(x_1,\dots,x_{\mu+1})$ (\ref{Bell-rec4}).
\begin{Lemma} Let $\mathcal{R}_{\mu,k}$ be the rational function given by (\ref{Bell-rec4}). Then we have
\begin{equation}\label{R-examples}
\begin{split}
\mathcal{R}_{1,1}&=-\frac{x_2}{x_1};\\
\mathcal{R}_{2,1}&=\frac{x_3}{x_1}-\frac{1}{2}\frac{x_2^2}{x_1^2};\\
\mathcal{R}_{2,2}&=\frac{x_3}{x_1^2}-\frac{x_2^2}{x_1^3};\\
\mathcal{R}_{3,1}&=\frac{x_4}{x_1}-\frac{x_2x_3}{x_1^2}+\frac{1}{3}\frac{x_2^3}{x_1^3};\\
\mathcal{R}_{3,2}&=\frac{x_4}{x_1^2}-2\frac{x_2x_3}{x_1^3}+\frac{x_2^3}{x_1^4};\\
\mathcal{R}_{3,3}&=\frac{x_4}{x_1^3}-3\frac{x_2x_3}{x_1^4}+2\frac{x_2^3}{x_1^5};\\
\mathcal{R}_{4,2}&=\frac{x_5}{x_1^2}-\frac{2x_2x_4+x_3^2}{x_1^3}+\frac{3x_2^2x_3}{x_1^4}-\frac{x_2^4}{x_1^5};\\
\mathcal{R}_{4,3}&=\frac{x_5}{x_1^3}-3\frac{x_2x_4}{x_1^4}-\frac{3}{2}\frac{x_3^2}{x_1^4}+6\frac{x_2^2x_3}{x_1^5}-\frac{5}{2}\frac{x_2^4}{x_1^6};\\
\mathcal{R}_{4,4}&=\frac{x_5}{x_1^4}-2\frac{2x_2x_4+x_3^2}{x_1^5}+10\frac{x_2^2x_3}{x_1^6}-5\frac{x_2^4}{x_1^7};\\
\mathcal{R}_{5,3}&=\frac{x_6}{x_1^3}-3\frac{x_2x_5+x_3x_4}{x_1^4}+6\frac{x_2^2x_4+x_2x_3^2}{x_1^5}-10\frac{x_2^3x_3}{x_1^6}+3\frac{x_2^5}{x_1^7};\\
\mathcal{R}_{5,4}&=\frac{x_6}{x_1^4}-4\frac{x_2x_5+x_3x_4}{x_1^5}+10\frac{x_2^2x_4+x_2x_3^2}{x_1^6}-20\frac{x_2^3x_3}{x_1^7}+7\frac{x_2^5}{x_1^8};\\
\mathcal{R}_{6,4}&=\frac{x_7}{x_1^4}-2\frac{2x_2x_6+2x_3x_5+x_4^2}{x_1^5}+\frac{10}{3}\frac{3x_2^2x_5
+6x_2x_3x_4+x_3^3}{x_1^6}-10\frac{2x_2^3x_4+3x_2^2x_3^2}{x_1^7}\\
&\quad +35\frac{x_2^4x_3}{x_1^8}-\frac{28}{3}\frac{x_2^6}{x_1^9}.
\end{split}
\end{equation}
\end{Lemma}
\emph{Proof:} These expressions are gotten by straightforward application of definition (\ref{Bell-rec4}) using relations (\ref{Bell-Examples}) and (\ref{Bell-rec0}) for partial Bell polynomials.

\fd

\begin{description}
\item[$\bullet$]\textbf{The Hurwitz-Frobenius manifold $\mathcal{H}_1(0,\dots,0)$}
\end{description}
We first start with the Hurwitz space of  $\mathbb{L}$-elliptic functions $\lambda$ having  simple poles $\infty^0,\dots,\infty^m$. That is, by (\ref{lambda-p}), $\lambda$ is given by
$$
\lambda(P)=c+\sum_{j=1}^mx^j\Big(\zeta(z_P-\infty^j)-\zeta(z_P-\infty^0)\Big),\quad \quad m\geq 1.
$$
In this case, the Frobenius manifold $\mathcal{H}_1(0,\dots,0)$ is of dimension $N=2+2m$ and  flat coordinates (\ref{FC-phi2}) of the metric $\eta(\phi)$ (\ref{eta-def}) are $u=\oint_a\lambda(P)\phi(P)$, $\tau=\oint_b\phi(P)$ and
$$
x^j=x_1(j)=\underset{\infty^j}{{\rm res}\ }\lambda(P)\phi(P), \quad \quad s^j=\infty^j-\infty^0,\quad \quad j=1,\dots,m.
$$
\begin{Prop} Assume that $n_0=\dots=n_m=0$. Then, with the preceding notation,  the WDVV prepotential $\mathbf{F}_{\phi}$ of the semi-simple Frobenius-Hurwitz manifold $\mathcal{H}_1(0,\dots,0)$ takes the following form
\begin{equation}\label{case0}
\begin{split}
\mathbf{F}_{\phi}&=\frac{u^2}{4{\rm i}\pi}\tau+u\sum_{j=1}^{m}x^js^j+\frac{1}{2}G_2(\tau)\bigg(\sum_{j=1}^{m}x^js^j\bigg)^2+\frac{1}{2}\sum_{i,j=1, i\neq j}^mx^ix^j\K(s^i-s^j|\tau)\\
&\quad -\bigg(\sum_{i=1}^mx^i\bigg)\sum_{j=1}^mx^j\K(s^j|\tau)+\frac{1}{2}\sum_{j=1}^{m}x^j\log(x^j)+\frac{1}{2}\bigg(\sum_{j=1}^mx^j\bigg)^2\log\bigg(\sum_{i=1}^mx^i\bigg)\\
&\quad +\log(-1)\bigg(\sum_{j=1}^mx^j\bigg)^2+\frac{\log(-1)}{4}\sum_{i,j=1, i\neq j}^mx^ix^j-\frac{3}{4}\sum_{i,j=1}^m\big(1+\delta_{ij}\big)x^ix^j,
\end{split}
\end{equation}
where $\K(v|\tau)=\log\sigma(v|\tau)$ and $\sigma$ is the Weierstrass function (\ref{sigma-function}). Moreover, the unit and Euler vector fields (\ref{E-phi} are
$$
e=\partial_{u},\quad \quad E=u\partial_{u}+\sum_{j=1}^mx^j\partial_{x^j}.
$$
\end{Prop}
\emph{Proof:} The result follows from Theorem \ref{Main result} and the relations
$$
\K(-v)=\log(-1)+\K(v),\quad \quad  x^0=-\big(x^1+\dots+x^m\big).
$$

\fd

\begin{Remark}
The WDVV solution (\ref{case0}) was obtained in \cite{Rejeb23} and written in terms of the Jacobi $\theta_1$-function. The case $m=1$ was first  worked out in \cite{Dubrovin2009, Miguel} using the Dubrovin bilinear method.
\end{Remark}

\begin{description}
\item[$\bullet$]\textbf{The Hurwitz-Frobenius manifold $\mathcal{H}_1(1,\dots,1)$}
\end{description}
Let us now deal with the WDVV solution $\mathbf{F}_{\phi}$ associated with the Hurwitz-Frobenius manifold parameterized by  elliptic functions of the following form:
\begin{align*}
\lambda(P)&=c+2(1-\delta_{m,0})\sum_{j=1}^mt^j\Big(\zeta(z_P-\infty^j)-\zeta(z_P-\infty^0)\Big)+\sum_{j=0}^m(x^j)^2\wp\big(z_P-\infty^j\big).
\end{align*}
This is nothing but the case where $n_j=1$ and
\begin{equation}\label{notation case1}
\begin{split}
&x_2(j):=\frac{1}{2}\underset{\infty^j}{{\rm res}\ }\lambda(P)\phi(P)=t^j;\\
&x_1(j):=\underset{\infty^j}{{\rm res}\ }\sqrt{\lambda(P)}\phi(P)=x^j,
\end{split}
\quad \quad\quad \forall\ j=0,\dots,m
\end{equation}
of formula (\ref{lambda-p}). The list of flat coordinates (\ref{FC-phi2}) reduces to
\begin{equation}\label{11-variables}
\Big\{u, \tau, s^1=\infty^1-\infty^0,\dots,s^m=\infty^m-\infty^0,  t^1,\dots,t^m, x^0,x^1,\dots,x^m\Big\}.
\end{equation}

\begin{Prop} Let $m\geq 0$. Then the prepotential of the $(3m+3)$-dimensional  semi-simple Frobenius-Hurwitz manifold $\mathcal{H}_1(1,\dots,1)$ is the function of variables (\ref{11-variables}) given  by
\begin{equation}\label{case1}
\begin{split}
&\mathbf{F}_{\phi}=\frac{u^2}{4{\rm i}\pi}\tau+2u\sum_{j=1}^{m}t^js^j+u\sum_{i=0}^{m}(x^i)^2+2G_2(\tau)\bigg(\sum_{j=1}^{m}t^js^j\bigg)
\bigg(\sum_{j=1}^{m}t^js^j+\sum_{i=0}^{m}(x^i)^2\bigg)\\
&+\frac{1}{2}G_2(\tau)\bigg(\sum_{j=0}^{m}(x^j)^2\bigg)^2+2\sum_{i,j=1, i\neq j}^mt^it^j\K(s^i-s^j|\tau)+2\sum_{i,j=1, i\neq j}^m(x^i)^2t^j\K'(s^i-s^j|\tau)\\
&-\frac{1}{2}\sum_{i,j=1, i\neq j}^{m}(x^ix^j)^2\K''(s^i-s^j|\tau)-4\bigg(\sum_{i=1}^mt^i\bigg)\bigg(\sum_{j=1}^mt^j\K(s^j|\tau)\bigg)
-2\bigg(\sum_{i=1}^mt^i\bigg)\bigg(\sum_{j=1}^{m}(x^j)^2\K'(s^j|\tau)\bigg)\\
&-2(x^0)^2\sum_{j=1}^mt^j\K'(s^j|\tau)-(x^0)^2\sum_{j=1}^{m}(x^j)^2\K''(s^j|\tau)+2\sum_{j=1}^{m}(t^j)^2\log(x^j)+2\bigg(\sum_{j=1}^mt^j\bigg)^2\log(x^0)\\
&+2\log(-1)\bigg(\sum_{j=1}^mt^j\bigg)^2+\log(-1)\sum_{i,j=1, i\neq j}^mt^it^j-3\bigg(\sum_{i=1}^mt^i\bigg)^2-4\sum_{j=1}^m(t^j)^2.
\end{split}
\end{equation}
The function $\mathbf{F}_{\phi}$ satisfies the WDVV equations (\ref{WDVV}), where the distinguished variable corresponding to the constant invertible matrix  is $u$ and $\partial_u$ plays the role of the unit vector field. Furthermore, we have $E.\mathbf{F}_{\phi}=2\mathbf{F}_{\phi}$, with $E$ being the following Euler vector field
$$
E=u\partial_{u}+\sum_{j=1}^mt^j\partial_{t^j}+\frac{1}{2}\sum_{j=0}^mx^j\partial_{x^j}.
$$
\end{Prop}
\emph{Proof:} Using Theorem \ref{Main result} and  the fact that all the poles of the coverings $\lambda$ are  double and taking into account  notation (\ref{notation case1}) as well as   expressions for the functions $\mathcal{R}_{1,1}$ and $\mathcal{R}_{2,2}$ listed in (\ref{R-examples}), we can see that the expression for the prepotential $\mathbf{F}_{\phi}$ can be written as follows:
\begin{align*}
&\mathbf{F}_{\phi}=\frac{u^2}{4{\rm i}\pi}\tau+2u\sum_{j=1}^{m}t^js^j+\frac{5}{6}u\sum_{i=0}^{m}(x^i)^2+2G_2(\tau)\bigg(\sum_{j=1}^{m}t^js^j\bigg)^2
+\frac{5}{3}G_2(\tau)\bigg(\sum_{j=1}^{m}t^js^j\bigg)\bigg(\sum_{i=0}^{m}(x^i)^2\bigg)\\
&+\frac{1}{3}G_2(\tau)\sum_{i,j=0}^{m}\big(x^ix^j\big)^2 +2\sum_{i,j=1, i\neq j}^mt^it^j\K(s^i-s^j)+\sum_{i,j=1, i\neq j}^m(x^i)^2t^j\K'(s^i-s^j)
-2\sum_{i,j=1}^mt^it^j\K(s^i)\\
&-\sum_{i,j=1}^m(x^i)^2t^j\K'(s^i)-2\bigg(\sum_{i=1}^mt^i\bigg)\bigg(\sum_{j=1}^mt^j\K(s^j)\bigg)
-(x^0)^2\sum_{j=1}^mt^j\K'(s^j)+\frac{2}{3}\sum_{i,j=1, i\neq j}^{m}(x^i)^2t^j\K'(s^i-s^j)\\
&-\frac{1}{3}\sum_{i,j=1, i\neq j}^{m}(x^ix^j)^2\K''(s^i-s^j)
-\frac{2}{3}\bigg(\sum_{i=1}^mt^i\bigg)\bigg(\sum_{i=1}^{m}(x^i)^2\K'(s^i)\bigg)-\frac{2}{3}(x^0)^2\sum_{j=1}^{m}(x^j)^2\K''(s^j)-\frac{2}{3}(x^0)^2\sum_{j=1}^{m}t^j\K'(s^j)\\
&+\frac{1}{3}\sum_{j=1}^{m}x^jx_3(j)+\frac{1}{3}x^0x_3(0)+2\sum_{j=1}^{m}(t^j)^2\log(x^j)+2\bigg(\sum_{j=1}^mt^j\bigg)^2\log(x^0)\\
&+2\log(-1)\bigg(\sum_{j=1}^mt^j\bigg)^2+\log(-1)\sum_{i,j=1, i\neq j}^mt^it^j-\frac{7}{2}\sum_{i,j=1}^m\big(1+\delta_{ij}\big)t^it^j.
\end{align*}
On the other hand, because of (\ref{x-alpha2}) we have
\begin{align*}
x_3(j)&=-\frac{1}{2}\big(x_1(j)\big)^{-1}\big(x_2(j)\big)^2+x_2(j)\mathcal{R}_{1,1}(j)+\frac{1}{2}x_1(j)\mathbf{f}_{0}(j)\\
&=-\frac{3}{2}\big(x_1(j)\big)^{-1}\big(x_2(j)\big)^2+\frac{1}{2}x_1(j)c+x_1(j)\sum_{i=0,i\neq j}^mx_2(i)\zeta(\infty^j-\infty^i)
-\frac{1}{2}x_1(j)\sum_{i=0,i\neq j}^m\big(x_1(i)\big)^2\zeta'(\infty^j-\infty^i)\\
&=-\frac{3}{2}\frac{(t^j)^2}{x^j}+\frac{1}{2}x^j\left(u+4\zeta(1/2)\sum_{i=1}^mt^is^i+2\zeta(1/2)\sum_{i=0}^m(x^i)^2\right)+x^j\sum_{i=1,i\neq j}^mt^i\K'(s^j-s^i)\\
&\quad -x^j\bigg(\sum_{i=1}^mt^i\bigg)\K'(s^j)-\frac{1}{2}x^j\sum_{i=1,i\neq j}^m(x^i)^2\K''(s^j-s^i)-\frac{1}{2}x^j(x^0)^2\K''(s^j),
\end{align*}
with $j=1,\dots,m$ and
\begin{align*}
x_3(0)&=-\frac{1}{2}\big(x_1(0)\big)^{-1}\big(x_2(0)\big)^2+x_2(0)\mathcal{R}_{1,1}(0)+\frac{1}{2}x_1(0)\mathbf{f}_{0}(0)\\
&=-\frac{3}{2}\big(x_1(0)\big)^{-1}\big(x_2(0)\big)^2+\frac{1}{2}x_1(0)c+x_1(0)\sum_{i=1}^mx_2(i)\zeta(\infty^0-\infty^i)
-\frac{1}{2}x_1(0)\sum_{i=1}^m\big(x_1(i)\big)^2\zeta'(\infty^0-\infty^i)\\
&=\frac{3}{2x^0}\bigg(\sum_{i=1}^mt^i\bigg)^2+\frac{x^0}{2}\left(u+4\zeta(1/2)\sum_{i=1}^mt^is^i+2\zeta(1/2)\sum_{i=0}^m(x^i)^2\right)
-x^0\sum_{i=1}^mt^i\K'(s^i)-\frac{1}{2}x^0\sum_{i=1}^m(x^i)^2\K''(s^i).
\end{align*}
Therefore, from equality (\ref{G2-zeta}) relating $\zeta(1/2)$ and the Eisenstein series $G_2$,  we can write
\begin{align*}
&\frac{1}{3}\sum_{j=1}^{m}x^jx_3(j)+\frac{1}{3}x^0x_3(0)\\
&=\frac{u}{6}\sum_{j=0}^{m}(x^j)^2+\frac{1}{3}G_2(\tau)\bigg(\sum_{j=0}^{m}(x^j)^2\bigg)\bigg(\sum_{i=1}^mt^is^i\bigg)
+\frac{1}{6}G_2(\tau)\bigg(\sum_{j=0}^{m}(x^j)^2\bigg)^2-\frac{1}{3}\sum_{i,j=1,i\neq j}^m(x^j)^2t^i\K'(s^i-s^j)\\
&\quad -\frac{1}{3}\bigg(\sum_{i=1}^mt^i\bigg)\sum_{j=1}^{m}(x^j)^2\K'(s^j)
-\frac{1}{3}(x^0)^2\sum_{i=1}^mt^i\K'(s^i)-\frac{1}{6}\sum_{i,j=1,i\neq j}^m(x^ix^j)^2\K''(s^i-s^j)\\
&\quad -\frac{1}{3}(x^0)^2\sum_{j=1}^m(x^j)^2\K''(s^j)+\frac{1}{2}\bigg(\sum_{i=1}^mt^i\bigg)^2-\frac{1}{2}\sum_{j=1}^m(t^j)^2.
\end{align*}
Substituting this into the fifth line of the above expression for $\mathbf{F}_{\phi}$,  we arrive at the desired explicit form (\ref{case1}).

\fd

\begin{description}
\item[$\bullet$]\textbf{The Hurwitz-Frobenius manifold $\mathcal{H}_1(n)$, with $n\geq 1$.}
\end{description}
Let us assume that the Hurwitz space is of type $\mathcal{H}_1(n)$, with $n\geq 1$. Let $[(\C/{\mathbb{L}}, \lambda)]$ be a point of $\mathcal{H}_1(n)$.
This means that the meromorphic function  $\lambda$ has a unique pole $\infty^0$ of order $n+1\geq 2$. Without loss of generality, we may assume that $z_{\infty^0}=0$.
Thus, the LG superpotential formula  (\ref{lambda-p}) and expression (\ref{u-a}) for $u:=\oint_a\lambda(P)\phi(P)$ imply that  $\lambda$ is given by
\begin{equation}\label{lambda-Hn}
\begin{split}
\lambda(P)&=c+(n+1)\sum_{k=2}^{n+1}\frac{(-1)^k}{k!}\mathcal{B}_{n+1,k}(x_1,\dots,x_{n-k+2})\wp^{(k-2)}\big(z_P\big)\\
&=u+\frac{n+1}{2}G_2(\tau)\sum_{k=1}^nx_kx_{n+1-k}\\
&\quad +(n+1)\sum_{k=2}^{n+1}\frac{(-1)^k}{k!}\mathcal{B}_{n+1,k}(x_1,\dots,x_{n-k+2})\wp^{(k-2)}\big(z_P\big),
\end{split}
\end{equation}
where, as in (\ref{phi-Taylor}), the coefficients $(x_r)_{r\geq 1}$ are defined by the following Taylor series of the holomorphic differential $\phi$ near the pole $z_P=0$ of $\lambda$:
$$
\phi(P)\underset{P\sim \infty^0}{=}\bigg(\sum_{r=1}^{\infty}rx_r\z(P)^{r-1}\bigg)d\z(P),\quad \quad \text{with}\quad \z(P):=\lambda(P)^{-\frac{1}{n+1}}.
$$
Note that
\begin{equation}\label{f-H(n)}
x_{n+1}:=\frac{1}{n+1}\underset{z_P=0}{{\rm res}\ }\lambda(P)dz_P=0
\end{equation}
and $\big\{x_{n+1+\alpha}\big\}_{\alpha\geq 1}$ are functions of the flat coordinates $\big\{ u,x_1,\dots,x_n,\tau\big\}$ of the flat metric $\eta(\phi)$ (\ref{eta-def}), given by  (\ref{x-alpha2}).\\
The general WDVV solution stated in Theorem \ref{Main result} amounts to the following result.
\begin{Prop}\label{Jacobi group} For any positive integer $n$, the WDVV solution associated with the $(n+2)$-dimensional $\phi$-Frobenius manifold structure on $\mathcal{H}_1(n)$ and denoted by $\mathbf{F}_{\phi,n}$ takes the following form:
\begin{equation}\label{Jacobi-An}
\begin{split}
\mathbf{F}_{\phi,n}&=\mathbf{F}_{\phi,n}(u,x_1,\dots,x_n,\tau)=\frac{u^2}{4{\rm i}\pi}\tau
+\frac{u}{2}\sum_{\alpha=1}^{n}\frac{(n+1-\alpha)(2n+2+\alpha)}{(n+1+\alpha)}x_{\alpha}x_{n+1-\alpha}\\
&\quad +\frac{n+1}{2}G_2(\tau)\sum_{\alpha=1}^{n}\sum_{\beta=1}^{n}\frac{(n+1-\alpha)(n+1-\beta)}{(n+1+\alpha)}x_{\alpha}x_{\beta}x_{n+1-\alpha}x_{n+1-\beta}\\
&\quad +\frac{(n+1)^2}{2}\sum_{\alpha=1}^{n}\sum_{k=2}^{n+1}\frac{(n+1-\alpha)}{k(n+1+\alpha)}x_{n+1-\alpha}\mathcal{B}_{n+1,k}\mathcal{R}_{\alpha+k-1,k}\\
&\quad +\frac{(n+1)^2}{2}\sum_{\alpha=1}^{n}\sum_{k=2, k+\alpha\geq 5}^{n+1}\Bigg\{
\left[(-1)^k\frac{(n+1-\alpha)}{k(n+1+\alpha)}x_{n+1-\alpha}\mathcal{B}_{n+1,k}\right]\\
&\quad \quad \times \bigg[\sum_{\ell=1, k+\ell\geq 5}^{\alpha}\binom{k+\ell-1}{\ell}\mathcal{B}_{\alpha,\ell}\frac{G_{k+\ell-1}(\tau)}{k+\ell-1}\bigg]\Bigg\},
\end{split}
\end{equation}
where $\mathcal{B}_{n+1,k}=\mathcal{B}_{n+1,k}\big(x_1,\dots,x_{n+2-k}\big)$ and $\mathcal{R}_{\alpha+k-1,k}=\mathcal{R}_{\alpha+k-1,k}\big(x_1,\dots,x_{\alpha+k}\big)$
are respectively the partial Bell polynomial (\ref{Bell-partial}) and the rational function (\ref{Bell-rec4}). The prepotential $\mathbf{F}_{\phi,n}$ (\ref{Jacobi-An}) is a quasi-homogeneous function of degree 2 with respect to the Euler vector field
$$
E=u\partial_u+\frac{1}{n+1}\sum_{k=1}^nkx_k\partial_{x_k}
$$
\end{Prop}

\medskip

We now provide a very compact form of the WDVV solution (\ref{Jacobi-An}) in the special cases where $n=1,2,3$.
The proof below  highlights the practicality of the aforementioned formulas, including those for $x_{n+1+\alpha}$ (\ref{x-alpha2}) and $\mathbf{f}_{\ell}$ (\ref{lambda-Laurent})-(\ref{f-function}) which appear respectively in the Laurent series of the differential $\phi$ and the elliptic function $\lambda$ (\ref{lambda-Hn}) near the pole $0$.
\begin{Cor}\label{A123} We have the following WDVV solutions
\begin{align}
\begin{split}\label{Jacobi-A1}
\mathbf{F}_{\phi,1}&=\frac{u^2}{4{\rm i}\pi}\tau+ux_1^2+\frac{x_1^4}{2}G_2;
\end{split}\\
\begin{split}\label{Jacobi-A2}
\mathbf{F}_{\phi,2}&=\frac{u^2}{4{\rm i}\pi}\tau+3ux_1x_2-\frac{3}{4}\frac{x_2^4}{x_1^2}+\frac{9}{2}x_1^2x_2^2G_2-\frac{3}{4}x_1^6G_4;
\end{split}\\
\begin{split}\label{Jacobi-A3}
\mathbf{F}_{\phi,3}&=\frac{u^2}{4{\rm i}\pi}\tau+u\Big(4x_1x_3+2x_2^2\Big)+\frac{8}{3}\frac{x_3^3}{x_1}-8\frac{x_2^2x_3^2}{x_1^2}+4\frac{x_2^4x_3}{x_1^3}
-\frac{2}{3}\frac{x_2^6}{x_1^4}\\
&\quad +\Big(8x_1^2x_3^2+8x_1x_2^2x_3+2x_2^4\Big)G_2+\Big(4x_1^5x_3-10x_1^4x_2^2\Big)G_4+\frac{5}{3}x_1^8G_6.
\end{split}
\end{align}

\end{Cor}
\emph{Proof:}  The WDVV solution (\ref{Jacobi-A1}) was first  obtained by Dubrovin  in \cite{Dubrovin2D} and reworked out in \cite{Bertola2,Rejeb23}. Here we are going to derive  it again using  (\ref{Jacobi-An}). Indeed, the particular case $n=1$ of (\ref{Jacobi-An})  reads
\begin{align*}
\mathbf{F}_{\phi,1}&=\frac{u^2}{4{\rm i}\pi}\tau+\frac{5}{6}ux_1^2+\frac{1}{3}x_1^4G_2+\frac{1}{3}x_1^3\mathcal{R}_{2,2}(x_1,x_2,x_3)\\
&=\frac{u^2}{4{\rm i}\pi}\tau+\frac{5}{6}ux_1^2+\frac{1}{3}G_2x_1^4+\frac{1}{3}x_1x_3.
\end{align*}
In addition, from (\ref{x-alpha2}) and (\ref{u-a}) it follows that
$$
x_3=\frac{x_1}{2}c=\frac{x_1}{2}\Big(u+x_1^2G_2\Big)
$$
and thus (\ref{Jacobi-A1}) holds. Note that if we consider  the case $m=0$ of (\ref{case1}), then we also arrive at (\ref{Jacobi-A1}). \\
Let us now deal with the case $n=2$.  Applying (\ref{Jacobi-An}) with $n=2$ and using (\ref{R-examples}) and the fact that $x_3=0$, we obtain
\begin{align*}
\mathbf{F}_{\phi,2}
&=\frac{u^2}{4{\rm i}\pi}\tau+\frac{51}{20}ux_1x_2+\frac{63}{20}x_1^2x_2^2G_2+\frac{9}{4}x_1x_2^2\mathcal{R}_{2,2}(x_1,x_2,x_3)+\frac{3}{4}x_2x_1^3
\mathcal{R}_{3,3}(x_1,x_2,x_3,x_4)\\
&\quad +\frac{9}{10}x_1^2x_2\mathcal{R}_{3,2}(x_1,x_2,x_3,x_4)+\frac{3}{10}x_1^4\mathcal{R}_{4,3}(x_1,\dots,x_{5})-\frac{9}{20}x_1^6G_4\\
&=\frac{u^2}{4{\rm i}\pi}\tau+\frac{51}{20}ux_1x_2+\frac{63}{20}x_1^2x_2^2G_2-\frac{3}{5}\frac{x_2^4}{x_1^2}
+\frac{3}{4}x_2x_4+\frac{3}{10}x_1x_5-\frac{9}{20}x_1^6G_4.
\end{align*}
Moreover, formula (\ref{x-alpha2}) and (\ref{u-a}) imply that
\begin{align*}
x_{4}&=\frac{1}{3}\sum_{k=1}^{2}\frac{1}{k+1}\binom{-3}{k}x_1^{-k}\mathcal{B}_{3,k+1}(x_2,\dots,x_{4-k})
+\sum_{k=1}^{2}\frac{1}{k}\mathcal{B}_{3,k}\mathcal{R}_{k,k}+\frac{1}{3}\mathbf{f}_{0}\mathcal{B}_{1,1}\\
&=-\frac{1}{3}\frac{x_2^3}{x_1^2}+\frac{c}{3}x_1=-\frac{1}{3}\frac{x_2^3}{x_1^2}+\frac{1}{3}ux_1+x_1^2x_2G_2
\end{align*}
and
\begin{align*}
x_5
&=-x_1^{-1}\mathcal{B}_{4,2}(x_2,x_3,x_4)+\frac{4}{3}x_1^{-2}\mathcal{B}_{4,3}(x_2,x_3)-\frac{5}{3}x_1^{-3}\mathcal{B}_{4,4}(x_2)
+\mathcal{B}_{3,2}\mathcal{R}_{3,2}+\frac{2}{3}cx_2+\frac{1}{3}x_1^2\mathbf{f}_{1}\\
&=\frac{1}{3}\frac{x_2^4}{x_1^3}+\frac{2}{3}ux_2+2x_1x_2^2G_2-x_1^5G_4.
\end{align*}
This shows (\ref{Jacobi-A2}). We now move on to prove (\ref{Jacobi-A3}).  We first calculate $x_5,x_6$ and $x_7$. By (\ref{f-H(n)}),  we have
\begin{align*}
&\mathbf{f}_0=c+x_1^4G_4=u+(4x_1x_3+2x_2^2)G_2+x_1^4G_4;\\
&\mathbf{f}_1=-4\mathcal{B}_{4,3}G_4=-12x_1^2x_2G_4;\\
&\mathbf{f}_2=6(2x_1x_3+x_2^2)G_4+10x_1^4G_6.
\end{align*}
Moreover, since $x_4=0$ and for all $\alpha\geq 1$
\begin{align*}
x_{\alpha+4}&=\frac{\alpha}{4}\sum_{k=1}^{\alpha+2}\frac{1}{k+1}\binom{-4}{k}x_1^{-k}\mathcal{B}_{\alpha+3,k+1}(x_2,\dots,x_{\alpha-k+4}) +\alpha\sum_{k=2}^{3}\frac{1}{k}\mathcal{B}_{4,k}\mathcal{R}_{\alpha+k-1,k}
+\frac{\alpha}{4}\sum_{\ell=0}^{\alpha-1}\frac{1}{\ell+1}\mathbf{f}_{\ell}\mathcal{B}_{\alpha,\ell+1},
\end{align*}
by (\ref{x-alpha2}), we can deduce after some computations based on (\ref{R-examples}) that
\begin{align*}
x_5&=\frac{1}{2}\frac{x_3^2}{x_1}-\frac{x_2^2x_3}{x_1^2}+\frac{1}{4}\frac{x_2^4}{x_1^3}
+\frac{x_1}{4}u+\Big(x_1^2x_3+\frac{1}{2}x_1x_2^2\Big)G_2+\frac{x_1^5}{4}G_4;\\
x_6&=-2\frac{x_2x_3^2}{x_1^2}+2\frac{x_2^3x_3}{x_1^3}-\frac{1}{2}\frac{x_2^5}{x_1^4}+\frac{x_2}{2}u
+(2x_1x_2x_3+x_2^3)G_2-\frac{5}{2}x_1^4x_2G_4;\\
x_7&=-\frac{1}{2}\frac{x_3^3}{x_1^2}+3\frac{x_2^2x_3^2}{x_1^3}-\frac{9}{4}\frac{x_2^4x_3}{x_1^4}+\frac{1}{2}\frac{x_2^6}{x_1^5} +\frac{3}{4}x_3u +\Big(3x_1x_3^2+\frac{3}{2}x_2^2x_3\Big)G_2\\
&\quad+\Big(\frac{15}{4}x_1^4x_3-\frac{15}{2}x_1^3x_2^2\Big)G_4+\frac{5}{2}x_1^7G_6.
\end{align*}
Therefore
\begin{align*}
\mathbf{F}_{\phi,3}&=\frac{u^2}{4{\rm i}\pi}\tau+\frac{u}{2}\sum_{\alpha=1}^{3}\frac{(4-\alpha)(8+\alpha)}{(4+\alpha)}x_{\alpha}x_{4-\alpha}
+2G_2\sum_{\alpha=1}^{3}\sum_{\beta=1}^{3}\frac{(4-\alpha)(4-\beta)}{(4+\alpha)}x_{\alpha}x_{\beta}x_{4-\alpha}x_{4-\beta}\\
&\quad +8\sum_{\alpha=1}^{3}\sum_{k=2}^{4}\frac{(4-\alpha)}{k(4+\alpha)}x_{4-\alpha}\mathcal{B}_{4,k}\mathcal{R}_{\alpha+k-1,k}\\
&\quad +8\sum_{\alpha=1}^{3}\sum_{k=2, k+\alpha\geq 5}^{4}\Bigg\{
\left[(-1)^k\frac{(n+1-\alpha)}{k(4+\alpha)}x_{4-\alpha}\mathcal{B}_{4,k}\right]
\bigg[\sum_{\ell=1, k+\ell\geq 5}^{\alpha}\binom{k+\ell-1}{\ell}\mathcal{B}_{\alpha,\ell}\frac{G_{k+\ell-1}}{k+\ell-1}\bigg]\Bigg\}\\
&=\frac{u^2}{4{\rm i}\pi}\tau+u\Big(\frac{122}{35}x_1x_3+\frac{5}{3}x_2^2\Big)+\Big(\frac{208}{35}x_1^2x_3^2+\frac{592}{105}x_1x_2^2x_3+\frac{4}{3}x_2^4\Big)G_2\\
&\quad +\frac{2}{7}x_1x_7+\frac{2}{3}x_2x_6+\frac{6}{5}x_3x_5+\frac{232}{105}\frac{x_3^3}{x_1}-\frac{664}{105}\frac{x_2^2x_3^2}{x_1^2}
+\frac{316}{105}\frac{x_2^4x_3}{x_1^3}-\frac{10}{21}\frac{x_2^6}{x_1^4}\\
&\quad +\Big(\frac{92}{35}x_1^5x_3-\frac{130}{21}x_1^4x_2^2\Big)G_4+\frac{20}{21}x_1^8G_6
\end{align*}
and this gives (\ref{Jacobi-A3}).

\fd

\begin{Remark}\emph{\textbf{(Jacobi forms for the Jacobi group of type $A_n$)}}
\end{Remark}
It is shown in \cite{Bertola1,Bertola2} that the Hurwitz-Frobenius manifold $\mathcal{H}_1(n)$ determined by the holomorphic differential $\phi$ can be identified with the Frobenius manifold structure on the orbit space of  the Jacobi group $\mathcal{J}(A_n)$ of type $A_n$ acting on the product $\mathcal{O}:=\C\times \C^n\times\mathbb{H}$, with $\mathbb{H}$ being the upper half-plane and $\C^n=\big\{(\xi_1,\dots,\xi_{n+1})\in\C^{n+1}:\quad \xi_1+\dots+\xi_{n+1}=0\big\}$. This correspondence is realized by the map (\ref{lambda-Jacobi}) below and  involves the so-called (weak) Jacobi forms for the Jacobi group $\mathcal{J}(A_n)$ which  are holomorphic functions in  $(\upsilon,\xi_1,\dots,\xi_{n+1},\tau)$, invariant under permutations of $\xi_1,\dots,\xi_{n+1}$ and have precise modular and  quasi-periodicity  properties \cite{Eichler, Wirthmuller, Bertola1}.
The theory of Jacobi forms was first investigated by  Eichler and  Zagier in \cite{Eichler} for the Jacobi group $\mathcal{J}(A_1)$  and generalized to the other cases of root systems by  Wirthm\"{u}ller in \cite{Wirthmuller} and can be viewed as an analogue of polynomials invariant theory for finite Coxeter groups. In particular, there is an analogue of Chevalley's theorem for the Jacobi group $\mathcal{J}(A_n)$ \cite{Wirthmuller} stating the following:\\
the algebra of  (weak) Jacobi forms is a polynomial ring (over the graded ring $\C[G_4,G_6]$ of $SL(2,\Z)$-modular forms) generated by $n+1$ algebraically independent Jacobi forms $\varphi_0,\varphi_2,\dots,\varphi_{n+1}$ of weight $0,-1,\dots,-(n+1)$, respectively. Here the weight $-k$ is an integer that determines the transformation properties of the Jacobi form under modular transformations, similar to the weight of modular forms  $G_{2k}$ (\ref{Eisenstein G})-(\ref{G-modular})  (see \cite{Wirthmuller, Bertola1} for a precise definition). \\
A set  of generators of the algebra of Jacobi forms  for the Jacobi group $\mathcal{J}(A_n)$ was obtained by Bertola in (\cite{Bertola1}, Theorem 1.4) using the following generating function:
\begin{equation}\label{lambda-Jacobi}
\begin{split}
\big(\varphi_0,\varphi_2,\dots,\varphi_{n+1},\tau)\in \mathcal{O}/\mathcal{J}(A_n)\longmapsto\lambda^{A_n}(P)
&=e^{2{\rm i}\pi\upsilon}\prod_{j=1}^{n+1}\frac{\theta_1(\mathfrak{z}_j-z_P|\tau)}{\theta_1(z_P|\tau)}\Bigg|_{\mathfrak{z}_1+\dots+\mathfrak{z}_{n+1}=0}\\
&=\varphi_0+\sum_{k=2}^{n+1}\frac{(-1)^k}{(k-1)!}\varphi_k\wp^{(k-2)}(z_P|\tau),
\end{split}
\end{equation}
where $\mathfrak{z}_1,\mathfrak{z}_2,\dots,\mathfrak{z}_{n+1}$ are the distinct zeros of the superpotential $\lambda^{A_n}$ and $\upsilon=\displaystyle \oint_a\log\big(\lambda(P)\big)\phi(P)$ and $\wp$ and $\theta_1$ are respectively the Weierstrass and Jacobi functions. The proof of the second equality in (\ref{lambda-Jacobi}) is based on   formula (5.102) in \cite{Dubrovin2D} together with the following expression for the lightest generator $\varphi_{n+1}$ constructed by Wirthm\"{u}ller \cite{Wirthmuller}:
\begin{equation*}\label{generator-n}
\varphi_{n+1}=e^{2{\rm i}\pi\upsilon}\prod_{j=1}^{n+1}\frac{\theta_1(\mathfrak{z}_j|\tau)}{\theta_1'(0|\tau)}\Bigg|_{\mathfrak{z}_1+\dots+\mathfrak{z}_{n+1}=0}.
\end{equation*}
Furthermore, in (\cite{Bertola2}, Corollary 1.1), Bertola  found   explicit polynomial expressions for the Jacobi forms  $\varphi_0$, $\varphi_2$, $\varphi_{n+1}$  by means of flat coordinates of the metric $\eta(\phi)$ (\ref{eta-def}) induced by the differential $\phi=dz_P$.\\
As an improvement of this result,  the upcoming proposition provides a complete and simple  description of the generators of the algebra of $\mathcal{J}(A_n)$-Jacobi forms in terms of coordinates $\big\{u,x_1,\dots,x_n,\tau\big\}$ related to $\phi$. This follows from (\ref{lambda-Jacobi}) and formula (\ref{lambda-Hn}) constructing LG superpotentials of the Hurwitz space $\mathcal{H}_1(n)$.
\begin{Prop}  Let $\big\{u,x_1,\dots,x_n,\tau\big\}$ be the flat coordinates on the Frobenius manifold $\mathcal{H}_1(n)$  associated with the holomorphic differential $dz_P$ determined by
$$
u:=\oint_a\lambda(P)dz_P,\quad x_{\alpha}=\frac{1}{\alpha}\underset{z_P=0}{\rm res}\lambda(P)^{\frac{\alpha}{n+1}}dz_P,\quad \quad \tau=\oint_bdz_P.
$$
Then the fundamental Jacobi forms $\varphi_0,\varphi_2,\dots,\varphi_{n+1}$  with respect to the Jacobi group $\mathcal{J}(A_n)$ defined by (\ref{lambda-Jacobi}) satisfy
\begin{equation}\label{Jacobi forms1}
\begin{split}
&\varphi_0=u+\frac{n+1}{2}\mathcal{B}_{n+1,2}(x_1,\dots,x_n)G_2(\tau);\\
&\varphi_k=\frac{(n+1)}{k}\mathcal{B}_{n+1,k}(x_1,\dots,x_{n+2-k}),\quad \forall\ k=2,\dots,n+1,
\end{split}
\end{equation}
where  $\mathcal{B}_{n+1,k}$ denotes  the homogeneous  Bell polynomial of degree $k$ (\ref{Bell-partial})-(\ref{Bell2}).\\
In particular, the ring of $\mathcal{J}(A_n)$-Jacobi forms is a subring of
$$
\C[G_2,G_4,G_6,u,x_1,\dots,x_n].
$$
\end{Prop}

\medskip

\begin{description}
\item[$\bullet$]\textbf{The Hurwitz-Frobenius manifold $\mathcal{H}_1(n,0)$, with $n\geq 0$.}
\end{description}
As in the preceding case, the $(n+4)$-dimensional Hurwitz-Frobenius manifold $\mathcal{H}_1(n,0)$ induced by the holomorphic primary differential $\phi$ can be identified with  the Frobenius manifold structure on  the orbit space of the $\widetilde{A}_{n+1}$-extended affine Jacobi group. This was recently studied  in \cite{Almeida} by adapting the scheme of the aforementioned $A_n$-Jacobi group case.\\
According to (\ref{lambda-p}) and (\ref{u-a}), the LG superpotentials of the Hurwitz space  $\mathcal{H}_1(n,0)$ are as follows:
\begin{equation}\label{Jacobi-A-superp}
\begin{split}
\lambda(P)
&=u+ysG_2(\tau)+\frac{n+1}{2}\mathcal{B}_{n+1,2}(x_1,\dots,x_n)G_2(\tau)+y\Big(\zeta(z_P-\infty^1)-\zeta(z_P-\infty^0)\Big)\\
&\quad +\big(1-\delta_{n,0}\big)(n+1)\sum_{k=2}^{n+1}\frac{(-1)^k}{k!}\mathcal{B}_{n+1,k}(x_1,\dots,x_{n+2-k})\wp^{(k-2)}\big(z_P-\infty^0\big),
\end{split}
\end{equation}
where $u, y, s, \tau, (x_k)_{k\geq1}$ are the parameters related to the holomorphic  differential $\phi=dz_P$ on the torus $\C/(\Z+\tau\Z)$ and given by
\begin{align*}
&\tau=\oint_b\phi,\quad \quad \quad u=\oint_a\lambda(P)\phi(P),\quad \quad  \quad s=\int_{\infty^0}^{\infty^1}\phi(P)=\infty^1-\infty^0,\\
&x_{k}=\frac{1}{k}\underset{\infty^0} {\rm res}\lambda(P)^{\frac{k}{n+1}}\phi(P),\quad \forall\ k\geq 1\quad \quad \quad \text{and}\quad \quad\quad
 y=\underset{\infty^1} {\rm res}\lambda(P)\phi(P).
\end{align*}
As we know from (\ref{FC-phi2}) and (\ref{Entries eta-phi}), the  variables
$$
\big\{u, y, s, \tau, x_1,\dots,x_n\big\}
$$
give a system of flat coordinates of the flat metric $\eta(\phi)$ defined on  $\mathcal{H}_1(n,0)$ by (\ref{eta-def}).\\
In particular, using results from \cite{Almeida} and reasoning  as in (\ref{Jacobi forms1}), it turns out that formula (\ref{Jacobi-A-superp}) allows us to obtain a  basis of generators of the algebra of extended Jacobi forms for the extended Jacobi group of type $\widetilde{A}_{n+1}$:
\begin{align*}
&\widetilde{\varphi}_0=u+ysG_2(\tau)+\frac{n+1}{2}\mathcal{B}_{n+1,2}(x_1,\dots,x_n)G_2(\tau);\\
&\widetilde{\varphi}_1=-y;\\
&\widetilde{\varphi}_k=\frac{(n+1)}{k}\mathcal{B}_{n+1,k}(x_1,\dots,x_{n+2-k}),\quad \forall\ k=2,\dots,n+1.
\end{align*}
Furthermore, by taking into account the fact that
$$
y=-(n+1)x_{n+1},
$$
we can obtain  from Theorem \ref{Main result} the following  explicit form of WDVV solution associated with the Frobenius manifold structure on  the orbit space of the $\widetilde{A}_{n+1}$-extended affine Jacobi group.
\begin{Prop}\label{Jacobi Affine} Let $n$ be a nonnegative integer and  $\big\{u,s,x_1,\dots,x_n,y,\tau\big\}$ be the variables defined above. Then the $\phi$-WDVV solution corresponding to the Hurwitz-Frobenius manifold $\mathcal{H}_1(n,0)$ is the function $\widetilde{\mathbf{F}}_{\phi,n}=\widetilde{\mathbf{F}}_{\phi,n}(u,s,x_1,\dots,x_n,y,\tau)$ determined  by
\begin{equation}\label{Jacobi-A-An}
\begin{split}
&\widetilde{\mathbf{F}}_{\phi,n}=\frac{u^2}{4{\rm i}\pi}\tau+uys
+\frac{u}{2}\sum_{\alpha=1}^{n}\frac{(n+1-\alpha)(2n+2+\alpha)}{(n+1+\alpha)}x_{\alpha}x_{n+1-\alpha}+\frac{1}{2}(ys)^2G_2(\tau)\\
&+\frac{n+1}{2}ysG_2(\tau)\sum_{\alpha=1}^{n}\left(\frac{n+1-\alpha}{n+1+\alpha}+\frac{n+1-\alpha}{n+1}\right)x_{\alpha}x_{n+1-\alpha}\\
&+\frac{n+1}{2}G_2(\tau)\sum_{\alpha=1}^{n}\sum_{\beta=1}^{n}\frac{(n+1-\alpha)(n+1-\beta)}{(n+1+\alpha)}x_{\alpha}x_{\beta}x_{n+1-\alpha}x_{n+1-\beta}\\
&-y^2\K(s|\tau)+\frac{n+1}{2}y\sum_{k=2}^{n+1}\frac{1}{k!}\mathcal{B}_{n+1,k}\K^{(k-1)}(-s|\tau)
+\frac{n+1}{2}y\sum_{\alpha=1}^{n}\frac{(n+1-\alpha)}{(n+1+\alpha)}x_{n+1-\alpha}\mathbf{L}_{\alpha}^{\phi}\big[\K(\cdot|\tau)\big](-s)\\
&-\frac{n+1}{2}y\sum_{\alpha=1}^{n}\left(\frac{n+1-\alpha}{n+1+\alpha}+\frac{n+1-\alpha}{n+1}\right)x_{n+1-\alpha}\mathcal{R}_{\alpha,1}
+\frac{(n+1)^2}{2}\sum_{\alpha=1}^{n}\sum_{k=2}^{n+1}\frac{(n+1-\alpha)}{k(n+1+\alpha)}x_{n+1-\alpha}\mathcal{B}_{n+1,k}\mathcal{R}_{\alpha+k-1,k}\\
&+\frac{(n+1)^2}{2}\sum_{\alpha=1}^{n}\sum_{k=1, k+\alpha\geq 5}^{n+1}\Bigg\{
\left[(-1)^k\frac{(n+1-\alpha)}{k(n+1+\alpha)}x_{n+1-\alpha}\mathcal{B}_{n+1,k}\right]
\bigg[\sum_{\ell=1, k+\ell\geq 5}^{\alpha}\binom{k+\ell-1}{\ell}\mathcal{B}_{\alpha,\ell}\frac{G_{k+\ell-1}(\tau)}{k+\ell-1}\bigg]\Bigg\}\\
&+\frac{n+1}{2}y\sum_{\ell=4}^{n}\frac{1}{\ell(\ell+1)}\mathcal{B}_{n+1,\ell+1}G_{\ell}(\tau)+\frac{1}{2}y^2\log(x_1y)
+\frac{\log(-1)}{2}y^2-\frac{3}{4}\frac{n+2}{n+1}y^2.
\end{split}
\end{equation}
Here, as in Theorem \ref{Main result}, $\K:=\log\sigma$, $\mathbf{L}_{\alpha}^{\phi}\big[\K\big]:=\sum_{k=1}^{\alpha}\frac{1}{k!}\mathcal{B}_{\alpha,k}\K^{(k)}$ and $\mathcal{B}_{\mu,\nu}$ and $\mathcal{R}_{\alpha+k-1,k}$ are  the functions of  the variables $(x_k)_k$ defined by (\ref{Bell-partial}) and   (\ref{Bell-rec4}), respectively.
\end{Prop}

Using expressions (\ref{Jacobi-An}) and (\ref{Jacobi-A-An}) for the $\phi$-WDVV solutions corresponding to the Hurwitz-Frobenius manifolds $\mathcal{H}_1(n)$ and $\mathcal{H}_1(n,0)$, we deduce that they are linked by the following limiting behavior.
\begin{Cor} Let $n$ be a positive integer. With the notation of Propositions \ref{Jacobi group} and \ref{Jacobi Affine}, the WDVV solutions $\mathbf{F}_{\phi,n}$ and $\widetilde{\mathbf{F}}_{\phi,n}$ given respectively by (\ref{Jacobi-An}) and (\ref{Jacobi-A-An}) satisfy
\begin{equation}\label{Jacobi link}
\lim_{y\to0}\widetilde{\mathbf{F}}_{\phi,n}(u,s,x_1,\dots,x_n,y,\tau)=\mathbf{F}_{\phi,n}(u,x_1,\dots,x_n,\tau).
\end{equation}
\end{Cor}
Note that when the considered Hurwitz space is of type  $\mathcal{H}_1(n,0)$, formulas (\ref{f-function}) and (\ref{x-alpha2}) read respectively as
\begin{align*}
\mathbf{f}_{\ell}&:=\delta_{\ell,0}\Big(u+ysG_2+\frac{n+1}{2}\mathcal{B}_{n+1,2}\big(x_1,\dots,x_n\big)G_2\Big)-\frac{(-1)^{\ell}}{\ell!}y\K^{(\ell+1)}(s)\\
&\quad +(n+1)\sum_{k=1, k+\ell\geq 4}^{n+1}\frac{(-1)^k}{k+\ell}\binom{k+\ell}{\ell}\mathcal{B}_{n+1,k}\big(x_1,\dots,x_{n+2-k}\big)G_{k+\ell}
\end{align*}
and
\begin{align*}
x_{n+1+\alpha}&=\frac{\alpha}{n+1}\sum_{k=1}^{n+\alpha-1}\frac{1}{k+1}\binom{-n-1}{k}x_1^{-k}\mathcal{B}_{n+\alpha,k+1}\big(x_2,\dots,x_{n+\alpha-k+1}\big)\\
&\quad +\alpha\sum_{k=1}^{n}\frac{1}{k}\mathcal{B}_{n+1,k}\big(x_1,\dots,x_{n+2-k}\big)\mathcal{R}_{\alpha+k-1,k}\big(x_1,\dots,x_{\alpha+k}\big)
+\frac{\alpha}{n+1}\sum_{\ell=0}^{\alpha-1}\frac{\mathbf{f}_{\ell}}{\ell+1}\mathcal{B}_{\alpha,\ell+1}\big(x_1,\dots,x_{\alpha-\ell}\big).
\end{align*}
By utilizing  this  and applying formula (\ref{Jacobi-A-An}) while adopting the scheme of the proof of Corollary \ref{A123},
we derive, after carrying  out some simplification, the following examples of $\phi$-WDVV solutions associated with the low-dimensional Frobenius manifolds $\mathcal{H}_1(n,0)$, $n=0,1,2,3$.
\begin{Cor}\label{AA123} We have the following  special cases $n=0,1,2,3$ of the WDVV solution (\ref{Jacobi-A-An}):
\begin{equation}\label{Jacobi-A-A0}
\widetilde{\mathbf{F}}_{\phi,0}=\frac{u^2}{4{\rm i}\pi}\tau+uys+y^2\log(y)+\frac{1}{2}(ys)^2G_2(\tau)-y^2\K(s|\tau)+\big(\log(-1)-3/2\big)y^2;
\end{equation}
\begin{equation}\label{Jacobi-A-A1}
\begin{split}
\widetilde{\mathbf{F}}_{\phi,1}&=\frac{u^2}{4{\rm i}\pi}\tau+u(ys+x^2)+\frac{1}{2}y^2\log(xy)+\frac{1}{2}\big(y^2s^2+2x^2ys+x^4\big)G_2(\tau)-y^2\K(s|\tau)\\
&\quad -x^2y\K'(s|\tau)+\frac{\log(-1)}{2}y^2-\frac{7}{4}y^2;
\end{split}
\end{equation}
\begin{equation}\label{Jacobi-A-A2}
\begin{split}
\widetilde{\mathbf{F}}_{\phi,2}&=\frac{u^2}{4{\rm i}\pi}\tau+u(ys+3x_1x_2)+2\frac{x_2^2y}{x_1}-\frac{3}{4}\frac{x_2^4}{x_1^2}+\frac{1}{2}y^2\log(x_1y)
+\frac{1}{2}\Big(y^2s^2+6x_1x_2ys+9x_1^2x_2^2\Big)G_2(\tau)\\
&\quad -\frac{3}{4}x_1^6G_4-y^2\K(s|\tau)-3x_1x_2y\K'(s|\tau)+\frac{1}{2}x_1^3y\K''(s|\tau) +\frac{\log(-1)}{2}y^2-\frac{3}{4}y^2;
\end{split}
\end{equation}
\begin{equation}\label{Jacobi-A-A3}
\begin{split}
\widetilde{\mathbf{F}}_{\phi,3}&=\frac{u^2}{4{\rm i}\pi}\tau+uys+4ux_1x_3+2ux_2^2+2\frac{x_2x_3y}{x_1}+\frac{2}{3}\frac{x_2^3y}{x_1^2}
+\frac{8}{3}\frac{x_3^3}{x_1}-8\frac{x_2^2x_3^2}{x_1^2}+8\frac{x_2^4x_3}{x_1^3}-\frac{2}{3}\frac{x_2^6}{x_1^4}\\
&\quad +\frac{1}{2}y^2\log(x_1y)+\Big(\frac{1}{2}y^2s^2+4x_1x_3ys+2x_2^2ys+8x_1^2x_3^2+8x_1x_2^2x_3+2x_2^4\Big)G_2(\tau)\\
&\quad +\Big(4x_1^5x_3-10x_1^4x_2^2\Big)G_4(\tau)+\frac{5}{3}x_1^8G_6(\tau) -y^2\K(s|\tau)-\Big(4x_1x_3y+2x_2^2y\Big)\K'(s|\tau)\\
&\quad +2x_1^2x_2y\K''(s|\tau)-\frac{1}{6}x_1^4y\K'''(s|\tau)+\frac{\log(-1)}{2}y^2-\frac{3}{4}y^2.
\end{split}
\end{equation}
\end{Cor}

\subsection{Applications}
According to formula (\ref{Hessian}),  all the prepotentials of the Hurwitz-Frobenius structures induced by Dubrovin's primary differentials (\ref{Primary-e}) share the same Hessian matrix. We shall employ this geometric  viewpoint  to  determine the derivatives of the Weierstrass functions with respect to the modulus $\tau$. Our proposed formulas take on a very concise form  using the introduced function $\mathscr{W}$ in (\ref{W-function}) and its derivatives. The obtained results can be found in \cite{Brezhnev} (Section 10) and the references therein.

\begin{Prop}\label{tau-derivation} Let $\sigma$ be the Weierstrass function (\ref{sigma-function}), $\K(v):=\log\sigma(v)$ and $G_2$ be the quasi-modular Eisenstein series  (\ref{G2-def}).
Then we have
\begin{align}
\begin{split}\label{K-tau1}
&\partial_{\tau}\Big(\K(v|\tau)-\frac{v^2}{2}G_2(\tau)\Big)=-\frac{1}{4{\rm{i}}\pi}\mathscr{W}(v);
\end{split}\\
\begin{split}\label{zeta-tau1}
&\partial_{\tau}\Big(\K'(v|\tau)-vG_2(\tau)\Big)=-\frac{1}{4{\rm{i}}\pi}\mathscr{W}'(v),
\end{split}
\end{align}
where $\mathscr{W}$ is  defined by
\begin{equation}\label{W-function}
\mathscr{W}(v)=\mathscr{W}(v|\tau):=\wp(v)-\Big(\zeta(v)-G_2(\tau)v\Big)^2-2G_2(\tau)
\end{equation}
and  $\wp$ and $\zeta$ are the Weierstrass functions (\ref{p-function})-(\ref{zeta-function}). \\
\end{Prop}

The function $\mathscr{W}$ is  even and multivalued. Indeed, because of the quasi-periodicity property of the function $\zeta$ and Legendre's identity (\ref{Legendre}), the function $\mathscr{W}$ (\ref{W-function})  satisfies:
\begin{align*}
&\mathscr{W}(v+1)-\mathscr{W}(v)=0;\\
&\mathscr{W}(v+\tau)-\mathscr{W}(v)=4{\rm i}\pi\Big(\zeta(v)-G_2(\tau)v\Big)-(2{\rm i}\pi)^2.
\end{align*}
Furthermore, making use of the differential equation $\wp''=6\wp^2-30G_4$ and observing that
\begin{align*}
2\big(\zeta-G_2v\big)\mathscr{W}'=4\wp^2-2\wp'\big(\zeta-G_2v\big)-4\big(\wp+G_2\big)\mathscr{W}-4G_2\wp-8G_2^2,
\end{align*}
then we can see that the function $\mathscr{W}$ (\ref{W-function}) obeys the following differential equation
\begin{equation}\label{W-diff eq}
\mathscr{W}''=2\big(\zeta-G_2v\big)\mathscr{W}'+4\big(\wp+G_2\big)\mathscr{W}-6(5G_4-G_2^2).
\end{equation}

To prove the preceding proposition,  we place ourselves in the context of the four dimensional Hurwitz-Frobenius manifold $\mathcal{H}_1(0,0)$ whose prepotential and LG superpotential are respectively  given by (\ref{Jacobi-A-A0}) and the case $n=0$ of (\ref{Jacobi-A-superp}), i.e.
\begin{equation}\label{model}
\begin{split}
&\lambda(P)=u+ysG_2(\tau)+y\Big(\zeta(z_P-\infty^1)-\zeta(z_P-\infty^0)\Big),\quad \quad \text{with}\\
&u=\oint_a\lambda(P)dz_P,\quad \quad  y=\underset{\infty^1}{\rm res}\ \lambda(P)dz_P\quad \text{and}\quad s=\infty^1-\infty^0.
\end{split}
\end{equation}
From (\ref{Primary-e}), we have four primary differentials given by
$$
\phi(P)=dz_P,\quad  \Omega_{\infty^0\infty^1}, \quad \underset{\infty^1}{\rm res}\lambda(Q)W(P,Q)
=\Big(\wp(z_P-\infty^1)+G_2(\tau)\Big)dz_P,\quad \quad \oint_a\lambda(Q)W(P,Q),
$$
where $W(P,Q)$ is the bidifferential (\ref{W-g1}) on the torus $\C/(\Z+\tau\Z)$. The last primary differential is described by the following useful lemma.
\begin{Lemma} Let $W(P,Q)$ be the genus one bidifferential (\ref{W-g1}), $\lambda$ be the meromorphic function (\ref{model}) and $\mathscr{W}$ be the multivalued function (\ref{W-function}). Then
\begin{equation}\label{jump}
\phi_{\mathbf{u}}(P):=\oint_a\lambda(Q)W(P,Q)=-\frac{y}{2}\Big(\mathscr{W}'(z_P-\infty^1)-\mathscr{W}'(z_P-\infty^0)\Big)dz_P.
\end{equation}
\end{Lemma}
\emph{Proof:} Putting $\alpha_0=-y$ and $\alpha_1=y$, we observe that the elliptic  function (\ref{model}) takes the following form
$$
\lambda(P)=u+ysG_2+\sum_{j=0}^1\alpha_j\zeta(z_P-\infty^j).
$$
Therefore  we can write
\begin{align*}
&\oint_a\lambda(Q)\frac{W(P,Q)}{dz_P}=\oint_a\lambda(Q)\Big(\wp(z_P-z_Q)+G_2\Big)dz_Q\\
&=G_2\oint_a\lambda(Q)dz_Q+\big(u+ysG_2\big)\oint_a\wp(z_P-z_Q)dz_Q+\sum_{j=0}^1\alpha_j\oint_a\zeta(z_Q-\infty^j)\wp(z_Q-z_P)dz_Q\\
&=-ysG_2^2+\sum_{j=0}^1\alpha_j\int_{x-z_P}^{x-z_P+1}\zeta(v+z_P-\infty^j)\wp(v)dv\\
&=-ysG_2^2+\bigg(\int_{x-z_P}^{x-z_P+1}\wp(v)dv\bigg)\bigg(\sum_{j=0}^1\alpha_j\zeta(z_P-\infty^j)\bigg)
+\sum_{j=0}^1\alpha_j\int_{x-z_P}^{x-z_P+1}\bigg(\frac{\wp'(v)-\wp'(z_P-\infty^j)}{2\big(\wp(v)-\wp(z_P-\infty^j)\big)}\bigg)\wp(v)dv\\
&=-G_2\sum_{j=0}^1\alpha_j\Big(\zeta(z_P-\infty^j)-G_2(\tau)(z_P-\infty^j)\Big)\\
&\quad +\sum_{j=0}^1\alpha_j\int_{x-z_P}^{x-z_P+1}\bigg(\frac{\wp'(v)-\wp'(z_P-\infty^j)}{2\big(\wp(v)-\wp(z_P-\infty^j)\big)}\bigg)
\bigg(\wp(v)-\wp(z_P-\infty^j)+\wp(z_P-\infty^j)\bigg)dv\\
&=-G_2\sum_{j=0}^1\alpha_j\Big(\zeta(z_P-\infty^j)-G_2(\tau)(z_P-\infty^j)\Big)-\frac{1}{2}\sum_{j=0}^1\alpha_j\wp'(z_P-\infty^j)\\
&\quad +\sum_{j=0}^1\alpha_j\wp(z_P-\infty^j)\int_{x-z_P}^{x-z_P+1}\frac{\wp'(v)-\wp'(z_P-\infty^j)}{2\big(\wp(v)-\wp(z_P-\infty^j)\big)}dv\\
&=-G_2\sum_{j=0}^1\alpha_j\Big(\zeta(z_P-\infty^j)-G_2(\tau)(z_P-\infty^j)\Big)-\frac{1}{2}\sum_{j=0}^1\alpha_j\wp'(z_P-\infty^j)\\
&\quad +\sum_{j=0}^1\alpha_j\wp(z_P-\infty^j)\int_{x-z_P}^{x-z_P+1}\bigg(\zeta(v+z_P-\infty^j)-\zeta(v)-\zeta(z_P-\infty^j)\bigg)dv\\
&=-\frac{1}{2}\sum_{j=0}^1\alpha_j\bigg(\wp'(z_P-\infty^j)+
2\Big[\wp(z_P-\infty^j)+G_2(\tau)\Big]\Big[\zeta(z_P-\infty^j)-G_2(\tau)\big(z_P-\infty^j)\Big]\bigg)\\
&=-\frac{1}{2}\sum_{j=0}^1\alpha_j\mathscr{W}'(z_P-\infty^j),
\end{align*}
where we used\\
-  expression (\ref{W-g1}) for the genus one bidifferential and the relations $G_2(\tau)=2\zeta(1/2)=-\oint_a\wp(v)dv$ in the first, third and fifth  equalities;\\
-  the  addition theorem (\ref{addition-zeta}) for the $\zeta$-function in the fourth and seventh equalities;\\
- the fact that $ys=y(\infty^1-\infty^0)=-\sum_{j=0}^1\alpha_j(z_P-\infty^j)$ in the fifth equality;\\
- the relation $\zeta(v)=\partial_v\log\big(\sigma(v)\big)$ (\ref{zeta-sigma}) and the quasi-periodicity properties (\ref{zeta-sigma periods}) of the function $\sigma$ in the eighth equality;\\
- expression (\ref{W-function}) for the function $\mathscr{W}$ in the last equality.

\fd

\noindent \emph{Proof of Proposition \ref{tau-derivation}:}
Although (\ref{zeta-tau1}) follows directly from (\ref{K-tau1}),   both relations come from the following two special cases of formulas  (\ref{a-periods-hessian}):
\begin{equation}\label{Hessian-Int}
\begin{split}
&\oint_a\lambda(P)\Omega_{\infty^0\infty^1}(P)=2{\rm i}\pi\partial_{\tau}\partial_{y}\widetilde{\mathbf{F}}_{\phi,0}(u,y,s,\tau);\\
&\oint_a\lambda(P)\bigg(\underset{\infty^1}{{\rm res}\ }\lambda(Q)W(P,Q)\bigg)=2{\rm i}\pi\partial_{\tau}\partial_{s}\widetilde{\mathbf{F}}_{\phi,0}(u,y,s,\tau),
\end{split}
\end{equation}
where $\widetilde{\mathbf{F}}_{\phi,0}$ is the WDVV solution (\ref{Jacobi-A-A0}). In other words, we have two ways to calculate the integrals of  right hand sides of (\ref{Hessian-Int}).\\
On the one hand, the function $\mathscr{W}$ defined by  (\ref{W-function}) is extendable to a holomorphic function inside a fundamental parallelogram $\mathrm{F}$ containing the origin  with
$$
\mathscr{W}(0):=\lim_{v\to0}\mathscr{W}(v)=0=\mathscr{W}'(0).
$$
Therefore, by interchanging the order of integration  and using step 1, we get
\begin{align*}
&\oint_a\lambda(P)\Omega_{\infty^0\infty^1}(P)=\oint_a\lambda(P)\bigg(\int_{\infty^0}^{\infty^1}W(P,Q)\bigg)
=\int_{\infty^0}^{\infty^1}\bigg(\oint_{P\in a}\lambda(P)W(P,Q)\bigg)\\
&=-\frac{y}{2}\int_{\infty^0}^{\infty^1}\Big(\mathscr{W}'(z_Q-\infty^1)-\mathscr{W}'(z_Q-\infty^0)\Big)dz_Q
=y\mathscr{W}(s)
\end{align*}
and
\begin{align*}
&\oint_a\lambda(P)\bigg(\underset{\infty^1}{{\rm res}\ }\lambda(Q)W(P,Q)\bigg)
=\underset{\infty^1}{{\rm res}\ }\bigg(\lambda(Q)\oint_{P\in a}\lambda(P)W(P,Q)\bigg)\\
&=-\frac{y}{2}\underset{\infty^1}{{\rm res}\ }\bigg(\lambda(Q)\Big(\mathscr{W}'(z_Q-\infty^1)-\mathscr{W}'(z_Q-\infty^0)\Big)dz_Q\bigg)
=\frac{y^2}{2}\mathscr{W}'(s).
\end{align*}
On the other hand, using expression (\ref{Jacobi-A-A0}) for the prepotential $\widetilde{\mathbf{F}}_{\phi,0}(u,y,s,\tau)$, we see that
\begin{align*}
&2{\rm i}\pi\partial_{\tau}\partial_{y}\widetilde{\mathbf{F}}_{\phi,0}(u,y,s,\tau)=-{4{\rm i}\pi}y\Big(\partial_{\tau}\K(s)-\frac{s^2}{2}\partial_{\tau}G_2\Big);\\
&2{\rm i}\pi\partial_{\tau}\partial_{s}\widetilde{\mathbf{F}}_{\phi,0}(u,y,s,\tau)=-{2{\rm i}\pi}y^2\Big(\partial_{\tau}\K'(s)-s\partial_{\tau}G_2(\tau)\Big).
\end{align*}
Thus comparing these two ways using (\ref{Hessian-Int}), we arrive at the stated relations  (\ref{K-tau1}) and (\ref{zeta-tau1}).

\fd

As corollary, we  give a simple  proof of Ramanujan's differential equations for all the non-normalized series $G_{2m}$, $m\geq 1$.
\begin{Cor} For any positive integer $m$, the Eisenstein series $G_{2m}$ defined by (\ref{Eisenstein G}), (\ref{G2-def}) and (\ref{Eisentein-Fourier}) satisfy the following Ramanujan differential equations:
\begin{equation}\label{Ramanujan-G}
\partial_{\tau}G_{2m}(\tau)=\frac{m}{2{\rm i}\pi}\bigg((2m+3)G_{2m+2}-\sum_{\ell=0}^{m-1}G_{2\ell+2}G_{2m-2\ell}\bigg).
\end{equation}
\end{Cor}
\emph{Proof:} The Laurent expansions (\ref{Laurend-p}) and (\ref{Laurent zeta}) for the Weierstrass functions $\wp$ and $\zeta$ and simple calculations enable us to obtain the following Taylor expansion for the function $\mathscr{W}$ (\ref{W-function}) near the origin:
\begin{align*}
\mathscr{W}(v)&=\wp(v)-\Big(\zeta(v)-G_2v\Big)^2-2G_2\\
&\underset{v\sim0}{=}\sum_{m=1}^{\infty}\bigg((2m+3)G_{2m+2}-\sum_{\ell=0}^{m-1}G_{2\ell+2}G_{2m-2\ell}\bigg)v^{2m}.
\end{align*}
In particular
\begin{align*}
\mathscr{W}'(v)&=\sum_{m=1}^{\infty}\bigg(2m(2m+3)G_{2m+2}-2m\sum_{\ell=0}^{m-1}G_{2\ell+2}G_{2m-2\ell}\bigg)v^{2m-1}.
\end{align*}
This together with the equality (\ref{zeta-tau1}) and $\tau$-derivation in the Laurent series (\ref{Laurent zeta}) of the function $\K'(v)-G_2v=\zeta(v)-G_2v$  show (\ref{Ramanujan-G}).

\fd

\begin{Remark}
\end{Remark}
$\bullet$ Let $E_{2m}:=G_{2m}/2\widetilde{\zeta}(2m)$ be the normalized Eisenstein series of weight $2m$, where $\widetilde{\zeta}$ denotes in our case the Riemann zeta function (to distinguish it with the Weierstrass zeta function). Then making use of the following  recursion formula for the Riemann zeta function that can be found in \cite{Williams53, Apostol73}):
\begin{equation}\label{Recursion formula}
(2m+3)\widetilde{\zeta}(2m+2)=2\sum_{k=1}^{m}\widetilde{\zeta}(2k)\widetilde{\zeta}(2m+2-2k),\quad \forall\ m\in \N, \ m\geq 1,
\end{equation}
we conclude that (\ref{Ramanujan-G}) can be expressed as follows:
$$
\frac{1}{2{\rm{i}}\pi}\partial_{\tau}E_{2m}=\frac{m}{\pi^2\widetilde{\zeta}(2m)}\sum_{k=0}^{m-1}
\Big(\widetilde{\zeta}(2k+2)\widetilde{\zeta}(2m-2k)\Big)\Big(E_{2k+2}E_{2m-2k}-E_{2m+2}\Big).
$$
This was worked out in (\cite{Shen}, Theorem D) and extends  the well-known Ramanujan's differential equations for the Eisenstein series $E_2,E_4$ and $E_6$ \cite{Ramanujan1916}. \\
$\bullet$ Relations (\ref{K-tau1}) and (\ref{zeta-tau1}) and the Ramanujan differential equation (\ref{Ramanujan-G}) for the quasi-modular form $G_2$ bring us to the following formulas for the $\tau$-derivative of Weierstrass functions $\sigma,\zeta,\wp, \wp^{(k)}$:
\begin{equation}\label{Weierstrass-tau1}
\begin{split}
&\partial_{\tau}\log\big(\sigma(v|\tau)\big)=-\frac{1}{4{\rm{i}}\pi}\Big(\mathscr{W}(v)-\big(5G_4-G_2^2\big)v^2\Big);\\
&\partial_{\tau}\zeta(v|\tau)=-\frac{1}{4{\rm{i}}\pi}\Big(\mathscr{W}'(v)-2\big(5G_4-G_2^2\big)v\Big);\\
&\partial_{\tau}\wp(v|\tau)=\frac{1}{4{\rm{i}}\pi}\Big(\mathscr{W}''(v)-2\big(5G_4-G_2^2\big)\Big);\\
&\partial_{\tau}\wp^{(k)}(u|\tau)=\frac{1}{4{\rm{i}}\pi}\mathscr{W}^{(k+2)}(v),\quad \quad \quad \forall\ k\geq 1.
\end{split}
\end{equation}
In addition,  expression (\ref{W-function}) for the function $\mathscr{W}$ and  differential equation (\ref{W-diff eq}) as well as the above relations lead to the following heat equation
$$
\partial_{\tau}\mathscr{W}(v|\tau)=\frac{1}{2{\rm{i}}\pi}\Big(\mathscr{W}''(v)-2\big(\wp(v)+G_2\big)\mathscr{W}\Big).
$$
This turns out to obtain the second order $\tau$-derivative of  Weierstrass functions by means of the function $\mathscr{W}$ (\ref{W-function}) and its derivatives.

\begin{Prop} Consider the function $\psi$ defined by
\begin{equation}\label{psi-function}
\psi(v):=\log\sigma(v)-\frac{G_2}{2}v^2=\log\bigg(\frac{\theta_1(v)}{\theta'_1(0)}\bigg),
\end{equation}
where $\sigma$ is  the Weierstrass function (\ref{sigma-function}), $G_2$ is the quasi-modular form (\ref{G2-def}) and $\theta_1$ is the Jacobi function (\ref{Jacobi}). Then the function $\psi$ solves the following  nonlinear  partial differential equations:
\begin{align}
\begin{split}\label{psi-diff eq}
&\psi^{(4)}+6\big(\psi''\big)^2+12G_2\psi''+6\big(G_2^2-5G_4)=0;
\end{split}\\
\begin{split}\label{psi-heat}
&(4{\rm{i}}\pi)\partial_{\tau}\psi=\psi''+\big(\psi'\big)^2+3G_2.
\end{split}
\end{align}
\end{Prop}
\emph{Proof:} Since
$$
\psi'(v)=\zeta(v)-G_2v \quad \quad \text{and} \quad\quad \psi''(v)=-\wp(v)-G_2,
$$
it follows that (\ref{psi-diff eq}) is an immediate consequence of the differential equation
$$
\wp''(v)=6\wp^2(v)-\frac{g_2}{2}=6\wp^2(v)-30G_4.
$$
Moreover,  the previous  expressions for $\psi'$ and $\psi''$ imply that
\begin{equation}\label{psi-W link}
\psi''+\big(\psi'\big)^2+3G_2=-\mathscr{W},
\end{equation}
with $\mathscr{W}$ being the function (\ref{W-function}).
Accordingly the heat equation (\ref{psi-heat}) follows directly from (\ref{K-tau1}) and (\ref{psi-W link}).

\fd

\medskip

In the next theorem, we show that the 4-dimensional Hurwitz-Frobenius manifold $\mathcal{H}_1(0,0)$ associated with the holomorphic differential $\phi=dz_P$ is a geometric realization of  partial differential equations (\ref{psi-diff eq})-(\ref{psi-heat}) satisfied by the function $\psi$ (\ref{psi-function}), in the sense that   the WDVV associativity equations (\ref{WDVV}) for the corresponding  prepotential $\widetilde{\mathbf{F}}_{\phi,0}$ (\ref{Jacobi-A-A0})  are equivalent to
(\ref{psi-diff eq})-(\ref{psi-heat}).

\begin{Thm}\label{PDE1} Let $F=F(t_0,t_1,t_2,t_3)$ be the function defined by
\begin{equation}\label{solutions-wdvv}
F=\frac{(t_3)^2}{2}t_0+t_1t_2t_3+(t_2)^2\log(t_2)-(t_2)^2\psi\big(t_1|2{\rm{i}}\pi t_0\big)+\quad \text{quadratic terms},
\end{equation}
for some $\psi=\psi(v|\tau)$  defined on an open subset of $\C\times \mathbb{H}$, with $\mathbb{H}$ being the upper half-plane.
Then the following statements are equivalent:
\begin{description}
\item[i)] $F$ satisfies the WDVV equations (\ref{WDVV});
\item[ii)] there is a function $f=f(\tau)$ such that $\psi$ is a solution to the following differential equations
\begin{equation}\label{Pde-CNS}
\left\{
\begin{array}{ll}
\psi^{(4)}+6\big(\psi''\big)^2+4f(\tau)\psi''-(4{\rm{i}}\pi)f'(\tau)=0;\\
\\
(4{\rm{i}}\pi)\partial_{\tau}\psi=\psi''+\big(\psi'\big)^2+f(\tau).
\end{array}
\right.
\end{equation}
\end{description}

\end{Thm}
\emph{Proof:} Let us denote by $\mathrm{H}$ the Hessian matrix of the function $F$ introduced in (\ref{solutions-wdvv}). Since the matrix $\partial_{t_3}H$ is constant and anti-diagonal, it follows that  the WDVV equations (\ref{WDVV}) for the function  $F$ can be written  as follows:
\begin{align*}
&\big(\partial_{t_j}\mathrm{H}\big)\big(\partial_{t_3}\mathrm{H}\big)\big(\partial_{t_k}\mathrm{H}\big)
=\big(\partial_{t_k}\mathrm{H}\big)\big(\partial_{t_3}\mathrm{H}\big)\big(\partial_{t_j}\mathrm{H}\big),\quad \quad \forall\ j,k=0,1,2.
\end{align*}
For all $j=0,1,2$, we see that the matrix $\partial_{t_j}\mathrm{H}$ has the following form:
$$
\partial_{t_j}\mathrm{H}=
\begin{pmatrix}
c_{j00} & c_{j01}      & c_{j02}     & 0 \\
c_{j10} & c_{j11}      & c_{j12}     & \delta_{j2} \\
c_{j20} & c_{j21}      & c_{j22}     & \delta_{j1} \\
  0     & \delta_{j2}  & \delta_{j1} & \delta_{j0}
\end{pmatrix},
\quad \quad \text{with}\quad \quad c_{jks}:=\partial_{t_j}\partial_{t_k}\partial_{t_s}F.
$$
This implies that
\begin{align*}
&\big(\partial_{t_j}\mathrm{H}\big)\big(\partial_{t_3}\mathrm{H}\big)=
\begin{pmatrix}
    0       &    c_{j02}  &    c_{j01} &  c_{j00} \\
\delta_{j2} &    c_{j12}  &    c_{j11} &  c_{j10} \\
\delta_{j1} &    c_{j22}  &    c_{j21} &  c_{j20} \\
\delta_{j0} & \delta_{j1} &\delta_{j2} &   0
\end{pmatrix}
\quad \quad \text{and}\quad \quad
&\big(\partial_{t_3}\mathrm{H}\big)\big(\partial_{t_j}\mathrm{H}\big)=
\begin{pmatrix}
  0     & \delta_{j2}  & \delta_{j1} & \delta_{j0}\\
c_{j20} & c_{j21}      & c_{j22}     & \delta_{j1} \\
c_{j10} & c_{j11}      & c_{j12}     & \delta_{j2} \\
c_{j00} & c_{j01}      & c_{j02}     & 0
\end{pmatrix}.
\end{align*}
As a consequence, we have
\begin{align*}
&\quad \quad \quad
\big(\partial_{t_j}\mathrm{H}\big)\big(\partial_{t_3}\mathrm{H}\big)\big(\partial_{t_k}\mathrm{H}\big)
=\big(\partial_{t_k}\mathrm{H}\big)\big(\partial_{t_3}\mathrm{H}\big)\big(\partial_{t_j}\mathrm{H}\big),\quad \quad\quad \quad \forall\ j,k=0,1,2;\\
\\
&\Longleftrightarrow\quad
\left\{
\begin{array}{llllllll}
c_{j02} c_{k11}+ c_{j01}c_{k21}+\delta_{k2}c_{j00}=\delta_{j2}c_{k00}+c_{k01}c_{j21}+c_{k02}c_{j11};\\
c_{j02} c_{k12}+ c_{j01}c_{k22}+\delta_{k1}c_{j00}=\delta_{j1}c_{k00}+c_{k01}c_{j22}+c_{k02}c_{j12};\\
\delta_{j2}c_{k02}+  c_{j11}c_{k22}+\delta_{k1}c_{j10}=\delta_{j1}c_{k10}+c_{k11}c_{j22}+\delta_{k2}c_{j02},
\end{array}
\quad \quad \forall\ j,k=0,1,2;
\right.\\
\\
&\Longleftrightarrow\quad
(S)\left\{
\begin{array}{lllllllllllllll}
c_{111}c_{222}=c_{112}c_{122}+2c_{012};\\
c_{011}c_{222}=c_{112}c_{022}+c_{002};\\
c_{002}c_{122}+c_{001}c_{222}=2c_{022}c_{012};\\
c_{011}c_{122}+c_{001}=c_{111}c_{022};\\
c_{002}c_{112}+c_{001}c_{122}+c_{000}=(c_{012})^2+c_{011}c_{022};\\
c_{002}c_{111}+c_{001}c_{112}=2c_{012}c_{011};\\
\end{array}
\right.
\end{align*}
Now, by using (\ref{solutions-wdvv}) we can deduce that the preceding sixth equations in the system (S) are nothing but the following partial differential equations satisfied  by the function $\psi$:
\begin{align*}
\begin{array}{lllllllll}
&\psi'''+2\psi''\psi'-2(2{\rm{i}}\pi)\big(\partial_{\tau}\psi\big)'=0;&\quad  \hbox{(E.1)}\\
&(2{\rm{i}}\pi)\partial_{\tau}^2\psi-2\psi''\big(\partial_{\tau}\psi\big)-\big(\partial_{\tau}\psi\big)''=0;&\quad  \hbox{(E.2)}\\
&\big(\partial_{\tau}^2\psi\big)'-2\big(\partial_{\tau}^2\psi\big)\psi'
+4\big(\partial_{\tau}\psi\big)\big(\partial_{\tau}\psi\big)'=0;&\quad  \hbox{(E.3)}\\
&(2{\rm{i}}\pi)\big(\partial_{\tau}^2\psi\big)'-2\big(\partial_{\tau}\psi\big)''\psi'
+2\psi'''\big(\partial_{\tau}\psi\big)=0;&\quad  \hbox{(E.4)}\\
&(2{\rm{i}}\pi)\partial_{\tau}^3\psi-4\Big(\big(\partial_{\tau}^2\psi\big)\psi''-\big[\big(\partial_{\tau}\psi\big)'\big]^2\Big)
+2\big(\partial_{\tau}\psi\big)\big(\partial_{\tau}\psi\big)''-2\psi'\big(\partial_{\tau}^2\psi\big)'=0;
&\quad  \hbox{(E.5)}\\
&\big(\partial_{\tau}^2\psi\big)\psi'''+\big(\partial_{\tau}^2\psi\big)'\psi''
-2\big(\partial_{\tau}\psi\big)'\big(\partial_{\tau}\psi\big)''=0. &\quad  \hbox{(E.6)}
\end{array}
\end{align*}
Here the notation  $\big(\partial_{\tau}^rQ\big)^{(m)}$ means that $\partial_v^m\partial_{\tau}^rQ(v|\tau)$.
At this level, we have proved that the WDVV equations for the function $F$ in (\ref{solutions-wdvv})  are equivalent to equations (E.1)-(E.6).\\
Let us move on to prove that $\textbf{ii)}\Rightarrow \textbf{i)}$. Differentiating with respect to the variable $v$ in  the second  equation of the system (\ref{Pde-CNS}) we see that (E.1) holds. Moreover, due to (\ref{Pde-CNS}),  we have
\begin{align*}
(4{\rm{i}}\pi)\big(\partial_{\tau}\psi\big)'&=\psi'''+2\psi'\psi'';\\
(4{\rm{i}}\pi)\big(\partial_{\tau}\psi\big)''&=\psi^{(4)}+2\psi'\psi'''+2\big(\psi''\big)^2;\\
(4{\rm{i}}\pi)^2\partial_{\tau}^2\psi
&=(4{\rm{i}}\pi)\big(\partial_{\tau}\psi\big)''+2(4{\rm{i}}\pi)\psi'\big(\partial_{\tau}\psi\big)'+ (4{\rm{i}}\pi)f'(\tau)\\
&=\psi^{(4)}+4\psi'\psi'''+2\big(\psi''\big)^2+4\big(\psi'\big)^2\psi''+ (4{\rm{i}}\pi)f'(\tau)\\
&=2\psi^{(4)}+4\psi'\psi'''+8\big(\psi''\big)^2+4\big(\psi'\big)^2\psi''+4f(\tau)\psi'';\\
(4{\rm{i}}\pi)^2\big(\partial_{\tau}^2\psi\big)'&
=2\Big(\psi^{(5)}+2\psi'\psi^{(4)}+10\psi''\psi'''
+2\big(\psi'\big)^2\psi'''+4\psi'\big(\psi''\big)^2+2f(\tau)\psi'''\Big)\\
&=4\Big(\psi'\psi^{(4)}-\psi''\psi'''+\big(\psi'\big)^2\psi'''
+2\psi'\big(\psi''\big)^2-f(\tau)\psi'''\Big).
\end{align*}
From this  we deduce that equations (E.2), (E.3) and (E.6) are satisfied.\\
Now if we differentiate with respect to $\tau$ in (E.1) and  with respect to $v$ in (E.2) and take the sum, then we arrive at equation (E.4). Similarly, equation (E.5) comes from (E.2) and (E.3). \\
Conversely,  assume that the function $\psi$ obeys equations (E.1)-(E.6). Given that $\psi$ solves (E.1), there exists a function $f(\tau)$ such that
$$
(4{\rm{i}}\pi)\partial_{\tau}\psi=\psi''+\big(\psi'\big)^2+f(\tau).
$$
In particular
\begin{align*}
(4{\rm{i}}\pi)^2\partial_{\tau}^2\psi&=\big((4{\rm{i}}\pi)\partial_{\tau}\psi\big)''+2\psi'\big((4{\rm{i}}\pi)\partial_{\tau}\psi\big)'
+(4{\rm{i}}\pi)f'(\tau)\\
&=\psi^{(4)}+4\psi'\psi'''+2\big(\psi''\big)^2+4\big(\psi'\big)^2\psi''+(4{\rm{i}}\pi)f'(\tau).
\end{align*}
On the other hand (E.2) yields  that
\begin{align*}
(4{\rm{i}}\pi)^2\partial_{\tau}^2\psi
&=2(4{\rm{i}}\pi)\big(\partial_{\tau}\psi\big)''+4(4{\rm{i}}\pi)\psi''\partial_{\tau}\psi\\
&=2\psi^{(4)}+4\psi'\psi'''+4\big(\psi''\big)^2
+4\psi''\Big(\psi''+\big(\psi'\big)^2+f(\tau)\Big)\\
&=2\psi^{(4)}+4\psi'\psi'''+8\big(\psi''\big)^2+4\big(\psi'\big)^2\psi''+4f(\tau)\psi''.
\end{align*}
Thus we must have
$$
\psi^{(4)}+6\big(\psi''\big)^2+4f(\tau)\psi''-(4{\rm{i}}\pi)f'(\tau)=0.
$$
This establishes that  $\textbf{i)}\Rightarrow \textbf{ii)}$ and finishes  the proof the theorem.

\fd

\begin{Remark}
Differential equations (\ref{psi-diff eq}) and  (\ref{psi-heat}) and Ramanujan's identity  (\ref{Ramanujan-G}) for the Eisenstein series $G_2$ enable us to conclude that the function $\psi$ defined by (\ref{psi-function}) obeys differential equations (\ref{Pde-CNS}) where we take $f(\tau)=3G_2(\tau)$.
 \end{Remark}


\section{WDVV solutions associated with the deformed Frobenius manifold structure}
This section explores a generalization of Theorem \ref{Main result} to the framework of $q$-deformed Frobenius manifold structures on the $N$-dimensional  Hurwitz space $\mathcal{M}=\mathcal{H}_1(n_0,\dots,n_m)$ in genus one, with $N=2+2m+\sum_{j=0}^mn_j$ and $q$ being a fixed complex parameter.\\
Deformed Hurwitz-Frobenius manifolds were constructed in  \cite{Vasilisa2} and their descriptions make use  of deformed bidifferentials defined on the underlying Riemann surfaces. Moreover, to calculate the corresponding WDVV prepotentials, a new method utilizing  a formula similar to that in  (\ref{Prepotential}) was derived in \cite{Rejeb23} (Section 5). This inspires us to adapt the scheme outlined  in Sections 3 and 4 as we work toward our goal.\\
We begin by extending the main results from Section 3 within the context of the $q$-deformation setting, where $q\in \C$ is fixed parameter. Let us then consider a lattice $\mathbb{L}=\Z+\tau\Z$, where $\tau$ is a complex number with  positive imaginary part, further satisfying  the  requirement
\begin{equation}\label{q-condition}
1+q\tau\neq 0
\end{equation}
and a meromorphic function $\lambda: \C/{\mathbb{L}}\longrightarrow \P^1$  having a fixed ramification profile over the point at infinity $\infty\in\P^1$, where
$\lambda^{-1}(\infty)=\big\{\infty^0,\dots,\infty^m \in \C/{\mathbb{L}}\big\}$ and the  order of the pole $\infty^j$ is $n_j+1$, $j=0,\dots,m$. Throughout this section, the meromorphic function $\lambda(P)$ under consideration will be assumed to be  the LG superpotential given by (\ref{lambda-zeta}) and we will mountain the notation used there
\begin{description}
\item[$\bullet$] \emph{Normalized holomorphic $q$-differential}
\end{description}
The  genus one $q$-bidifferential on the torus $\C/{\mathbb{L}}$ is the $q$-deformation of the bidifferential $W(P,Q)$ (\ref{W-g1}) and defined by
\begin{equation}\label{q-W-g1}
\begin{split}
W_q(P,Q)&=W(P,Q)-\frac{{2{\rm i}\pi}q}{1+q\tau}\phi(P)\phi(Q)\\
&=\Big(\wp(z_P-z_Q)+2\zeta(1/2)-\frac{{2{\rm i}\pi}q}{1+q\tau}\Big)dz_Pdz_Q,
\end{split}
\quad \quad P,Q\in \C/{\mathbb{L}},
\end{equation}
where $\phi(P)=dz_P$ is the normalized holomorphic differential and $\wp,\zeta$ are the Weierstrass functions. The $q$-bidifferential $W_q(P,Q)$ reduces to the bidifferential $W(P,Q)$ when $q=0$.  Motivated by  (\ref{phi-def}), the $q$-deformation  of the  usual normalized holomorphic differential  is defined by means of the $q$-bidifferential $W_{q}(P,Q)$ as
\begin{equation}\label{q-phi-def}
\phi_q(P):=\frac{1}{2{\rm i}\pi}\oint_bW_q(P,Q)=\frac{1}{1+q\tau}\phi(P)=\frac{dz_P}{1+q\tau},
\end{equation}
with the second equality being an immediate consequence of expression (\ref{q-W-g1}) for  $W_q(P,Q)$. The periods of the holomorphic differential $\phi_q$ (\ref{q-phi-def})
are given by
\begin{equation}\label{q-phi-periods}
\oint_a\phi_q=\frac{1}{1+q\tau}\quad \quad \text{and}\quad \quad \tau_q:=\oint_b\phi_q=\frac{\tau}{1+q\tau}
\end{equation}
and then it satisfies the  normalization condition:
\begin{equation}\label{q-normalization}
\oint_a\phi_q+q\oint_b\phi_q=1.
\end{equation}
Analogously to (\ref{phi-Taylor}), for $j=0,1,\dots,m$, let us write the Taylor expansion of the holomorphic $q$-differential $\phi_q$ (\ref{q-phi-def}) near the pole $\infty^j$ of $\lambda$ in the following form
\begin{equation}\label{q-phi-Taylor}
\phi_q(P)\underset{P\sim \infty^j}{=}\bigg(\sum_{r=1}^{\infty}rx_{q,r}(j)\mathbf{z}_j(P)^{r-1}\bigg)d\z_j(P),
\quad \quad \quad \quad \lambda(P)\underset{P\sim \infty^j}{=}\z_j(P)^{-n_j-1}.
\end{equation}
In particular, because of (\ref{q-phi-def}) and (\ref{q-phi-Taylor}), the coefficients $x_{q,r}(j)$  and $x_r(j)$ in (\ref{phi-Taylor})-(\ref{Evaluation}) are related by
\begin{equation}\label{q-evaluation}
x_{q,r}(j)=\frac{1}{r}\underset{\infty^j} {\rm res}\  \lambda(P)^{\frac{r}{n_j+1}}\phi_q(P)=\frac{x_r(j)}{1+q\tau},\quad \quad \forall\ r\geq1.
\end{equation}
For later use and in a similar manner as in the non-deformed case (i.e. when $q=0$),  we shall introduce the following objects and notation involving the coefficients  $\big\{x_{q,r}(j)\}$ (\ref{q-phi-Taylor})-(\ref{q-evaluation}) associated with the $q$-differential $\phi_q$:
\begin{equation}\label{Bell-R-Lq}
\begin{split}
&\mathcal{B}_{\mu,k}^q(j):=\mathcal{B}_{\mu,k}\big(x_{q,1}(j),\dots,x_{q,\mu-k+1}(j)\big);\\
&\mathcal{R}_{\mu,k}^q(j):=\mathcal{R}_{\mu,k}\big(x_{q,1}(j),\dots,x_{q,\mu+1}(j)\big);\\
&\mathbf{L}^{\phi_q}_{j,\alpha}:=\sum_{k=1}^{\alpha}\frac{1}{k!}\mathcal{B}_{\alpha,k}^q(j)\partial_v^k,
\end{split}
\quad \quad \quad j=0,1,\dots,m,
\end{equation}
where $\mathcal{B}_{\mu,k}$ and $\mathcal{R}_{\mu,k}$ are respectively partial Bell polynomial (\ref{Bell-partial}) and the related rational function (\ref{Bell-rec4}) and
$\mathbf{L}^{\phi_q}_{j,\alpha}$ is the $q$-analogue of Bell operator (\ref{Bell operator}).
\begin{description}
\item[$\bullet$] \emph{Abelian $q$-differentials of the third kind}
\end{description}
For $j=1,\dots,m$, define the Abelian $q$-differential of the third kind $\Omega_{q,\infty^0\infty^j}(P)$ by
\begin{equation}\label{q-Third-def}
\begin{split}
\Omega_{q,\infty^0\infty^j}(P)&:=\int_{\infty^0}^{\infty^j}W_q(P,Q)=\Omega_{\infty^0\infty^j}(P)-\frac{{2{\rm i}\pi}q}{1+q\tau}s^j\phi(P)\\
&=\bigg(\zeta(z_P-\infty^j)-\zeta(z_P-\infty^0)+2\zeta(1/2)s^j-\frac{{2{\rm i}\pi}q}{1+q\tau}s^j\bigg)\phi(P),
\end{split}
\end{equation}
where $W_q(P,Q)$ is the $q$-bidifferential (\ref{q-W-g1}), $\Omega_{\infty^0\infty^j}$ is the normalized differential of the third kind (\ref{Third kind}) and $s^j=\int_{\infty^0}^{\infty^j}\phi=\infty^j-\infty^0$. A straightforward calculation using (\ref{Third kind-periods}) and the first line in (\ref{q-Third-def}) shows that the Abelian $q$-differential of the third kind $\Omega_{q,\infty^0\infty^j}(P)$  is normalized by the property
$$
\oint_a\Omega_{q,\infty^0\infty^j}(P)+q\oint_b\Omega_{q,\infty^0\infty^j}(P)=0.
$$
Let $i=0,\dots,m$. Then, formula (\ref{q-Third-def}) implies that  the Laurent expansion of the $q$-differential $\Omega_{q,\infty^0\infty^j}(P)$ near the point $\infty^i$ takes the following form
$$
\Omega_{q,\infty^0\infty^j}(P)\underset{P\sim \infty^i}{=}\bigg(\frac{\delta_{ij}}{z_i(P)}-\frac{\delta_{i0}}{z_i(P)}
+\sum_{\alpha=1}^{\infty}\Big(\mathcal{I}_{i,\alpha}[\Omega_{q,\infty^0\infty^j}]\Big)\z_i(P)^{\alpha-1}\bigg)d\z_i(P),
\quad \quad \lambda(P)\underset{P\sim \infty^i}{=}\z_i(P)^{-n_i-1},
$$
where the coefficients $\mathcal{I}_{i,\alpha}[\Omega_{q,\infty^0\infty^j}]$ are the $q$-analogue of $\mathcal{I}_{i,\alpha}[\Omega_{\infty^0\infty^j}]$ (\ref{rho-ij}) and given by
\begin{equation}\label{q-Third-alpha-ij}
\begin{split}
\mathcal{I}_{i,\alpha}[\Omega_{q,\infty^0\infty^j}]&:=\underset{\infty^i}{{\rm res}}\ \lambda(P)^{\frac{\alpha}{n_i+1}}\Omega_{q,\infty^0\infty^j}(P)\\
&=\mathcal{I}_{i,\alpha}[\Omega_{\infty^0\infty^j}]-\frac{{2{\rm i}\pi}q}{1+q\tau}{\alpha}x_{\alpha}(i)s^j,
\end{split}
 \quad \quad \forall\ \alpha\geq 1
\end{equation}
Therefore, expressions (\ref{Third-alpha-ij})-(\ref{Third-alpha-0j}) for the terms $\mathcal{I}_{i,\alpha}[\Omega_{\infty^0\infty^j}]$  and formula (\ref{q-Third-alpha-ij}) show that the coefficients $\mathcal{I}_{i,\alpha}[\Omega_{q,\infty^0\infty^j}]$ are explicitly  describable in terms of the variables $\tau$, $s^j=\infty^j-\infty^0$ and $x_r(j)$ (\ref{Evaluation}) associated with the usual normalized holomorphic differential $\phi(P)=dz_P$.\\
As another consequence of (\ref{q-Third-def}), we mention that
\begin{equation}\label{q-s-ij}
\begin{split}
\mathbf{s}^k\big(\Omega_{q,\infty^0\infty^j}\big)&:=p.v. \int_{\infty^0}^{\infty^k}\Omega_{q,\infty^0\infty^j}(P)\\
&=\mathbf{s}^k\big(\Omega_{\infty^0\infty^j}\big)-\frac{{2{\rm i}\pi}q}{1+q\tau}s^js^k,
\end{split}
\quad \quad\quad  \forall\ k,j=1,\dots,m,
\end{equation}
where, as above, the  principal value is defined by subtracting  the divergent part of the integral as a function of the local parameters $\z_0(P):=\lambda(P)^{-\frac{1}{n_0+1}}$ and  $\z_j(P):=\lambda(P)^{-\frac{1}{n_j+1}}$ near the points $\infty^0$ and $\infty^j$, respectively and $\mathbf{s}^k\big(\Omega_{\infty^0\infty^j}\big)=\mathbf{s}^k\big(\phi_{\mathbf{s}^j}\big)$ is explicitly determined by (\ref{sij})-(\ref{sjj}).

\begin{description}
\item[$\bullet$] \emph{Abelian $q$-differentials of the second kind}
\end{description}
By following (\ref{Psi-ialpha-def}), for fixed $j=0,\dots,m$ and fixed positive integer $\alpha$, consider the meromorphic $q$-differential
\begin{equation}\label{q-Psi-alphad-def}
\Psi_{q,j,\alpha}(P):=\underset{\infty^j} {\rm res}\  \lambda(Q)^{\frac{\alpha}{n_j+1}}W_q(P,Q)=\Psi_{j,\alpha}(P)-\frac{{2{\rm i}\pi}q}{1+q\tau}\alpha x_{\alpha}(j)\phi(P),
\end{equation}
where $W_q(P,Q)$ denotes  the $q$-bidifferential (\ref{q-W-g1}), $\Psi_{j,\alpha}$ is the Abelian differential of the second kind (\ref{Psi-ialpha-def}) and as above $\alpha x_{\alpha}(j)$ are defined by Taylor's series (\ref{phi-Taylor}) for the differential $\phi$ near the pole $\infty^j$ of $\lambda$. Formula (\ref{q-Psi-alphad-def}) implies that the meromorphic $q$-differential $\Psi_{q,j,\alpha}(P)$ has the same singularity structure (\ref{Psi-i-alpha-PP}) as the Abelian differential $\Psi_{j,\alpha}(P)$. In addition, expression (\ref{Psi-i-alpha}) for  $\Psi_{j,\alpha}$ turns out to the following explicit for its $q$-analogue
$$
\Psi_{q,j,\alpha}(P)=\bigg(2{\alpha}x_{\alpha}(j)\zeta(1/2)-\frac{{2{\rm i}\pi}q}{1+q\tau}\alpha x_{\alpha}(j)-\alpha
\sum_{k=1}^{\alpha}\frac{(-1)^k}{k!}\mathcal{B}_{\alpha,k}(j)\wp^{(k-1)}\big(z_P-\infty^j\big)\bigg)\phi(P).
$$
On the other hand, as in (\ref{q-Third-alpha-ij}), the following holds true for all $i=0,\dots,m$ and all $\beta\geq 1$:
\begin{equation}\label{q-chi-ij}
\begin{split}
\mathcal{I}_{i,\beta}[\Psi_{q,j,\alpha}]&:=\underset{\infty^i}{{\rm res}}\ \lambda(P)^{\frac{\beta}{n_i+1}}\Psi_{q,j,\alpha}(P)\\
&=\mathcal{I}_{i,\beta}[\Psi_{j,\alpha}]-\frac{{2{\rm i}\pi}q}{1+q\tau}{\alpha\beta}x_{\alpha}(j)x_{\beta}(i).
\end{split}
\end{equation}
\begin{description}
\item[$\bullet$] \emph{Primary q-differentials and flat coordinates of the corresponding flat metrics}
\end{description}
In the deformation setting depending on the fixed  parameter $q\in\C$, we also have  $N$ semi-simple Frobenius manifold structures on the $N$-dimensional Hurwitz space $\mathcal{H}_1(n_0,\dots,n_m)$ in genus one, with $N=2+2m+\sum_{j=0}^mn_j$. These structures are associated with  the following list of $N$ primary $q$-differentials defined on a surface $\C/\Z+\tau\Z$, with $\Im\tau>0$ and $1+q\tau\neq0$ (see \cite{Vasilisa2, Rejeb23} for details):
\begin{equation}\label{q-Primary-e}
\begin{split}
\phi_{q,\mathbf{t}^{i,\alpha}}(P)&:=\frac{\sqrt{n_i+1}}{\alpha}\underset{\infty^i}{{\rm res}} \ \lambda(Q)^{\frac{\alpha}{n_i+1}}W_q(P,Q)=\frac{\sqrt{n_i+1}}{\alpha}\Psi_{q, i,\alpha}(P),\quad
\begin{array}{ll}
&i=0,\dots,m,\\
&\alpha=1,\dots,n_i;
\end{array}\\
\phi_{q,\mathbf{v}^i}(P)&:=\underset{\infty^i}{{\rm res}\ }\lambda(Q)W_q(P,Q)=\Psi_{q,i,n_i+1}(P), \quad \quad\quad  i=1,\dots,m;
\\
\phi_{q,\mathbf{s}^i}(P)&:=\int_{\infty^0}^{\infty^i}W_q(P,Q)=\Omega_{q,\infty^0\infty^i}(P), \quad \quad \quad \quad i=1,\dots,m;
\\
\phi_{q,\mathbf{r}}(P)&:=\frac{1}{2{\rm{i}}\pi}\oint_{b}W_q(P,Q)=\phi_{q}(P);
\\
\phi_{q,\xi}(P)&:=\oint_a\lambda(Q)W_q(P,Q)+q\oint_b\lambda(Q)W_q(P,Q)\\
&=\phi_{\mathbf{u}}(P)+q\phi_{\widehat{\mathbf{u}}}(P)-\frac{{2{\rm i}\pi}q}{1+q\tau}\big(u+q\widehat{u}\big)\phi(P),
\end{split}
\end{equation}
where $\lambda(P)$ is the meromorphic function given by  (\ref{LG}) and satisfying the conditions indicated in  Remark \ref{Rk-Hurwitz}, $\Psi_{q, j,\mu}(P)$ is the Abelian $q$-differential of the second kind (\ref{q-Psi-alphad-def}), $\Omega_{q,\infty^0\infty^i}(P)$ is the Abelian $q$-differential of the third kind (\ref{q-Third-def}),  $\phi_q(P)$ is the holomorphic $q$-differential (\ref{q-phi-def}), $\phi_{\mathbf{u}}(P)$ is the multivalued differential in (\ref{Primary-e}) and
$$
\phi_{\widehat{\mathbf{u}}}(P):=\oint_b\lambda(P)W(P,Q),\quad \quad u=\oint_a\lambda(P)dz_P, \quad \quad \widehat{u}=\oint_b\lambda(P)dz_P.
$$
Apart from  the multivalued $q$-differential $\phi_{q,\xi}$, the $q$-differential (\ref{q-Primary-e}) have the same appearance of Dubrovin's primary differentials (\ref{Primary-e}), but with respect to the $q$-bidifferential $W_q(P,Q)$ (\ref{q-W-g1}). In addition, from \cite{Rejeb23}, all the $N$  primary $q$-differentials in (\ref{q-Primary-e}) are also quasi-homogeneous and  their  degrees of quasi-homogeneity occur in a  manner similar to (\ref{Int-rep-W})-(\ref{degrees})  and represent the power of $\lambda(Q)$ in their integral representation formulas relative to the $q$-bidifferential $W_q(P,Q)$. \\
Now, let us assume that $\omega_{q,0}$ is one of previously mentioned  primary $q$-differentials (\ref{q-Primary-e}) of quasi-homogeneous degree $d_0\geq 0$. Then from \cite{Vasilisa2} and as in (\ref{eta-def}), we know that the formula
\begin{equation}\label{eta-def-q}
\eta(\omega_{q,0}):=\frac{1}{2}\sum_{j}\big(\omega_{q,0}(P_j)\big)^2(d\lambda_j)^2
\end{equation}
defines a Darboux-Egoroff metric (which is the metric of the Frobenius manifold) on the open set $\mathcal{M}_{\omega_{q,0}}$ of equivalence classes $[(\C/(\Z+\tau\Z), \lambda)]\in \mathcal{H}_1(n_0,\dots,n_m)$  such that
$$
1+q\tau\neq 0\quad \quad \text{and}\quad \quad \omega_{q,0}(P_j):=\frac{\omega_{q,0}(P)}{d\mathbf{x}_j(P)}\Big|_{P=P_j}\neq 0,\quad \quad \forall\ j,
$$
where $P_1,\dots,P_N$ are the simple ramification points of the covering $\big(\C/(\Z+\tau\Z), \lambda\big)$ and $\mathbf{x}_j(P)=\sqrt{\lambda(P)-\lambda_j}$ is the standard local parameter near $P_j$. Note that, since the genus one $q$-bidifferential $W_q(P,Q)$ (\ref{q-W-g1}) turns out to satisfy variational formulas which look like those for the bidifferential (\ref{Rauch}), the rotation coefficients $\beta_{ij}(q)$ of the metric $\eta(\omega_{q,0})$ are of the form:
$$
\beta_{ij}(q)=\frac{1}{2}W_q(P_i,P_j):=\frac{1}{2}\frac{W_q(P,Q)}{d\mathbf{x}_i(P)d\mathbf{x}_j(Q)}\bigg|_{P=P_i,Q=P_j},\quad \quad i\neq j.
$$
In addition, the following set is a system of flat coordinates (satisfying analogous duality relations as (\ref{duality-picture})) for the metric $\eta(\omega_{q,0})$:
$$
\Big\{\mathbf{t}^{i,\alpha}(\omega_{q,0}), \mathbf{v}^i(\omega_{q,0}), \mathbf{s}^i(\omega_{q,0}), \mathbf{r}(\omega_{q,0}), \xi(\omega_{q,0})\Big\},
$$
where $\mathbf{t}^{i,\alpha}$, $\mathbf{v}^i$, $\mathbf{s}^i$, $\mathbf{r}$ are the same operations defined by (\ref{Flat basis}) and $\xi$ is the following one
\begin{equation}\label{xi-operation}
\xi(\omega_{q,0}):=\oint_a\lambda(P)\omega_{q,0}(P)+q\oint_b\lambda(P)\omega_{q,0}(P)
\end{equation}
which plays the role of the last flat coordinate $\mathbf{u}(\omega_0)$ in (\ref{Flat basis}) and in particular, the coordinates $\xi(\omega_{q,0})$ and $\mathbf{r}(\omega_{q,0})$ are dual to each other with respect to the metric $\eta(\omega_{q,0})$, where by this we mean (as above) that
$$
\eta(\omega_{q,0})\big(\partial_{\mathbf{t}^A(\omega_{q,0})}, \partial_{\xi(\omega_{q,0})}\big)=\delta_{\mathbf{t}^A,\mathbf{r}}, \quad \quad \forall\ \mathbf{t}^A.
$$

\begin{description}
\item[$\bullet$]\emph{Prepotentials of deformed semi-simple Hurwitz-Frobenius manifolds:}
\end{description}
According to Theorem 5.1 in \cite{Rejeb23}, the following function depending on the parameter $q\in \C$ and looking exactly as (\ref{Prepotential}) is the prepotential of the deformed  Hurwitz-Frobenius manifold structure determined by  the primary and quasi-homogeneous $q$-differential $\omega_{q,0}$:
\begin{equation}\label{q-Prepotential}
\begin{split}
&\mathbf{F}_{\omega_{q,0}}=\frac{1}{2(1+d_0)}\sum_{A,B}\frac{\big((d_0+d_A)\mathbf{t}^A(\omega_{q,0})+\rho_{\omega_{q,0},\phi_{q,\mathbf{t}^A}}\big)
\big((d_0+d_B)\mathbf{t}^B(\omega_{q,0})+\rho_{\omega_{q,0},\phi_{q,\mathbf{t}^B}}\big)}{1+d_0+d_{A^{\prime}}}\mathbf{t}^{A^{\prime}}(\phi_{q,\mathbf{t}^{B^{\prime}}})\\
&\quad +\frac{\log(-1)}{4}\sum_{i,j=1, i\neq j}^m\mathbf{v}^i(\omega_{q,0})\mathbf{v}^j(\omega_{q,0})
-\frac{3}{4(1+d_0)}\sum_{i,j=1}^m\bigg(\frac{1}{n_0+1}+\delta_{ij}\frac{1}{n_i+1}\bigg)\mathbf{v}^i(\omega_{q,0})\mathbf{v}^j(\omega_{q,0}),
\end{split}
\end{equation}
where $d_0=deg\big(\omega_{q,0}\big)$, $d_A=deg\big(\phi_{q,\mathbf{t}^A}\big)$, $d_{A^{\prime}}=deg\big(\phi_{\mathbf{t}^{A^{\prime}}}\big)$,  $\mathbf{t}^{A^{\prime}}$ is the dual coordinate of $t^A$ in according with the following relations
\begin{align*}
\mathbf{t}^{({i,\alpha})^{\prime}}=\mathbf{t}^{i,n_i+1-\alpha},\quad \quad \mathbf{v}^{{i}^{\prime}}=\mathbf{s}^i,\quad \quad \mathbf{s}^{{i}^{\prime}}=\mathbf{v}^i,
\quad \quad \mathbf{r}^{\prime}=\xi,\quad \quad \xi^{\prime}=\mathbf{r}
\end{align*}
 and $\rho_{\omega_{q,0},\phi_{q,\mathbf{t}^A}}$ is the $q$-analogue of the constant (\ref{r-AB}) determined by:
\begin{equation*}\label{q-r-AB}
\rho_{\omega_{q,0},\phi_{q,\mathbf{t}^A}}=\sum_{i,j=1}^m\delta_{\omega_{q,0},\phi_{q,\mathbf{s}^i}}\delta_{\phi_{q,\mathbf{t}^A},\phi_{q,\mathbf{s}^j}}
\Big(\frac{1}{n_0+1}+\delta_{ij}\frac{1}{n_i+1}\Big)=\rho_{\phi_{q,\mathbf{t}^A},\omega_{q,0}}.
\end{equation*}

\begin{description}
\item[$\bullet$]\emph{WDVV solution associated with the normalized holomorphic $q$-differential}
\end{description}
Let us now turn our attention to the prepotential of the  Hurwitz-Frobenius manifold $\mathcal{M}_{\phi_q}\subset \mathcal{H}_1(n_0,\dots,n_m)$ determined by the $q$-deformed holomorphic differential $\phi_q$ (\ref{q-phi-def}). Here, we mention that the differential $\phi_q$ has no zero and then  the open set $\mathcal{M}_{\phi_q}$ is only determined by condition (\ref{q-condition}). Moreover, the two flat metrics $\eta(\phi)$ (\ref{eta-def}) and $\eta(\phi_q)$ (\ref{eta-def-q}) corresponding respectively  to the usual normalized holomorphic differential $\phi(P)=dz_P$ and to its $q$-deformation $\phi_q(P)=dz_P/(1+q\tau)$ (\ref{q-phi-def}) are conformally related to each other, in the sense that
\begin{equation}\label{eta-eta}
\eta(\phi_q)=\frac{1}{(1+q\tau)^2}\eta(\phi).
\end{equation}
Moreover, as a local  system of flat coordinates for the flat metric $\eta(\phi_q)$, we can take the following one which is closely linked to that  (\ref{FC-phi2}) for the
metric $\eta(\phi)$:
 \begin{equation}\label{FC-phi-q}
\begin{split}
&x_{q,\alpha}(j)=\frac{1}{\sqrt{n_j+1}}\mathbf{t}^{j,\alpha}(\phi_q):=\frac{1}{\alpha} \underset{\infty^j}{{\rm res}}\ \lambda(P)^{\frac{\alpha}{n_j+1}}\phi_q(P)=\frac{x_{\alpha}(j)}{1+q\tau},\quad
\begin{array}{ll}
&j=0,\dots,m,\\
&\alpha=1,\dots,n_j;
\end{array}\\
&x_{q,n_j+1}(j)=(n_j+1)\mathbf{v}^j(\phi)=\underset{\infty^j}{{\rm res}\ }\lambda(P)\phi_q(P)
=\frac{x_{n_j+1}(j)}{1+q\tau},\hspace{1.8cm} j=1,\dots,m;\\
&s_q^j:=\mathbf{s}^j(\phi_q):=\int_{\infty^0}^{\infty^i}\phi_q(P)=\frac{\infty^j-\infty^0}{1+q\tau}=\frac{s^j}{1+q\tau},\hspace{2.8cm} j=1,\dots,m;\\
&\tau_q:={2{\rm{i}}\pi}\mathbf{r}(\phi_q):=\oint_{b}\phi_q(P)=\frac{\tau}{1+q\tau};\\
&\xi_q:=\xi(\phi_q):=\oint_a\lambda(P)\phi_q(P)+q\oint_b\lambda(P)\phi_q(P).
\end{split}
\end{equation}

The latter coordinate $\xi_q$ is described in the upcoming lemma  as a function of flat coordinates (\ref{FC-phi2}) related to the holomorphic differential $\phi(P)=dz_P$.
\begin{Lemma} Let $\big(\C/(\Z+\tau\Z), \lambda\big)$ be the covering from Theorem \ref{LG}, with $\Im\tau>0$ and $1+q\tau\neq0$, $\phi$ and $\phi_q$ be respectively the usual normalized holomorphic differential (\ref{phi-def}) and its $q$-deformation (\ref{q-phi-def}).
Then, with the notation of  Theorem \ref{LG}, we have
\begin{equation}\label{xi-phi-q}
\begin{split}
\xi_q&:=\oint_a\lambda(P)\phi_q(P)+q\oint_b\lambda(P)\phi_q(P)\\
&=u+\frac{{2{\rm i}\pi}q}{1+q\tau}\sum_{j=1}^m(n_j+1)x_{n_j+1}(j)s^j+\frac{{{\rm i}\pi}q}{1+q\tau}\sum_{j=0}^m(n_j+1)\mathcal{B}_{n_j+1,2}(j),
\end{split}
\end{equation}
where $\mathcal{B}_{n_j+1,2}(j):=\big(1-\delta_{n_j,0}\big)\sum_{\beta=1}^{n_j}(n_j+1)x_{\beta}(j)x_{n_j+1-\beta}(j)$, $s^j=\infty^j-\infty^0$ and $u:=\oint_a\lambda(P)\phi(P)$.
Conversely, the $a$-period $u$ of the Abelian differential $\lambda(P)\phi(P)$ can be expressed in terms of  flat coordinates (\ref{FC-phi-q}) as follows:
\begin{equation}\label{u-xi-q}
\begin{split}
&u:=\oint_a\lambda(P)\phi(P)=\xi_q-\frac{{2{\rm i}\pi}q}{1-q\tau_q}\sum_{j=1}^m(n_j+1)x_{q,n_j+1}(j)s_q^j-\frac{{{\rm i}\pi}q}{1-q\tau_q}
\sum_{j=0}^m(n_j+1)\mathcal{B}_{n_j+1,2}^q(j),
\end{split}
\end{equation}
with $\mathcal{B}_{n_j+1,2}^q(j):=\big(1-\delta_{n_j,0}\big)\sum_{\beta=1}^{n_j}x_{q,\beta}(j)x_{q,n_j+1-\beta}(j)$.
\end{Lemma}
\emph{Proof:} Using expression (\ref{q-phi-def}) for the holomorphic $q$-differential $\phi_q$ along  with formulas (\ref{u-a}) and (\ref{u-b})  that provide  the periods $u$ and $\widehat{u}$ of the differential $\lambda(P)\phi(P)$, we deduce that
\begin{align*}
\xi_q&:=\oint_a\lambda(P)\phi_q(P)+q\oint_b\lambda(P)\phi_q(P)=\frac{u+q\widehat{u}}{1+q\tau}\\
&=c-2\Big(\zeta(1/2)-{{\rm i}\pi}\frac{q}{1+q\tau}\Big)\sum_{j=1}^m(n_j+1)x_{n_j+1}(j)s^j
-\Big(\zeta(1/2)-{{\rm i}\pi}\frac{q}{1+q\tau}\Big)\sum_{j=0}^m)(n_j+1)\mathcal{B}_{n_j+1,2}(j)\\
&=u+\frac{{2{\rm i}\pi}q}{1+q\tau}\sum_{j=1}^m(n_j+1)x_{n_j+1}(j)s^j+\frac{{{\rm i}\pi}q}{1+q\tau}\sum_{j=0}^m(n_j+1)\mathcal{B}_{n_j+1,2}(j),
\end{align*}
where in the last equality we have applied (\ref{u-a}) once again to rewrite $c$ in terms of variables $\{u,\tau,s^j,x_{\alpha}(j)\}$ corresponding to the holomorphic differential $\phi(P)=dz_P$. On the hand, since the partial Bell polynomial $\mathcal{B}_{n_j+1,2}$ is homogeneous of degree $2$, equality (\ref{u-xi-q}) follows from (\ref{FC-phi-q}) and (\ref{xi-phi-q}).

\fd

\medskip

Note that, formulas (\ref{FC-phi-q}), (\ref{xi-phi-q}) and (\ref{u-xi-q}) define a change of coordinates transformation $\mathbf{T}_q$ from  $\eta(\phi)$-flat coordinates (\ref{FC-phi2}) to their deformations (\ref{FC-phi-q}), where $\mathbf{T}_q$ is given by
\begin{equation}\label{T-q}
\begin{split}
&\mathbf{T}_q\Big(\tau, x_1(0),\dots,x_{n_0}(0),x_1(j),\dots,x_{n_j+1}(j),s^j,u\Big)\\
&=\frac{1}{1+q\tau}\bigg(\tau, x_1(0),\ \dots, x_{n_0}(0), x_1(j),\ \dots, x_{n_j+1}(j), s^j, (1+q\tau)u\bigg)\\
&\quad +\frac{{{\rm i}\pi}q}{1+q\tau}\bigg(0, \ \dots\, 0, 2\sum_{j=1}^m(n_j+1)x_{n_j+1}(j)s^j+\sum_{j=0}^m(n_j+1)\mathcal{B}_{n_j+1,2}(j)\bigg)\\
&=\Big(\tau_q, x_{q,1}(0),\dots,x_{q,n_0}(0),x_{q,1}(j),\dots,x_{q,n_j+1}(j),s_q^j,\xi_q\Big)
\end{split}
\end{equation}
and its inverse is such that  $\mathbf{T}_q^{-1}=\mathbf{T}_{-q}$. \\

Our main result here states that the WDVV solution  $\mathbf{F}_{\phi_q}$ corresponding to the holomorphic $q$-differential $\phi_q$ can be conveniently described  in terms of its non-deformed counterpart  $\mathbf{F}_{\phi}$, previously  obtained in Theorem \ref{Main result} and  the following homogeneous polynomial function
\begin{equation}\label{Q-function}
\begin{split}
&\mathbf{Q}=\mathbf{Q}\big(\tau,x_1(0),\dots,x_{n_0}(0),x_1(j),\dots,x_{n_j+1}(j), s^j,u\big)\\
&:=\frac{u^2}{4{\rm i}\pi}\tau+u\sum_{j=1}^{m}(n_j+1)x_{n_j+1}(j)s^j
+\frac{u}{2}\sum_{i=0}^{m}\sum_{\alpha=1}^{n_i}\frac{(n_i+1-\alpha)(2n_i+2+\alpha)}{(n_i+1+\alpha)}x_{\alpha}(i)x_{n_i+1-\alpha}(i)
\end{split}
\end{equation}
which is nothing but the first line in the expression for the prepotential $\mathbf{F}_{\phi}$ (see Theorem \ref{Main result}).

\begin{Thm}\label{Main result-q} Consider the deformed semi-simple Frobenius manifold $\mathcal{M}_{\phi_q}\subset \mathcal{H}_1(n_0,\dots,n_m)$ determined by \\
- the flat metric $\eta(\phi_q)$ induced by the holomorphic $q$-differential $\phi_q$ (\ref{q-phi-def}) and given by (\ref{eta-def-q}) and (\ref{eta-eta});\\
- the system $\Big\{\tau_q,x_{q,1}(0),\dots,x_{q,n_0}(0),x_{q,1}(j),\dots,x_{q,n_j+1}(j), s_q^j,\xi_q\Big\}$ of flat coordinates (\ref{FC-phi-q}) of the flat metric $\eta(\phi_q)$;\\
- the unit vector field $e=\partial_{\xi_q}$;\\
- the Euler vector field as the $q$-analogue of (\ref{E-phi}):
\begin{equation}\label{E-phi-q}
E_q=\xi_q\partial_{\xi_q}+\sum_{j=1}^mx_{q,n_j+1}(j)\partial_{x_{q,n_j+1}(j)}
+\sum_{j=0}^m\sum_{\alpha=1}^{n_j}\frac{\alpha}{n_j+1}x_{q,\alpha}(j)\partial_{x_{q,\alpha}(j)}.
\end{equation}
Then the WDVV solution associated with the  deformed $\phi_q$-Frobenius manifold structure is as follows:
\begin{equation}\label{Prepo-phi-q}
\begin{split}
\mathbf{F}_{\phi_q}&=\frac{\xi_q^2}{4{\rm i}\pi}\tau_q+\xi_q\sum_{j=1}^{m}(n_j+1)x_{q,n_j+1}(j)s_q^j
+\frac{\xi_q}{2}\sum_{i=0}^{m}\sum_{\alpha=1}^{n_i}\frac{(n_i+1-\alpha)(2n_i+2+\alpha)}{(n_i+1+\alpha)}x_{q,\alpha}(i)x_{q,n_i+1-\alpha}(i)\\
&+(1-q\tau_q)^2\mathbf{F}_{\phi}\circ \mathbf{T}_{-q}\big(\tau_q,x_{q,1}(0),\dots,x_{q,n_0}(0),x_{q,1}(j),\dots,x_{q,n_j+1}(j), s_q^j,\xi_q\big)\\
&-(1-q\tau)^2\mathbf{Q}\circ \mathbf{T}_{-q}\big(\tau_q,x_{q,1}(0),\dots,x_{q,n_0}(0),x_{q,1}(j),\dots,x_{q,n_j+1}(j), s_q^j,\xi_q\big)\\
&-\frac{{{\rm i}\pi}q}{1-q\tau_q}\bigg(\sum_{j=1}^m(n_j+1)x_{q,n_j+1}(j)s_q^j\bigg)^2\\
&-\frac{{{\rm i}\pi}q}{1-q\tau_q}\bigg(\sum_{j=1}^m(n_j+1)x_{q,n_j+1}(j)s_q^j\bigg)
\bigg(\sum_{i=0}^{m}\sum_{\alpha=1}^{n_i}\frac{(n_i+1-\alpha)(2n_i+2+\alpha)}{n_i+1+\alpha}x_{q,n_i+1-\alpha}(i)x_{q,\alpha}(i)\bigg)\\
&-\frac{{{\rm i}\pi}q}{1-q\tau_q}\sum_{i,j=0}^{m}\sum_{\alpha=1}^{n_i}\sum_{\beta=1}^{n_j}\frac{(n_i+1-\alpha)(n_j+1-\beta)}{(n_i+1+\alpha)}
(n_i+1)x_{q,\alpha}(i)x_{q,\beta}(j)x_{q,n_i+1-\alpha}(i)x_{q,n_j+1-\beta}(j),
\end{split}
\end{equation}
where $\mathbf{F}_{\phi}$ is the WDVV prepotential from Theorem \ref{Main result} induced by the usual normalized holomorphic differential, $\mathbf{T}_{-q}$ is the inverse of the map $\mathbf{T}_q$ defined by (\ref{T-q}) and $\mathbf{Q}$ is the polynomial function given  by (\ref{Q-function}).\\
The function $\mathbf{F}_{\phi_q}$ is quasi-homogeneous of degree 2 with respect to Euler's vector field (\ref{E-phi-q}):
$$
E_q.\mathbf{F}_{\phi_q}=2\mathbf{F}_{\phi,q}
$$
Furthermore,  the Hessian matrix of $\partial_{\xi_q}\mathbf{F}_{\phi_q}$ coincides the Gram constant matrix of the metric $\eta(\phi_q)$ (\ref{eta-eta}) in flat coordinates (\ref{FC-phi-q}).
\end{Thm}
\emph{Proof:} Making use of formula (\ref{q-Prepotential}) and reasoning as in the proof of Proposition \ref{Preliminary prep}, we arrive at the following expression for the prepotential $\mathbf{F}_{\phi_q}$ which bears a resemblance to (\ref{Prep-phi1}):
\begin{align*}
&\mathbf{F}_{\phi_q}=\frac{\xi_q^2}{4{\rm i}\pi}\tau_q
+\xi_q\sum_{j=1}^{m}(n_j+1)x_{q,n_j+1}(j)s_q^j+\frac{\xi_q}{2}\sum_{i=0}^{m}\sum_{\alpha=1}^{n_i}\frac{(n_i+1-\alpha)(2n_i+2+\alpha)}{(n_i+1+\alpha)}
x_{q,\alpha}(i)x_{q,n_i+1-\alpha}(i)\\
&+\frac{1}{2}\sum_{i=1}^{m}\sum_{j=1}^m(n_i+1)(n_j+1)x_{q,n_i+1}(i)x_{q,n_j+1}(j)\mathbf{s}^i\big(\Omega_{q,\infty^0\infty^j})\\
&+\frac{1}{2}\sum_{j=1}^m\sum_{i=0}^{m}\sum_{\alpha=1}^{n_i}\left(\frac{n_i+1-\alpha}{n_i+1+\alpha}+\frac{n_i+1-\alpha}{n_i+1}\right)
(n_i+1)(n_j+1)x_{q,n_i+1-\alpha}(i)x_{q,n_j+1}(j)\Big(\frac{1}{\alpha}\mathcal{I}_{i,\alpha}[\Omega_{q,\infty^0\infty^j}]\Big)\\
&+\frac{1}{2}\sum_{i,j=0}^{m}\sum_{\alpha=1}^{n_i}\sum_{\beta=1}^{n_j}\frac{(n_i+1-\alpha)(n_j+1-\beta)}{(n_i+1+\alpha)}
(n_i+1)x_{q,n_i+1-\alpha}(i)x_{q,n_j+1-\beta}(j)\Big(\frac{1}{\alpha\beta}\mathcal{I}_{i,\alpha}\big[\Psi_{q,j,\beta}\big]\Big)\\
&+\frac{\log(-1)}{4}\sum_{i,j=1, i\neq j}^m(n_i+1)(n_j+1)x_{q,n_i+1}(i)x_{q,n_j+1}(j)\\
&-\frac{3}{4}\sum_{i,j=1}^m\bigg(\frac{(n_i+1)(n_j+1)}{n_0+1}+\delta_{ij}(n_j+1)\bigg)x_{q,n_i+1}(i)x_{q,n_j+1}(j),
\end{align*}
where $\mathbf{s}^i\big(\Omega_{q,\infty^0\infty^j})$, $\mathcal{I}_{i,\alpha}[\Omega_{q,\infty^0\infty^j}]$ and $\mathcal{I}_{i,\alpha}\big[\Psi_{q,j,\beta}\big]$ are respectively given by (\ref{q-s-ij}), (\ref{q-Third-alpha-ij}) and (\ref{q-chi-ij}). \\
Now, due to the change of variables (\ref{FC-phi-q}), we easily see that  formulas (\ref{q-s-ij}), (\ref{q-Third-alpha-ij}) and (\ref{q-chi-ij}) can be rewritten as
\begin{align*}
&\mathbf{s}^i\big(\Omega_{q,\infty^0\infty^j}\big)=\mathbf{s}^i\big(\Omega_{\infty^0\infty^j}\big)-\frac{{2{\rm i}\pi}q}{1-q\tau_q}s_q^is_q^j,
\quad\quad \quad  \quad i,j=1,\dots,m;\\
&\mathcal{I}_{i,\alpha}[\Omega_{q,\infty^0\infty^j}]=\mathcal{I}_{i,\alpha}[\Omega_{\infty^0\infty^j}]-\frac{{2{\rm i}\pi}q}{1-q\tau_q}{\alpha}x_{q,\alpha}(i)s_q^j,
\quad\quad  \quad i,j=0,1,\dots,m\quad \text{and}\quad j\neq 0;\\
&\mathcal{I}_{i,\alpha}[\Psi_{q,j,\beta}]=\mathcal{I}_{i,\alpha}[\Psi_{j,\beta}]-\frac{{2{\rm i}\pi}q}{1-q\tau_q}\alpha\beta x_{q,\alpha}(i)x_{q,\beta}(j),
\quad\quad  \quad i,j=0,1,\dots,m.
\end{align*}
These relations together with the equality  $x_{q,\alpha}(j)=x_{\alpha}(j)/(1+q\tau)$ (\ref{q-evaluation}) bring us to the following expression involving the prepotential $\mathbf{F}_{\phi}$ (\ref{Prep-phi1}) relative to the usual holomorphic differential $\phi$:
\begin{align*}
&\mathbf{F}_{\phi_q}=\frac{\xi_q^2}{4{\rm i}\pi}\tau_q
+\xi_q\sum_{j=1}^{m}(n_j+1)x_{q,n_j+1}(j)s_q^j+\frac{\xi_q}{2}\sum_{i=0}^{m}\sum_{\alpha=1}^{n_i}\frac{(n_i+1-\alpha)(2n_i+2+\alpha)}{(n_i+1+\alpha)}
x_{q,\alpha}(i)x_{q,n_i+1-\alpha}(i)\\
&-\frac{1}{(1+q\tau)^2}\Bigg\{\frac{u^2}{4{\rm i}\pi}\tau
+u\sum_{j=1}^{m}(n_j+1)x_{n_j+1}(j)s^j+\frac{u}{2}\sum_{i=0}^{m}\sum_{\alpha=1}^{n_i}\frac{(n_i+1-\alpha)(2n_i+2+\alpha)}{(n_i+1+\alpha)}
x_{\alpha}(i)x_{n_i+1-\alpha}(i)\Bigg\}\\
&+\frac{1}{(1+q\tau)^2}\mathbf{F}_{\phi}\big(\tau,x_1(0),\dots,x_{n_0}(0),x_1(j),\dots,x_{n_j+1}(j), s^j,u\big)\\
&-\frac{{{\rm i}\pi}q}{1-q\tau_q}\sum_{i=1}^{m}\sum_{j=1}^m(n_i+1)(n_j+1)x_{q,n_i+1}(i)x_{q,n_j+1}(j)s_q^js_q^k\\
&-\frac{{{\rm i}\pi}q}{1-q\tau_q}\sum_{j=1}^m\sum_{i=0}^{m}\sum_{\alpha=1}^{n_i}\left(\frac{n_i+1-\alpha}{n_i+1+\alpha}+\frac{n_i+1-\alpha}{n_i+1}\right)
(n_i+1)(n_j+1)x_{q,n_i+1-\alpha}(i)x_{q,\alpha}(i)x_{q,n_j+1}(j)s_q^j\\
&-\frac{{{\rm i}\pi}q}{1-q\tau_q}\sum_{i,j=0}^{m}\sum_{\alpha=1}^{n_i}\sum_{\beta=1}^{n_j}\frac{(n_i+1-\alpha)(n_j+1-\beta)}{(n_i+1+\alpha)}
(n_i+1)x_{q,\alpha}(i)x_{q,\beta}(j)x_{q,n_i+1-\alpha}(i)x_{q,n_j+1-\beta}(j)\\
&=\frac{\xi_q^2}{4{\rm i}\pi}\tau_q+\xi_q\sum_{j=1}^{m}(n_j+1)x_{q,n_j+1}(j)s_q^j
+\frac{\xi_q}{2}\sum_{i=0}^{m}\sum_{\alpha=1}^{n_i}\frac{(n_i+1-\alpha)(2n_i+2+\alpha)}{(n_i+1+\alpha)}x_{q,\alpha}(i)x_{q,n_i+1-\alpha}(i)\\
&-(1-q\tau)^2\mathbf{Q}\circ \mathbf{T}_q^{-1}\big(\tau_q,x_{q,1}(0),\dots,x_{q,n_0}(0),x_{q,1}(j),\dots,x_{q,n_j+1}(j), s_q^j,\xi_q\big)\\
&+(1-q\tau_q)^2\mathbf{F}_{\phi}\circ \mathbf{T}_q^{-1}\big(\tau_q,x_{q,1}(0),\dots,x_{q,n_0}(0),x_{q,1}(j),\dots,x_{q,n_j+1}(j), s_q^j,\xi_q\big)\\
&-\frac{{{\rm i}\pi}q}{1-q\tau_q}\bigg(\sum_{j=1}^m(n_j+1)x_{q,n_j+1}(j)s_q^j\bigg)^2\\
&-\frac{{{\rm i}\pi}q}{1-q\tau_q}\bigg(\sum_{j=1}^m(n_j+1)x_{q,n_j+1}(j)s_q^j\bigg)
\bigg(\sum_{i=0}^{m}\sum_{\alpha=1}^{n_i}\frac{(n_i+1-\alpha)(2n_i+2+\alpha)}{n_i+1+\alpha}x_{q,n_i+1-\alpha}(i)x_{q,\alpha}(i)\bigg)\\
&-\frac{{{\rm i}\pi}q}{1-q\tau_q}\sum_{i,j=0}^{m}\sum_{\alpha=1}^{n_i}\sum_{\beta=1}^{n_j}\frac{(n_i+1-\alpha)(n_j+1-\beta)}{(n_i+1+\alpha)}
(n_i+1)x_{q,\alpha}(i)x_{q,\beta}(j)x_{q,n_i+1-\alpha}(i)x_{q,n_j+1-\beta}(j).
\end{align*}

\fd

\begin{description}
\item[$\bullet$] \emph{Deformed Eisenstein series and Weierstrass functions}
\end{description}
It would be more convenient during the rest of this section  to utilize the  $q$-deformed version of Eisenstein series and Weierstrass functions defined below.
In particular, formula (\ref{Prepo-phi-q}) from Theorem \ref{Main result-q} and these $q$-deformed special functions enable us  to present the deformed  WDVV solution $\mathbf{F}_{\phi_q}$ (\ref{Prepo-phi-q})  in a form closely resembling that of  the prepotential $\mathbf{F}_{\phi}$ from Theorem \ref{Main result}.
This  provides an efficient approach for producing  the $q$-analogues of explicit  WDVV solutions previously obtained in Section 4.\\
Let $\tau$ be as in (\ref{q-condition}). For $\ell\geq 2$, define the $q$-deformation of the Eisenstein series $G_{\ell}$ (\ref{Eisentein-Fourier}) by
\begin{equation}\label{G-q}
G_{q,\ell}(\tau_q):=(1-q\tau_q)^{-\ell}G_{\ell}\Big(\frac{\tau_q}{1-q\tau_q}\Big)-\delta_{\ell,2}\frac{{2{\rm i}\pi}q}{1-q\tau_q},
\quad \quad \tau_q:=\frac{\tau}{1+q\tau}.
\end{equation}
Note that the following $q$-analogue of Ramanujan's differential equations (\ref{Ramanujan-G}) are satisfied:
\begin{equation}\label{Ramanujan-G-q}
\forall\ m\geq 1,\quad \quad  \partial_{\tau_q}G_{q,2m}(\tau_q)=\frac{m}{2{\rm i}\pi}\bigg((2m+3)G_{q,2m+2}(\tau_q)-\sum_{\ell=0}^{m-1}G_{q,2\ell+2}(\tau_q)G_{q,2m-2\ell}(\tau_q)\bigg).
\end{equation}
Let $\sigma$ be the Weierstrass function  (\ref{sigma-function}). We shall  define the $q$-deformation of the function $\K(v|\tau):=\log\sigma(v|\tau)$ by
\begin{equation}\label{K-q}
\K_q(v)=\K_q(v|\tau_q):=\K\Big(\frac{v}{1-q\tau_q}\Big|\frac{\tau_q}{1-q\tau_q}\Big)+\log(1-q\tau_q),\quad \quad \tau_q:=\frac{\tau}{1+q\tau}.
\end{equation}
In particular, for any positive integer $k$ we have
\begin{equation}\label{K-q-derivative}
\K_q^{(k)}(v):=\partial_v^k\K_q(v|\tau_q)=\frac{1}{(1-q\tau_q)^k}\K^{(k)}\Big(\frac{v}{1-q\tau_q}\Big|\frac{\tau_q}{1-q\tau_q}\Big).
\end{equation}
Let us point out that  $q$-Eisenstein series appear in the Laurent expansion of the function $\K_q'(v)$  (which is the $q$-analogue of the Weierstrass zeta function) near $v=0$, in like manner as
(\ref{Laurent zeta}):
\begin{equation}\label{Laurent-Kq}
\K_q'(v)=\frac{1}{(1-q\tau_q)}\zeta\Big(\frac{v}{1-q\tau_q}\Big|\tau\Big)\underset{v\sim0}{=}\frac{1}{v}-\sum_{\ell=3}^{\infty}G_{q,\ell+1}(\tau_q)v^{\ell}.
\end{equation}
In addition,  when $q$ is an integer, the $b$-period $\tau_q=\tau/(1+q\tau)$ (\ref{q-phi-periods}) of the holomorphic $q$-differential $\phi_q$ has also a positive imaginary part. Thus, owing to the modularity properties of Eisenstein series  (\ref{G-modular})-(\ref{G2-modular}) and those  of Weierstrass functions (\ref{psz-modular}), formulas (\ref{G-q}) and (\ref{K-q}) read  respectively as
$$
G_{q,\ell}(\tau_q)=G_{\ell}(\tau_q)\quad \text{and}\quad \K_q(v|\tau_q)=\K(v|\tau_q).
$$

The next result presents an alternative  description of the LG superpotential (\ref{lambda-zeta}) from Theorem \ref{LG} involving deformed Eisenstein series and Weierstrass functions (\ref{G-q})-(\ref{K-q-derivative}) along with  flat coordinates (\ref{FC-phi-q}) associated with the $q$-differential $\phi_q$. Let us emphasize that the obtained $q$-version (\ref{lambda-q}) below  looks exactly as expression (\ref{lambda-zeta}) which is nothing but the case $q=0$ of  (\ref{lambda-q}).
\begin{Prop} Let $\mathbb{L}$ be  the  lattice $\mathbb{L}=\Z+\tau\Z$, with  $\Im\tau>0$ and $1+q\tau\neq 0$.
Let $\lambda$  be a meromorphic function on the torus $\C/{\mathbb{L}}$ given by (\ref{lambda-zeta}).
Then, with notation (\ref{G-q}), (\ref{K-q-derivative}) and (\ref{Bell-R-Lq}), the meromorphic function  $\lambda$ takes the following form:
\begin{equation}\label{lambda-q}
\begin{split}
\lambda(P)&=\xi_q+G_{q,2}(\tau_q)\sum_{j=1}^m(n_j+1)x_{q,n_j+1}(j)s_q^j+\frac{1}{2}G_{q,2}(\tau_q)\sum_{j=0}^m(n_j+1)\mathcal{B}^q_{n_j+1,2}(j)\\
&\quad -\sum_{j=0}^m(n_j+1)\left(\sum_{k=1}^{n_j+1}\frac{(-1)^k}{k!}\mathcal{B}_{n_j+1,k}^q(j)\big[\partial_v^k\K_q(\cdot|\tau_q)\big]
\big((1-q\tau_q)(z_P-\infty^j)\big)\right),
\end{split}
\end{equation}
where $\xi_q$ and $s_q^j$ are given by (\ref{FC-phi-q}) and (\ref{xi-phi-q}). Moreover, for any $j=0,\dots,m$, the function $\lambda$ (\ref{lambda-q}) has the following Laurent expansion  near its pole $\infty^j$ of order $n_j+1$:
\begin{equation}\label{lambda-q-Laurent}
\lambda(P)\underset{P\sim\infty^j}=(n_j+1)\sum_{k=1}^{n_j+1}\frac{\mathcal{B}^q_{n_j+1,k}(j)}{k(1-q\tau_q)^{k}(z_P-\infty^j)^k}
+\sum_{\ell=0}^{\infty}\mathbf{f}_{q,\ell}(j)(1-q\tau_q)^{\ell}(z_P-\infty^j)^{\ell},
\end{equation}
where $\mathbf{f}_{q,\ell}(j)$ is the quantity  defined by
\begin{equation}\label{f-function-q}
\begin{split}
\mathbf{f}_{q,\ell}(j)&=\delta_{\ell,0}\bigg(\xi_q+G_{q,2}(\tau_q)\sum_{j=1}^m(n_j+1)x_{q,n_j+1}(j)s_q^j
+\frac{1}{2}G_{q,2}(\tau_q)\sum_{j=0}^m(n_j+1)\mathcal{B}^q_{n_j+1,2}(j)\bigg)\\
& \quad -\frac{1}{\ell!}\sum_{i=0,i\neq j}^m(n_i+1)
\left(\sum_{k=1}^{n_i+1}\frac{(-1)^k}{k!}\mathcal{B}^q_{n_i+1,k}(i)\big[\partial_v^{k+\ell}\K_q(\cdot|\tau_q)\big](s_q^j-s_q^i)\right)\\
&\quad +(n_j+1)\sum_{k=1, k+\ell\geq 4}^{n_j+1}\frac{(-1)^k}{k+\ell}\binom{k+\ell}{\ell}\mathcal{B}^q_{n_j+1,k}(j)G_{q,k+\ell}(\tau_q).
\end{split}
\end{equation}
Here, according to convention (\ref{convention}), the term containing the sum $\sum_{j=1}^m$ in formulas (\ref{lambda-q}) and (\ref{f-function-q}) is omitted  for $m=0$. Similarly, we should understand the sum $\sum_{k=1, k+\ell\geq 4}^{n_j+1}$ in (\ref{f-function-q}).
\end{Prop}
\emph{Proof:} By (\ref{lambda-zeta}) and (\ref{u-a}) and the equalities  $\zeta(v)=\K'(v)$ and $2\zeta(1/2|\tau)=G_2(\tau)$ (\ref{G2-zeta}), we can write
\begin{align*}
\lambda(P)&=u+\big(1-\delta_{m,0}\big)G_2(\tau)\sum_{j=1}^m(n_j+1)x_{n_j+1}(j)s^j+\frac{1}{2}G_2(\tau)\sum_{j=0}^m(n_j+1)\mathcal{B}_{n_j+1,2}(j)\\
&\quad -\sum_{j=0}^m(n_j+1)\left(\sum_{k=1}^{n_j+1}\frac{(-1)^k}{k!}\mathcal{B}_{n_j+1,k}\big(x_1(j),\dots,x_{n_j+2-k}(j)\big)\K^{(k)}\big(z_P-\infty^j|\tau\big)\right)\\
&=\xi_q+\big(1-\delta_{m,0}\big)\bigg((1-q\tau)^{-2}G_2\Big(\frac{\tau_q}{1-q\tau_q}\Big)-\frac{{2{\rm i}\pi}q}{1-q\tau_q}\bigg)\sum_{j=1}^m(n_j+1)x_{q,n_j+1}(j)s_q^j\\
&\quad +\frac{1}{2}\bigg((1-q\tau)^{-2}G_2\Big(\frac{\tau_q}{1-q\tau_q}\Big)-\frac{{2{\rm i}\pi}q}{1-q\tau_q}\bigg)\sum_{j=0}^m(n_j+1)\mathcal{B}^q_{n_j+1,2}(j)\\
&\quad -\sum_{j=0}^m(n_j+1)\left(\sum_{k=1}^{n_j+1}\frac{(-1)^k}{k!(1-q\tau_q)^k}\mathcal{B}_{n_j+1,k}\big(x_{q,1}(j),\dots,x_{q,n_j+2-k}(j)\big)
\K^{(k)}\Big(z_P-\infty^j\Big|\frac{\tau_q}{1-q\tau_q}\Big)\right),
\end{align*}
where, in the second equality, we used the change of coordinates $\mathbf{T}_{-q}$ (\ref{T-q}) and the homogeneity property (\ref{Bell-homog}) of Bell polynomials.
Thus, due to relations (\ref{G-q}), (\ref{K-q-derivative}) and notation  (\ref{Bell-R-Lq}), we conclude that (\ref{lambda-q}) holds. To get the Laurent expansion (\ref{lambda-q-Laurent}), it suffices to use the $q$-deformation version (\ref{lambda-q}) of $\lambda(P)$ and the Laurent series for the derivatives of the function $\K_q(v)$ and proceed as in the proof of (\ref{lambda-Laurent}).

\fd

\medskip

Due to the homogeneity property (\ref{Bell-homog}) of partial Bell polynomial and the relation $x_{q,r}(j)=(1-q\tau_q)x_r(j)$ from (\ref{q-evaluation}), we have
$$
\mathcal{B}^q_{n_j+1,k}(j):=\mathcal{B}_{\mu,k}\big(x_{q,1}(j),\dots,x_{q,\mu-k+1}(j)\big)=(1-q\tau_q)^{k}\mathcal{B}_{n_j+1,k}(j).
$$
Therefore, the principal parts in the two Laurent series (\ref{lambda-q-Laurent}) and (\ref{lambda-Laurent})  coincide.
Now, the uniqueness of the Laurent series expansion for the LG superpotential $\lambda(P)$ implies the following link between the coefficients of their regular parts.
\begin{Cor} Let $\lambda$ be the meromorphic function on $\C/(\Z+\tau\Z)$, with  $\Im\tau>0$ and $1+q\tau\neq 0$ and given by formula (\ref{lambda-zeta}) and its $q$-analogue (\ref{lambda-q}). Let $j=0,\dots,m$ and  $\mathbf{f}_{\ell}(j)$ and $\mathbf{f}_{q,\ell}(j)$ be the coefficients defined respectively by (\ref{f-function}) and (\ref{f-function-q}) and appeared in the Laurent series (\ref{lambda-Laurent}) and (\ref{lambda-q-Laurent}) of $\lambda(P)$.
Then $\mathbf{f}_{\ell}(j)$ and $\mathbf{f}_{q,\ell}(j)$ are linked by
\begin{equation}\label{f-fq-link}
\forall\ \ell\geq 0,\quad \quad \mathbf{f}_{\ell}(j)=(1-q\tau)^{\ell}\mathbf{f}_{q,\ell}(j).
\end{equation}
\end{Cor}

As another consequence, we obtain the $q$-analogue of formula (\ref{x-alpha2}).
\begin{Cor} For $j=0,1,\dots,m$, let $\big\{rx_{q,r}(j)\big\}_r$ be the coefficients of Taylor's series (\ref{q-phi-Taylor}) of the holomorphic $q$-differential $\phi_q$ (\ref{q-phi-def})  near the pole $\infty^j$ of $\lambda(P)$ (\ref{lambda-q}). Then for any positive integer $\alpha$, we have
\begin{equation}\label{q-x-alpha2}
\begin{split}
x_{q,n_j+1+\alpha}(j)&=\frac{\alpha}{n_j+1}\sum_{k=1}^{n_j+\alpha-1}\frac{1}{k+1}\binom{-n_j-1}{k}\big(x_{q,1}(j)\big)^{-k}
\mathcal{B}_{n_j+\alpha,k+1}\big(x_{q,2}(j),\dots,x_{q,n_j+\alpha-k+1}(j)\big)\\
&\quad +\alpha\sum_{k=1}^{n_j}\frac{1}{k}\mathcal{B}^q_{n_j+1,k}(j)\mathcal{R}^q_{\alpha+k-1,k}(j)
+\frac{\alpha}{n_j+1}\sum_{\ell=0}^{\alpha-1}\frac{1}{\ell+1}\mathbf{f}_{q,\ell}(j)\mathcal{B}^q_{\alpha,\ell+1}(j),
\end{split}
\end{equation}
where $\mathcal{B}^q_{n,k}(j)$ and $\mathcal{R}^q_{\mu,k}(j)$ are defined (\ref{Bell-R-Lq}) and $\mathbf{f}_{q,\ell}(j)$  is given  by (\ref{f-function-q}).
\end{Cor}
\emph{Proof:} From (\ref{q-evaluation}), we know that
$$
x_{q,\mu}(j)=(1-q\tau_q)x_{\mu}(j),\quad \quad \forall\ \mu\geq 1.
$$
This and the homogeneity properties (\ref{Bell-homog}) and (\ref{R-homog}) of the functions $\mathcal{B}_{\mu,k}$ and $\mathcal{R}_{\mu,k}$  imply that
\begin{equation}\label{Bell-R-q2}
\begin{split}
&\mathcal{B}_{\mu,k}(j):=\mathcal{B}_{\mu,k}\big(x_1(j),\dots,x_{\mu-k+1}(j)\big)=(1-q\tau_q)^{-k}\mathcal{B}_{\mu,k}^q(j);\\
&\mathcal{R}_{\mu,k}(j):=\mathcal{R}_{\mu,k}\big(x_1(j),\dots,x_{\mu+1}(j)\big)=(1-q\tau_q)^{k-1}\mathcal{R}_{\mu,k}^q(j).
\end{split}
\end{equation}
Therefore, we get (\ref{q-x-alpha2}) by using its  non-deformed case (\ref{x-alpha2}), relations (\ref{Bell-R-q2}) as well as the link (\ref{f-fq-link}).

\fd

\begin{Remark}
\end{Remark}
$\bullet$ Let $\mathbf{L}_{j,\alpha}^{\phi}$ and $\mathbf{L}_{j,\alpha}^{\phi_q}$ be respectively the Bell operators (\ref{Bell operator}) and (\ref{Bell-R-Lq}) related to the holomorphic differentials $\phi$ (\ref{phi-def}) and $\phi_q$ (\ref{q-phi-def}), with $j=0,\dots,m$ and $\alpha$ a positive integer.  Note that (\ref{K-q-derivative}) and (\ref{Bell-R-q2}) imply that
\begin{equation}\label{Bell-q}
\begin{split}
\mathbf{L}_{j,\alpha}^{\phi}\big[\K^{(k)}(\cdot|\tau)\big]\Big(\frac{v}{1-q\tau_q}\Big)
&=\sum_{r=1}^{\alpha}\frac{1}{r!}\frac{\mathcal{B}^q_{\alpha,r}(j)}{(1-q\tau_q)^r}\K^{(k+r)}\Big(\frac{v}{1-q\tau_q}\Big|\frac{\tau_q}{1-q\tau_q}\Big)\\
&=(1-q\tau_q)^k\mathbf{L}_{j,\alpha}^{\phi_q}\big[\K_q^{(k)}(\cdot|\tau_q)](v),
\end{split}
\end{equation}
where $\K:=\log\sigma$ is the Weierstrass function and $\K_q$ is its $q$-deformation (\ref{K-q}). \\
$\bullet$ Applying formula (\ref{Prepotential}), we can see that to write the $q$-analogue of the general WDVV solution $\mathbf{F}_{\phi}$ from Theorem \ref{Main result}, it suffices to replace $\phi$-flat coordinates, Eisenstein series, the Weierstrass function $\K(v)=\log\sigma(v)$ and its derivatives, partial Bell  polynomials $\mathcal{B}_{\mu,k}(j)$, the rational function $\mathcal{R}_{\mu,k}(j)$ and Bell operators $\mathbf{L}^{\phi}_{j,\alpha}\big[\K^{(k)}(\cdot|\tau)\big]$ by their $q$-deformations listed respectively in (\ref{FC-phi-q}), (\ref{G-q}), (\ref{K-q})-(\ref{K-q-derivative}),  (\ref{Bell-R-q2}) and (\ref{Bell-q}).\\
This will be illustrated in the next example.
\begin{Example}
\end{Example}
We shall discuss the already described WDVV solution (\ref{Prepo-phi-q}) when the considered Hurwitz space is of type $\mathcal{H}_1(n,0)$, with $n\in \N$. In other words, we are going to give  the $q$-analogue of the WDVV prepotential  $\widetilde{\mathbf{F}}_{\phi,n}$ (\ref{Jacobi-A-An}), denoted by $\widetilde{\mathbf{F}}_{\phi_q,n}$.\\
According to the LG superpotential given  by (\ref{Jacobi-A-superp}) and the notation used there, we deduce that the $q$-deformed flat coordinates (\ref{FC-phi-q})-(\ref{xi-phi-q}) are
\begin{align*}
&\tau_q:=\oint_b\phi_q=\frac{\tau}{1+q\tau};\\
&x_{q,k}:=\frac{1}{k}\underset{\infty^0} {\rm res}\lambda(P)^{\frac{k}{n+1}}\phi_q(P)=\frac{x_k}{1+q\tau}, \quad k=1,\dots,n;\\
&y_q:=\underset{\infty^1} {\rm res}\lambda(P)\phi_q(P)=\frac{y}{1+q\tau};\\
&s_q:=\int_{\infty^0}^{\infty^1}\phi_q=\frac{s}{1+q\tau};\\
&\xi_q:=\oint_a\lambda(P)\phi_q(P)+q\oint_b\lambda(P)\phi_q(P)\\
&\quad =u+\frac{{2{\rm i}\pi}q}{1+q\tau}ys+\frac{{{\rm i}\pi}q}{1+q\tau}\big(1-\delta_{n,0}\big)(n+1)\sum_{\beta=1}^{n}x_{\beta}x_{n+1-\beta}.
\end{align*}
In particular, the inverse of the map  (\ref{T-q}) reduces to
\begin{align*}
&\mathbf{T}_{-q}\Big(\tau_q, x_{q,1},\dots,x_{q,n},y_q,s_q,\xi_q\Big)\\
&=\frac{1}{1-q\tau_q}\Big(\tau_q, x_{q,1},\dots,x_{q,n},y_q,s_q,(1-q\tau_q)\xi_q-2{{\rm i}\pi}qy_qs_q-{{\rm i}\pi}q(n+1)\mathcal{B}_{n+1,2}^q\Big),
\end{align*}
with $\mathcal{B}_{n+1,2}^q:=\big(1-\delta_{n,0}\big)\sum_{k=1}^{n}x_{q,k}x_{q,n+1-k}$. Moreover, by (\ref{lambda-q}), the $q$-deformation of the LG superpotential (\ref{Jacobi-A-superp}) reads as follows
\begin{equation}\label{Jacobi-superq}
\begin{split}
\lambda(P)&=\xi_q+y_qs_qG_{q,2}(\tau_q)+\frac{n+1}{2}\mathcal{B}_{n+1,2}^qG_{q,2}(\tau_q)\\
&-(n+1)\sum_{k=1}^{n+1}\frac{(-1)^k}{k!}\mathcal{B}_{n+1,k}^q\big[\partial_v^k\K_q(\cdot|\tau_q)\big]\big((1-q\tau_q)(z_P-\infty^j)\big),
\end{split}
\end{equation}
where $\mathcal{B}_{n+1,k}^q=\mathcal{B}_{n+1,k}(x_{q,1},\dots,x_{q,n+2-k})$ and $\K_q$ is defined by (\ref{K-q}).\\
On the other hand, formula (\ref{Prepo-phi-q}) implies that the $q$-WDVV solution $\widetilde{\mathbf{F}}_{\phi_q,n}$ has the following form involving the function
$\widetilde{\mathbf{F}}_{\phi,n}$ (\ref{Jacobi-A-An}):
\begin{align*}
&\widetilde{\mathbf{F}}_{\phi_q,n}\Big(\tau_q, x_{q,1},\dots,x_{q,n},y_q,s_q,\xi_q\Big)\\
&=\frac{\xi_q^2}{4{\rm i}\pi}\tau_q+\xi_qy_qs_q
+\frac{\xi_q}{2}\sum_{\alpha=1}^{n}\frac{(n+1-\alpha)(2n+2+\alpha)}{(n+1+\alpha)}x_{q,\alpha}x_{q,n+1-\alpha}\\
&+(1-q\tau_q)^2\widetilde{\mathbf{F}}_{\phi,n}\Big(\frac{\tau_q}{1-q\tau_q}, \frac{x_{q,1}}{1-q\tau_q},\dots,\frac{x_{q,n}}{1-q\tau_q},
\frac{y_q}{1-q\tau_q},\frac{s_q}{1-q\tau_q},\xi_q-\frac{2{{\rm i}\pi}q}{1-q\tau_q}y_qs_q-\frac{{{\rm i}\pi}q}{1-q\tau_q}(n+1)\mathcal{B}_{n+1,2}^q\Big)\\
&-(1-q\tau)^2\mathbf{Q}\Big(\frac{\tau_q}{1-q\tau_q}, \frac{x_{q,1}}{1-q\tau_q},\dots,\frac{x_{q,n}}{1-q\tau_q},
\frac{y_q}{1-q\tau_q},\frac{s_q}{1-q\tau_q},\xi_q-\frac{2{{\rm i}\pi}q}{1-q\tau_q}y_qs_q-\frac{{{\rm i}\pi}q}{1-q\tau_q}(n+1)\mathcal{B}_{n+1,2}^q\Big)\\
&-\frac{{{\rm i}\pi}q}{1-q\tau_q}\big(y_qs_q\big)^2-\frac{{{\rm i}\pi}q}{1-q\tau_q}y_qs_q
\sum_{\alpha=1}^{n}\frac{(n+1-\alpha)(2n+2+\alpha)}{n+1+\alpha}x_{q,n+1-\alpha}x_{q,\alpha}\\
&-\frac{{{\rm i}\pi}q}{1-q\tau_q}(n+1)\sum_{\alpha,\beta=1}^{n}\frac{(n+1-\alpha)(n+1-\beta)}{(n+1+\alpha)}x_{q,\alpha}x_{q,\beta}x_{q,n+1-\alpha}x_{q,n+1-\beta}.
\end{align*}
Now, using expression (\ref{Jacobi-A-An}) for $\widetilde{\mathbf{F}}_{\phi,n}$, $q$-Eisenstein series (\ref{G-q}), Weierstrass function $\K_q$ (\ref{K-q})-(\ref{K-q-derivative})  and bearing in mind the homogeneity properties  (\ref{Bell-R-q2})  of the functions $\mathcal{B}_{\mu,k}$ and $\mathcal{R}_{\mu,k}$, we can conclude, after carrying out some computation,  that
\begin{align*}
&\widetilde{\mathbf{F}}_{\phi_q,n}\Big(\tau_q, x_{q,1},\dots,x_{q,n},y_q,s_q,\xi_q\Big)=\frac{\xi_q^2}{4{\rm i}\pi}\tau_q+\xi_qy_qs_q
+\frac{1}{2}\xi_q\sum_{\alpha=1}^{n}\frac{(n+1-\alpha)(2n+2+\alpha)}{(n+1+\alpha)}x_{q,\alpha}x_{q,n+1-\alpha}\\
&+\frac{1}{2}(y_qs_q)^2G_{q,2}(\tau_q)
+\frac{1}{2}y_qs_qG_{q,2}(\tau_q)\sum_{\alpha=1}^{n}\frac{(n+1-\alpha)(2n+2+\alpha)}{n+1+\alpha}x_{q,\alpha}x_{q,n+1-\alpha}\\
&+\frac{n+1}{2}G_{q,2}(\tau_q)\sum_{\alpha=1}^{n}\sum_{\beta=1}^{n}\frac{(n+1-\alpha)(n+1-\beta)}{(n+1+\alpha)}x_{q,\alpha}x_{q,\beta}x_{q,n+1-\alpha}x_{q,n+1-\beta}\\
&-y_q^2\K_q(s_q)+\frac{n+1}{2}y_q\sum_{k=2}^{n+1}\frac{1}{k!}\mathcal{B}^q_{n+1,k}\K_q^{(k-1)}(-s_q)
+\frac{n+1}{2}y_q\sum_{\alpha=1}^{n}\frac{(n+1-\alpha)}{(n+1+\alpha)}x_{q,n+1-\alpha}\mathbf{L}_{\alpha}^{\phi_q}\big[\K_q\big](-s_q)\\
&-\frac{n+1}{2}y_q\sum_{\alpha=1}^{n}\left(\frac{n+1-\alpha}{n+1+\alpha}+\frac{n+1-\alpha}{n+1}\right)x_{q,n+1-\alpha}\mathcal{R}_{\alpha,1}^q
+\frac{(n+1)^2}{2}\sum_{\alpha=1}^{n}\sum_{k=2}^{n+1}\frac{(n+1-\alpha)}{k(n+1+\alpha)}x_{q,n+1-\alpha}\mathcal{B}^q_{n+1,k}\mathcal{R}^q_{\alpha+k-1,k}\\
&+\frac{(n+1)^2}{2}\sum_{\alpha=1}^{n}\sum_{k=1, k+\alpha\geq 5}^{n+1}\Bigg\{
\left[(-1)^k\frac{(n+1-\alpha)}{k(n+1+\alpha)}x_{q,n+1-\alpha}\mathcal{B}^q_{n+1,k}\right]\bigg[\sum_{\ell=1, k+\ell\geq 5}^{\alpha}\binom{k+\ell-1}{\ell}
\mathcal{B}^q_{\alpha,\ell}\frac{G_{q,k+\ell-1}(\tau_q)}{k+\ell-1}\bigg]\Bigg\}\\
&+\frac{n+1}{2}y_q\sum_{\ell=4}^{n}\frac{1}{\ell(\ell+1)}\mathcal{B}^q_{n+1,\ell+1}G_{q,\ell}(\tau_q)
+\frac{1}{2}y_q^2\log(x_{q,1}y_q)+\frac{\log(-1)}{2}y_q^2-\frac{3}{4}\frac{n+2}{n+1}y_q^2.
\end{align*}
The two  special cases $n=0$ and $n=1$ read as 
\begin{equation}\label{Jacobi-A-A0-q}
\widetilde{\mathbf{F}}_{\phi_q,0}=\frac{\xi_q^2}{4{\rm i}\pi}\tau_q+\xi_qy_qs_q+y_q^2\log(y_q)+\frac{1}{2}(y_qs_q)^2G_{q,2}(\tau_q)
-y_q^2\K_q(s_q|\tau_q)+\big(\log(-1)-3/2\big)y_q^2;
\end{equation}
\begin{equation}\label{Jacobi-A-A1-q}
\begin{split}
\widetilde{\mathbf{F}}_{\phi_q,1}&=\frac{\xi_q^2}{4{\rm i}\pi}\tau_q+\xi_q(y_qs_q+x_q^2)+\frac{1}{2}y_q^2\log(x_qy_q)
+\frac{1}{2}\big(y_q^2s_q^2+2x_q^2y_qs_q+x_q^4\big)G_{q,2}(\tau_q)\\
&\quad -y_q^2\K_q(s_q|\tau_q)-x_q^2y_q\K_q'(s_q|\tau_q)+\frac{\log(-1)}{2}y_q^2-\frac{7}{4}y_q^2.
\end{split}
\end{equation}
Moreover, as in Theorem \ref{PDE1}, the WDVV equations for the deformed  prepotential $\widetilde{\mathbf{F}}_{\phi_q,0}$ are equivalent to the following $q$-counterpart of PDEs (\ref{psi-diff eq})-(\ref{psi-heat}):
\begin{align*}
&\psi_q^{(4)}+6\big(\psi_q''\big)^2+12G_{q,2}(\tau_q)\psi_q''+6\big(G_{q,2}^2(\tau_q)-5G_{q,4}(\tau_q)\big)=0;\\
&(4{\rm{i}}\pi)\partial_{\tau_q}\psi_q=\psi_q''+\big(\psi_q'\big)^2+3G_{q,2}(\tau_q),
\end{align*}
where $\psi_q$ is  the function
\begin{equation*}
\psi_q(v):=\K_q(v|\tau_q)-\frac{1}{2}G_{q,2}(\tau_q)v^2,
\end{equation*}
and $G_{q,2}$ and $\K_q$ are respectively the $q$-deformed Eisenstein series (\ref{G-q}) of weight 2  and the Weierstrass function (\ref{K-q}).


\end{document}